\documentclass{aa}  

\usepackage{graphicx}
\usepackage{txfonts}
\usepackage{lipsum}
\usepackage{subcaption}         
\usepackage{lscape}             
\usepackage{placeins}           
                                
\newcommand{\xmm}{{\it XMM-Newton}}
\newcommand{\sw}{{\it Swift}}
\newcommand{\ch}{{\it Chandra}}
\newcommand{\kms}{km\,s$^{-1}$}
\newcommand{\gc}{$\gamma$\,Cas}
\newcommand{\sgc}{SS\,397}
\newcommand{\vs}{HD\,162718}
\newcommand{\cl}{NGC 6649 9}

\begin{document}

   \title{Long-term investigation of \gc\ analogs}

   \author{Ya\"el Naz\'e\inst{1}\fnmsep\thanks{F.R.S.-FNRS Senior Research Associate}
     \and Gregor Rauw\inst{1} \and Robbie Webbe\inst{2} \and Myron A. Smith\inst{3} \and Christian Motch\inst{4}
        }

   \institute{Groupe d'Astrophysique des Hautes Energies, STAR, Universit\'e de Li\`ege, B5c, All\'ee du 6 Ao\^ut 19c, B-4000 Li\`ege, Belgium
              \email{ynaze@uliege.be}
   \and
            Institute for Research in Astrophysics and Planetology, CNRS, Toulouse, 31400, France
   \and
            Catholic University of America, 620 Michigan St., NE, Washington, DC, 20064, USA
    \and
            Universit\'e de Strasbourg, CNRS, Observatoire astronomique de Strasbourg, UMR 7550, F-67000 Strasbourg, France
  }

  \abstract
   {The subcategory of \gc\ analogs gathers Be stars with bright and hard X-ray emission.}
   {This paper aims to enlarge the sample of \gc\ analogs studied on long timescales. Long-term variations are expected in such objects for two reasons. First, their Be disk builds and dissipates, leading to changes in optical emission lines, broad-band photometry and possibly even in the X-ray spectrum. Second, such stars are suspected long-period binaries hence velocity variations can be expected, helping to characterize the companions. } 
   {Seven targets are analysed in this paper: five of them benefit from a spectroscopic monitoring in the visible (ESO, TIGRE, and amateur data) and three of them have been repeatedly observed at X-ray wavelengths (using \xmm, \ch, and Swift). Broad-band photometric data, sampling both long and short timescales, are also examined in support of the optical and X-ray spectroscopy.}
   {We confirm the binary status of five targets (HD\,44458, HD\,110432, HD\,119682, HD\,161103, and \vs) and propose first orbital solutions for all of them (they remain preliminary for two cases). Their long periods (59--322\,d) and small velocity amplitudes ($K\sim5$\,\kms) imply low-mass ($\sim1$\,M$_{\odot}$) companions. With such properties, these systems appear similar to other Be binaries in general and \gc\ analogs in particular. They also agree with expectations from binary interaction models. In parallel, variations of the X-ray flux are detected in all three targets with a large dataset of X-ray observations. For \cl\ and \vs, these changes are modest (a factor of three) and uncorrelated to simultaneous optical broad-band photometry (which remains rather stable). In contrast, \sgc\ varies by nearly one dex and the largest and best monitored X-ray changes correlate well with optical variations. At minimum flux, \sgc\ keeps a hard X-ray spectrum despite a nearly normal $L_X/L_{\rm BOL}$ ratio, which has not been seen yet among \gc\ analogs. Finally, the photometric behaviours on short timescales of HD\,161103, \sgc, and \cl\ appear linked to broad frequency groups, as typically found for Be stars (including \gc\ analogs). The frequency spectrum of \vs\ displays a complex mix of (isolated) periodicities with the main one at 6.658\,d$^{-1}$. This target is thus one of the rare \gc\ analogs to display a strong high-frequency signal typical of $\beta$\,Cep activity. }
   {}

   \keywords{binaries: spectroscopic -- Stars: massive -- X-rays: stars}

   \maketitle
 \nolinenumbers

   \section{Introduction}

   Among massive stars, Oe and Be stars gather objects displaying Balmer emission lines. In the classical cases, these lines arise from a disk in Keplerian rotation surrounding the (fast-rotating) star and are fed regularly by its mass ejections \citep{riv13,lab25}. Over the last two decades, a non-negligible fraction ($10-15$\%) of early-type Oe and Be stars was found to display atypical X-ray emissions. The recorded peculiarities are fourfold: brightness tens to hundreds of times larger than usually seen, very high plasma temperatures, large short-term variations, and the presence of (iron) fluorescence lines. These objects are named \gc\ analogs, after the first discovered case (for a review, see \citealt{smi16}).
   
   To better understand these stars, long-term monitoring plays a key role. First, it can reveal binarity signatures. Indeed, growing evidence indicates that most OBe stars are products of binary interactions (for a review, see \citealt{lab25b}). Binary evolution simulations have shown that these systems should feature low-mass companions on long orbital periods (from tens to a few hundreds of days; see, e.g. \citealt{shao14}). The binarity can therefore be assessed from the detection of slow and shallow velocity shifts. While detecting such shifts clearly represents a challenge for these notoriously variable stars, several studies were successful in this endeavour thanks to high-quality long-term spectroscopic monitoring, including for several \gc\ analogs: \gc\ \citep{har00,nem12,smi12}, $\pi$\,Aqr \citep{bjo02}, $\zeta$\,Tau \citep{ada05,rud09}, and a few additional ones \citep{naz22}. Enlarging this small group would provide better constraints for evolutionary models.
   
   Second, the disks of OBe stars, including \gc\ analogs, are dynamic structures. They build and dissipate over months or years, leading to photometric and spectroscopic variations. In parallel, such disks are also expected to play a role in the generation of X-rays, directly (by magnetically interacting with the star) or indirectly (by providing material to an accreting companion). Thus, a correlation may be expected between optical and X-ray emissions. A few \gc\ systems have been studied to assess this connection. Overall, no link could be found between X-ray variations and orbital phase or between X-rays and H$\alpha$ emission strength \citep{rau22,naz19,naz22evol,naz24}. However, changes in broad-band optical photometry have been found to correlate with X-ray flux variations \citep{mot15,rau18,naz22evol}. The most spectacular case is HD\,45314, the hottest \gc\ analog, which experienced a substantial reduction in its hard X-rays when its disk was dissipating \citep{rau18}. In contrast, the X-ray emission of $\pi$\,Aqr was only slightly less bright and hard at disk minimum, and X-rays remained within the \gc\ criteria even when the H$\alpha$ emission was very weak for HD\,119682 and V767\,Cen \citep{naz22evol}. However, few objects have been studied over long intervals, impairing an in-depth understanding of the relationships between X-ray emission and disk. The sample clearly needs to be enlarged.
   
   This paper is a step towards these goals, with seven more \gc\ analogs studied on long timescales. Section 2 presents these targets, together with the data and methods used to analyse them. Section 3 reports the results of these analyses, while Section 4 summarizes our results.

\section{Sample and data}

\subsection{The \gc\ targets}

For our long-term spectroscopic investigation we revisit the five \gc\ analogs identified as binary candidates through radial velocity (RV) shifts in \citet{naz22}: HD\,44458 (FR\,CMa), HD\,110432 (BZ\,Cru), HD\,119682, HD\,161103 (V3892\,Sgr), and \vs\ (V771\,Sgr). For long-term X-ray monitoring, we focus on three \gc\ analogs with new or unpublished data: \cl, \sgc, and again \vs. All these stars were identified as \gc\ analogs several years ago \citep{smi06,rak06,lop06,mot07,naz18,naz20}.

\begin{table}
  \begin{center}
  \scriptsize
  \caption{Target properties.
 \label{target}}
  \begin{tabular}{lcccc}
    \hline
    Name & sp. type & $d$(pc) & $E(B-V)$ & $\log(L_{\rm BOL}/L_{\odot})$ \\
    \hline
HD\,44458  & B1.5IVe &  534$\pm$16 & 0.10 & 4.23$\pm$0.03 \\
HD\,110432 & B0.5IVpe&  444$\pm$17 & 0.29 & 4.56$\pm$0.03 \\
HD\,119682 & B0Ve    & 1582$\pm$65 & 0.22 & 4.60$\pm$0.04 \\
HD\,161103 & B0.5IVe & 1216$\pm$41 & 0.59 & 4.28$\pm$0.03 \\
\vs\       & B2Ve    &             & 0.77 &               \\
\sgc\      & B0.5Ve  & 1500$\pm$200& 1.76 & 4.64$\pm$0.12 \\
\cl\       & B0Ve    & 1943$\pm$114& 1.37 & 4.66$\pm$0.05 \\
    \hline
  \end{tabular}
  \end{center}
\end{table}

Table \ref{target} lists their main properties. Spectral types were taken from the literature. For \vs, the usually quoted type is uncertain (B3/5) but since spectroscopy was at hand (see below), we could re-derive it. No He\,{\sc ii} lines are present and Si\,{\sc iv}$\lambda$4089 appears extremely weak (possibly blended with O\,{\sc ii}), which rules out the earliest B-types. On the other hand, Mg\,{\sc ii}$\lambda$4481\AA\ remains small while Si\,{\sc ii}$\lambda$4128--30\AA\ and Si\,{\sc iii}$\lambda$4552\AA\ are present. The profile of Ti\,{\sc ii} and Fe\,{\sc ii} emissions appear double-peaked but not shell-like, suggesting a high but not equatorial inclination (60--75$^{\circ}$?). In addition, the presence of Stark broadened wings in Balmer lines and He\,{\sc i} lines suggests a main sequence luminosity class. The spectral type of \vs\ should then be revised to B(2$\pm$1)Ve.

Distances were taken from \citet{bai21} if the \textit{Gaia}-DR3 data were of good quality (i.e.  the parallax error is small compared to its value, or $R_{plx}=\pi/\sigma_\pi>5$, and its re-normalized unit weight error is small, $RUWE<1.4$) . In addition, a distance of $1.5\pm0.2$\,kpc was proposed for \sgc\ by \citet{riq12}. The distance of \vs, unfortunately, remains unknown. Its derived \textit{Gaia} parallax truly seems unreliable as it varied by a factor of two between the DR2 and DR3 versions of the \textit{Gaia }catalogue. 

Reddenings were estimated from dust maps for all but two targets, using the position and \textit{Gaia }distance of the targets (tools {\sc stilism} and {\sc g-tomo}\footnote{https://explore-platform.eu/ - this tool provides $A_V$, which were converted to $E(B-V)$ using the typical value $R_V=3.1$.}). The reddening of \sgc\ was estimated from its spectral type (B0.5V) and its photometry \citep{riq12}, leading to $E(B-V)=1.76$. Considering the revised B2 spectral type, the reddening of \vs\ was estimated from its Simbad photometry to $E(B-V)=0.77$, only slightly below the old value of 0.89 from \citet{koz85}. In the dust maps, such a reddening appears in line with that expected for the DR3 distance, but this combination would lead to a bolometric luminosity more than one dex too high.

The bolometric luminosities $\log(L_{\rm BOL}/L_{\odot})$ were then derived in the usual way from the V-magnitudes listed in Simbad (except for \sgc, for which $V=12.3$ from \citealt{riq12} was used). They appear in line with those expected for the spectral types. For \cl\ an alternative spectral type of B1III was also proposed \citep{alo20}, which would lead to slightly reduced luminosity ($\log[L_{\rm BOL}/L_{\odot}]=4.44$) but still in line with that expected for this spectral type and luminosity class.

\subsection{X-ray observations}

The three X-ray targets were observed under various conditions and by various facilities. \vs\ was observed twice by \xmm, the first time as the main target of the observation and the second time as a field target. It was also covered by several \sw\ exposures. Being close to the supernova remnant G21.5-00.9, \sgc\ and \cl\ have been observed many times by \xmm\ and \ch\ (\sw\ data of the same supernova proved to be unreliable for these stars, even after combining several exposures, due to the lack of correct centring). Not all datasets resulted in a spectral extraction, however, as the sources appear off-axis (hence they may sometimes fall in a gap or outside the field of view) and variable (hence they are sometimes too faint for deriving a spectrum). Table \ref{journal} provides the list of datasets used in this paper for each target.

The \xmm\ data were downloaded from the European Space Agency archives\footnote{https://www.cosmos.esa.int/web/xmm-newton/xsa} and processed locally using Science Analysis Software {\sc SAS} v21.0.0. After pipeline processing, the event files were filtered to keep only the high-quality data ({\sc pattern} 0--12 for MOS and 0--4 with null flag for pn). Light curves above 10\,keV were built to assess the presence of background proton flares and, when they are present, the event files were further filtered to eliminate the time intervals corresponding to these flares. Spectra and their response matrices were also built using nearby regions devoid of sources as background reference. A binning was applied to get at least a signal-to-noise ratio of 3 and an oversampling factor of maximum 5. All EPIC spectra were fitted simultaneously. Exposure times were typically in the range of 8--32\,ks. 

The \ch\ data were downloaded from National Aeronautics and Space Administration archives\footnote{https://cda.harvard.edu/chaser/} and re-processed locally if their processing version was old (ASCDSVER$<$8.4.2). Spectral extraction was done in a standard way using {\sc specextract} and a nearby empty region as background reference. Weighted response matrices were created by the same task. A binning was also applied to get at least 10 counts per spectral bin. We note that the oldest data (ObsID=1230, 158, 159, 160, 161, 165) required the input of a specific energy range (0.3--9.8\,keV with a step of 0.01\,keV) to be processed. Exposure times were typically in the range of 7--15\,ks.

The \sw\ data of \vs\ consist of multiple short exposures, usually too short to gather enough counts for individual analyses. However, the \sw\ data cover basically three epochs: September 2012, February 2018, and May 2025. Since we are interested in long-term variations, datasets taken a few days apart can reasonably be combined. Using the \sw\ online tool\footnote{https://www.swift.ac.uk/user\_objects/ - the selection was done using the target ID field, as the three epochs used different target IDs (43742/9 in Sept. 2012, 10533 in Feb. 2018, and 89907 in May 2025).}, individual exposures in each epoch were merged to get a single spectrum and associated response matrices. The combined exposure times typically reach $\sim$2\,ks.

X-ray spectral fitting was done using {\sc Xspec} v12.11.1 with solar abundances from \citet{asp09}. The X-ray spectra were fitted by absorbed optically-thin thermal plasma models, as is adequate for \gc\ analogs: $phabs(ISM)\times phabs \times apec$. The first absorption was fixed to the interstellar medium (ISM) value, derived from the reddening $E(B-V)$ (Table \ref{target}) and the relationship of \citet{gud12}. It amounts to 0.47, 0.84, and $1.08\times 10^{22}$\,cm$^{-2}$ for \vs, \cl, and \sgc, respectively. The second absorption represents potential circumstellar absorption. A single emission model was sufficient to fit the recorded spectra. Most spectra of \sgc\ and \cl\ suffered from a low number of counts, which sometimes rendered the fitting erratic if there were too many free parameters. We therefore decided to partially constrain the spectral shapes by fixing temperature but letting absorption vary. To this end, the spectra were merged (separately for \xmm\ and \ch\ data) and then fitted by the chosen model. Best-fit 'average' temperatures of 8.5 and 14\,keV were derived for \sgc\ and \cl, respectively. Individual fits were then performed with the temperatures fixed to these values. For \vs, the temperature found in the first \xmm\ observation was used when fitting the lower quality \sw\ data (using the larger value of the second \xmm\ exposure does not significantly change the results). Fitting results can be found in Table \ref{journal}. We have compared the derived fluxes with the aperture-photometry broad-band fluxes of \sgc\ and \cl\ listed in the CSC v2.1.1 catalogue (queried using task {\sc search\_csc}) and the broad-band count rates of the same objects in the 4XMM-DR14 catalogue. The trends seen in these catalogue values (based on counts) are similar to those derived from our more detailed spectral fitting. This demonstrates the consistency of the results.

\subsection{Optical spectroscopy}
Following our preliminary investigation in 2020 and 2021 \citep{naz22}, four of the five binary candidates continued to be monitored every two weeks in 2022 with the Ultraviolet and Visual Echelle Spectrograph (UVES, \citealt{dek00}), installed on the second Unit Telescope (UT2) at Cerro Paranal of the European Southern Observatory (ESO), for our program 109.22V6. UVES was used in dichroic mode, allowing simultaneous access to the 3300--4560\AA\ and 4730--6830\AA\ regions with resolutions of 70\,000--100\,000. As before, the new UVES spectra, taken in service mode, were not always checked for saturation so that the H$\alpha$ emission line is sometimes unusable. Lower resolution, but broader band, spectra were also obtained for us at ESO with the Xshooter instrument \citep{ver11} as part of our program ObsID 105.204D. The ESO archive\footnote{https://archive.eso.org/} also provided additional spectra, from the same or other ESO facilities (Espresso), for HD\,110432 and HD\,119682. All of these data, reduced in a standard way using the ESO pipeline, can be downloaded from the ESO archive website. Signal-to noise ratios were in the range of 100 to 400.

The last binary candidate, HD\,44458, continued to be monitored with the fully robotic 1.2 m TIGRE telescope located in central Mexico \citep{sch14}, equipped with the Heidelberg Extended Range Optical Spectrograph (HEROS) echelle spectrograph. This instrument covers the bands 3800--5700\AA\ and 5800--8800\AA\ with a resolution of 20\,000. Data reduction was performed with the dedicated TIGRE/HEROS reduction pipeline \citep{mit10}.

The five targets were also observed by amateur astronomers, whose spectra are available in the Be star spectra (BeSS) database\footnote{http://basebe.obspm.fr/basebe/} \citep{nei11}. However, we did not use the lowest ($R<5000$) resolution data. In addition, all spectra covered the H$\alpha$ line, but only a few also covered H$\beta$. The typical S/N amounted to 50 (range: 20--150). For \vs, one of the four BeSS spectra, the one taken in 2023, had to be discarded for RV measurements as it clearly displayed a shift in the interstellar feature at 6613\AA\ (see also Fig. \ref{lineprof}).

Telluric lines near H$\alpha$ were corrected using the template of \citet{hin00} and normalization was performed using low-order splines in selected spectral windows. Velocities of Be stars are notoriously difficult to derive because of the shallow photospheric absorptions (if any). The strong emission lines are instead analysed, but this requires some caution. As in \citet{naz22}, RVs were measured using three techniques: first order moment, mirror method, and double-Gaussian method. The first order moment $M_1=\sum (F_i-1) \times v_i/ \sum (F_i-1)$ (with $v_i$ and $F_i$ the velocity and normalized flux of the ith spectral bin, respectively) derives the line centroid. Equivalent widths (EWs) of these emission lines were estimated from the zeroth order moment ($\sum (F_i-1)$) in the same velocity ranges as those used for the centroiding. As $M_1$ probes the whole line profile, it is sensitive to its variations, which are quite common in Be stars, especially in the line cores. The two other methods therefore rely instead on the line wings, which are closely connected to the inner disk of the Be star, hence should provide more reliable values for the stellar velocity. However, they are not immune to global profile changes (see below). The mirror method \citep{nem12} compares the blue wing to the mirrored red wing, for several velocity shifts. The line velocity corresponds to the shift that yields the least difference between wings. The double-Gaussian method \citep{smi12} correlates the line profile with a function composed of two Gaussians of identical widths but of reversed amplitudes and centre velocities (chosen to sample the line profiles near half amplitude). The correlation is calculated for several shifts of the function; the RV then corresponds to the shift at which the correlation reaches zero. Tables \ref{rv}, \ref{rv110432},  \ref{rv44458}, and  \ref{rv119682} provide the EWs and the RVs (measured using the double-Gaussian method - the mirror method provides similar results, while the first-order moments display more scatter due to the core contamination). 

For HD\,119682, the emission was very much reduced in 2019--2021 \citep{naz22evol}, but the disk seems to have been rebuilt since then. Thus, RV measurements on H$\alpha$ emissions could only be performed on data from 2022 onwards. As this limits the number of data points, we have also considered analysing other lines. Many spectral lines, for example from He\,{\sc i}, are affected by the disk emission, hence their profiles change much between epochs, rendering them useless. However, a set of small absorptions from N\,{\sc ii}, O\,{\sc ii}, C\,{\sc ii}, and S\,{\sc iii} over 4233--4324\AA\ appear stable. To derive RVs, the spectra in this range were cross-correlated (1) with a synthetic TLUSTY spectrum interpolated from BSTAR2006 database \citep{lan07} for $T_{eff}=28$\,kK, $\log(g)=4$, and $v\sin(i)=200$\,\kms; and (2) with the XShooter spectrum. Comparing the methods revealed a global agreement between their results. Global differences of --15\,\kms, on average, between RVs obtained from the XShooter correlation and those from the TLUSTY correlation are measured, and the difference amounts to +15\,\kms\ between H$\alpha$ velocities and the same reference. However, the scatter is about 5\,\kms. This is much higher than for the strong emission lines analysed for other stars. This lower quality is explained by the broad $\chi^2$ distribution obtained when comparing the spectra with a reference, mostly due to the faintness of the lines. Table \ref{rv119682} provides the RVs measured with respect to the synthetic spectrum in the column entitled `blue'.

\subsection{Optical photometry}
As we wished to examine the correlation between optical and X-ray brightness, we searched whether broad-band photometry was available for the three X-ray targets. All three were observed in the All-Sky Automated Survey (ASAS\footnote{https://www.astrouw.edu.pl/asas/?page=aasc}, \citealt{poj97}) and All-Sky Automated Survey for SuperNovae (ASAS-SN\footnote{https://asas-sn.osu.edu/}, \citealt{shap14}). For ASAS, only the highest quality data (grade `A') were kept. \vs\ was also observed by the optical monitor camera (OMC, \citealt{gim97})\footnote{https://sdc.cab.inta-csic.es/omc/} onboard Integral. If taken on the same day and of the highest quality (flag `0'), the OMC data were averaged since we were interested in long-term variations. In addition, \sgc\ and \cl\ were also observed by the Zwicky transient facility (ZTF\footnote{https://irsa.ipac.caltech.edu/cgi-bin/Gator/nph-dd}, \citealt{bel19}) and Asteroid Terrestrial-impact Last Alert System (ATLAS\footnote{https://fallingstar-data.com/forcedphot/}, \citealt{hei18}). Finally, data are available from the Kamogata/Kiso/Kyoto Wide-field Survey (KWS, \citealt{mae14})\footnote{http://kws.cetus-net.org/$\sim$maehara/VSdata.py} but the error bars and data dispersions for \cl\ and \sgc\ are too large to be usable: only the \vs\ photometry was used.

The short-term photometric behaviours of \gc\ analogs were examined in \citet{naz20tess}, but no Transiting Exoplanet Survey Satellite ({\it TESS}) data were available at the time for HD\,161103, \vs, \sgc, and \cl. We therefore decided to complement the previous study with the new photometric data. {\it TESS} observed the first two stars in Sectors 91 and 92 and the last two stars in Sector 80. Targets HD\,161103, \vs, and \cl\ were preselected. Therefore, their light curves with 2-min cadence are directly available from the MAST portal\footnote{https://mast.stsci.edu}. In this case, simple aperture photometry time series and conditioned light curves are available. We selected the latter ones as they appear of better quality (smaller long-term trends) and include additional corrections (crowding, etc.). For \sgc, we extracted the light curve from image cutouts of 51$\times$51 pixels using aperture photometry performed with the Python package Lightkurve\footnote{See https://lightkurve.github.io/lightkurve/ - in particular, we selected data with high quality thanks to the option quality\_bitmask=`hard' in task {\sc search.tesscut}.}. The background mask was defined by pixels with fluxes below the median flux (i.e. below the null threshold). The background was then estimated in that region using either a principal component analysis with five components or a simple median. The latter method yielded larger long-term trends; hence we kept the light curve extracted with the former approach. 

\begin{figure}
  \begin{center}
    \includegraphics[width=8cm]{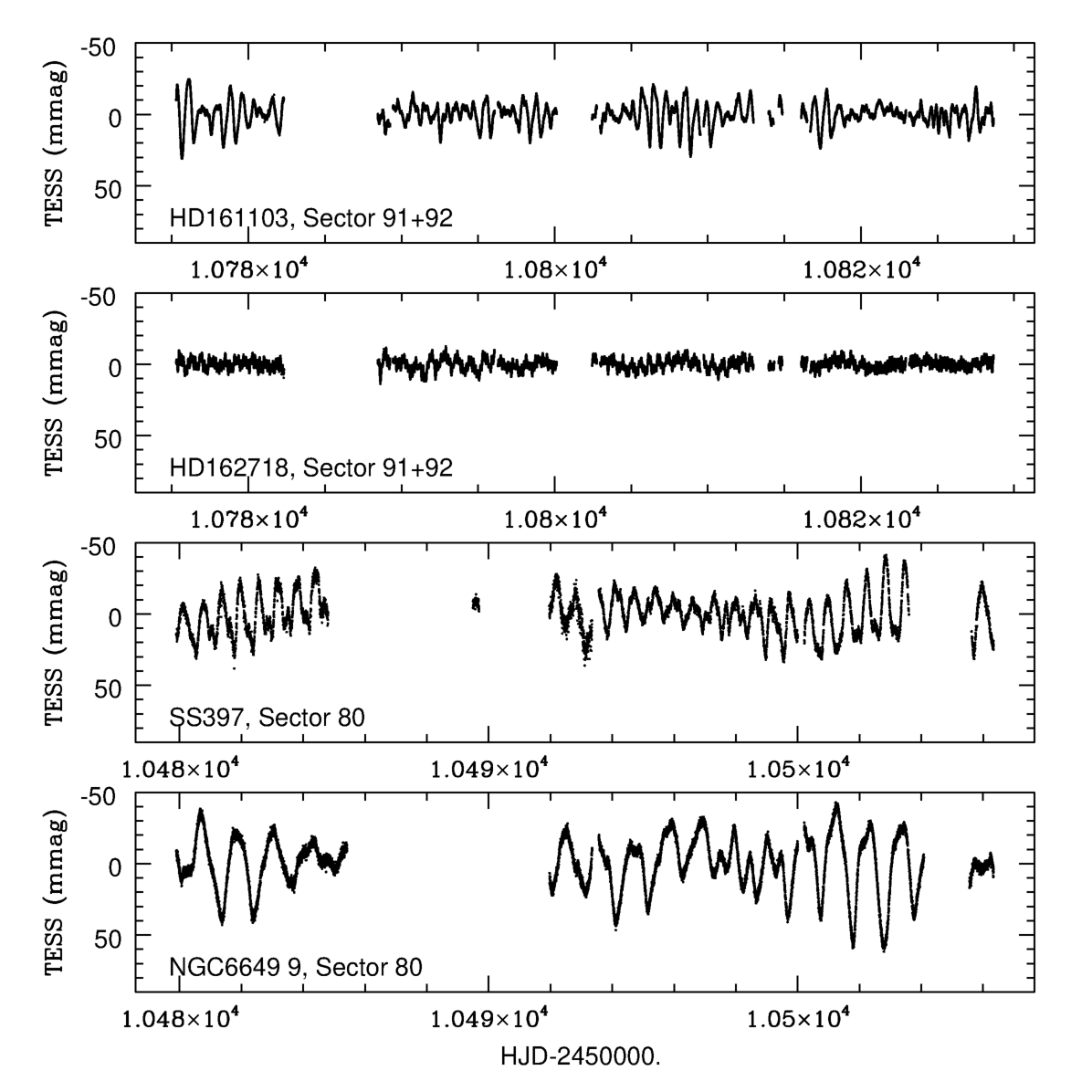}
    \includegraphics[width=8cm]{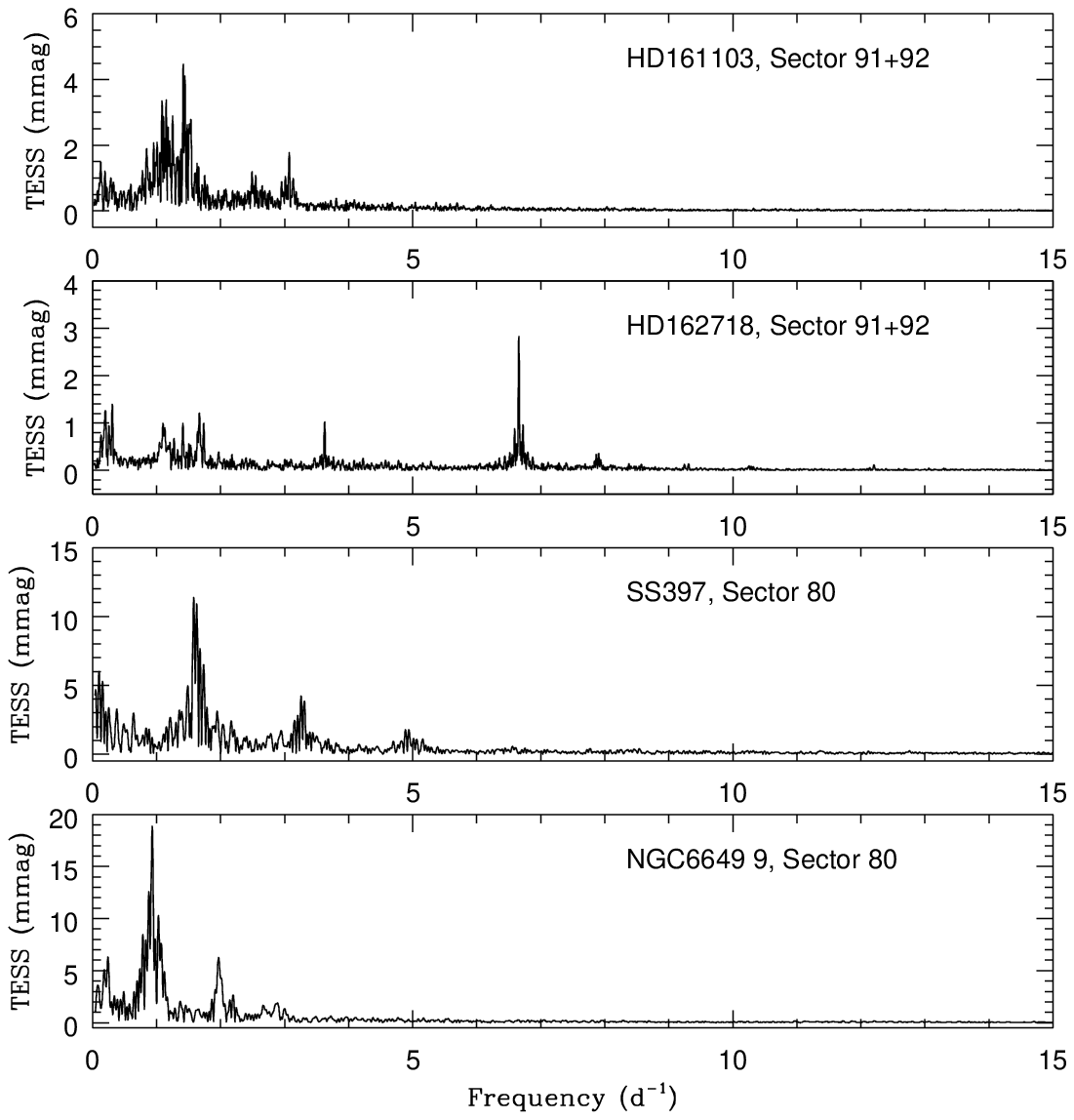}
  \end{center}  
  \caption{{\it Top panels:} {\it TESS} light curves for four \gc\ analogs. {\it Bottom panels:} Frequency spectra of these light curves. \label{tess}}
\end{figure}

The {\it TESS} pixels are large (21\arcsec) and the photometry extraction is made over several pixels, so contamination may be a problem. Fortunately, the four targets appear to be rather isolated. In the \textit{Gaia} DR3 catalogue, the second brightest object within 1\arcmin\ of HD\,161103, \vs, and \sgc\ has at least a difference $\Delta(G)$=3\,mag compared to our targets. For \cl, an object with similar brightness lies 0.9\arcmin\ away, but this is far enough away to limit the contamination. The {\it TESS} fluxes were then converted into magnitudes and the average magnitude was subtracted from each individual light curve. The light curves from consecutive sectors were also combined. Finally, a period search was done on all light curves to assess their frequency content \citep{hmm,gos01}.

\section{Results and discussion}
\subsection{TESS light curves}
\citet{naz20tess} reported on the short-term behaviours of 15 \gc\ analogs. They were found to be very similar to those of other Be stars with the same spectral types. Indeed, the frequency spectra often display broad frequency groups and/or narrow isolated peaks superimposed on red noise. Only a few cases harbour significant peaks above 5\,d$^{-1}$. 

The top panels in Fig. \ref{tess} show the {\it TESS} light curves of HD\,161103, \vs, \sgc, and \cl. The photometric variations appear very different from one target to the other, but the frequency spectra, shown in the bottom panels of Fig. \ref{tess}, actually display the typical features found in Be stars and \gc\ analogs \citep{naz20tess}. First, no significant signal is detected at very high frequencies (>15\,d$^{-1}$). Second, frequency groups are present, often with lower-amplitude harmonics and a faint group at very low frequencies, a usual group configuration for Be stars. The main groups are located at 1.4, 1.6, and 0.95\,d$^{-1}$ for HD\,161103, \sgc, and \cl, respectively. HD\,161103 has a second group close to the first one, at 1.1\,d$^{-1}$.

Finally, the frequency spectrum of the least variable target, \vs, appears complex. It is dominated by a 2.8\,mmag signal at 6.658$\pm$0.002\,d$^{-1}$, suggesting a $\beta$\,Cep behaviour. A subharmonic at a quarter of this frequency is also present. Another isolated signal appears near 3.626$\pm$0.002\,d$^{-1}$, a value slightly too low to be a subharmonic of the main signal. Additional low-amplitude peaks, apparently unrelated to the signals previously mentioned, are found near 12.2, 10.3, 9.3, 7.9, 1.4\,d$^{-1}$, plus 1.1 and 0.3\,d$^{-1}$ (appearing mostly in Sector 91) or 0.2\,d$^{-1}$ (mostly from Sector 92). It may be noted in this context that the profiles of the bluest (hence less affected by emissions) He\,{\sc i} lines recorded in our spectra seem to display some profile changes which might be linked to this pulsational activity, in addition to those linked to variations in the disk emissions. This renders the use of He\,{\sc i} lines difficult for RV determination.

\subsection{RV monitoring}
With the RVs at hand (see Appendix), we can now examine whether an orbital solution can be found for the five binary candidates identified in \citet{naz22}. The proposed orbital solutions are summarized in Table \ref{bin}.

\begin{table*}
  \begin{center}
  \scriptsize
  \caption{Binary solution of \gc\ analogs derived in this work and published in the literature.
 \label{bin}}
  \begin{tabular}{lcccccccccl}
    \hline
    Name & $P$(d) & $T_0$ & $\gamma$(\kms) &  $K$(\kms) & $f(m)$(M$_{\odot}$) & $M_1$(M$_{\odot}$) & $M_2$(M$_{\odot}$) & $L_X$(erg\,s$^{-1}$) & $\log(L_X/L_{\rm BOL})$ & Ref. \\
    \hline
HD\,44458  &  322.1$\pm$1.7 & 61\,132.969$\pm$2.050 & $19.7\pm0.2$  & 5.1$\pm$0.3 & $(4.42\pm0.78)\times10^{-3}$ & 10   & 0.8--1.3 & 8.90e31 & --6.02 & [8] \\
HD\,110432 &  215.1$\pm$3.0 & 61\,048.566$\pm$2.800 & $-0.7\pm0.3$  & 5.1$\pm$0.4 & $(2.95\pm0.70)\times10^{-3}$ & 15   & 0.9--1.4 & 2.80e32 & --5.69 & [9] \\
HD\,119682 &   59.3$\pm$0.3 & 60\,551.898$\pm$3.000 & $-20.8\pm1.0$ & 3.6$\pm$1.0 & $(2.86\pm2.39)\times10^{-4}$ & 17.7 & 0.5--0.7 & 7.70e32 & --5.56 & [9] \\
HD\,161103 &  224.7$\pm$2.9 & 61\,208.480$\pm$2.100 & $0.7\pm0.2$   & 5.9$\pm$0.3 & $(4.78\pm0.73)\times10^{-3}$ & 15   & 1.1--1.7 & 3.90e32 & --5.68 & [9] \\
\vs\       &  266.7$\pm$4.0 & 61\,092.425$\pm$2.670 & $-13.8\pm0.3$ & 5.3$\pm$0.3 & $(4.11\pm0.70)\times10^{-3}$ & 7.3  & 0.6--1.0 &         & --4.7  & [10] \\
\hline
\gc\         [1]   & 203.252$\pm$0.356  & & &4.01$\pm$0.09& $(1.36\pm0.09)\times10^{-3}$ & 17.7 & 0.77--1.22 & 1.10e33 & --5.29 & [9]   \\
V782\,Cas    [2]   & 122.0$\pm$1.5      & & & 5.2$\pm$0.9 & $(1.78\pm0.92)\times10^{-3}$ & 6    & 0.42--0.67 & 3.00e32 & --5.25 & [9]   \\
$\zeta$\,Tau [3,4] & 132.987$\pm$0.050  & & & 7.4$\pm$0.8 & $(5.58\pm1.81)\times10^{-3}$ & 11   & 0.93--1.49 & 3.90e31 & --5.74 & [4,11]\\
HD\,45995    [2]   & 103.1$\pm$1.0      & & & 6.7$\pm$0.4 & $(3.21\pm0.58)\times10^{-3}$ & 7.3  & 0.58--0.94 & 5.70e31 & --5.86 & [8]   \\
V750\,Ara    [5]   & 95.23$\pm$0.07     & & & 6.3$\pm$0.2 & $(2.40\pm0.20)\times10^{-3}$ & 11   & 0.69--1.10 & 3.20e32 & --5.57 & [4]   \\
V558\,Lyr    [2]   &  83.3$\pm$1.8      & & & 8.2$\pm$1.1 & $(4.76\pm1.91)\times10^{-3}$ & 5.4  & 0.55--0.89 & 7.60e31 & --5.37 & [8]   \\
V1256\,Cyg   [2]   & 126.6$\pm$2.0      & & & 5.5$\pm$0.7 & $(2.18\pm0.83)\times10^{-3}$ & 10   & 0.63--1.00 & 7.50e31 & --5.30 & [9]   \\
$\pi$\,Aqr   [6,7] &  84.1$\pm$0.02     & & & 8.1$\pm$1.4 & $(4.63\pm2.40)\times10^{-3}$ & 11   & 0.87--1.39 & 1.10e32 & --5.42 & [9,12]\\
V810\,Cas    [2]   &  75.8$\pm$0.7      & & & 6.4$\pm$0.7 & $(2.06\pm0.68)\times10^{-3}$ & 11   & 0.65--1.04 & 4.60e32 & --5.17 & [9]   \\
    \hline
  \end{tabular}
  \end{center}
  \tablefoot{$T_0$ is expressed as HJD-2\,400\,000 and corresponds to the time of conjunction with the Be star in front of its companion. The masses for the Be stars are typical ones for their spectral types\footnote{See https://www.pas.rochester.edu/$\sim$emamajek/EEM\_dwarf\_UBVIJHK\_colors\_Teff.txt}, while the secondary masses are derived from the mass function considering the primary mass and an inclination of $i=40-90^{\circ}$. For \vs, $\gamma$ is $-10.7\pm0.2$\,\kms\ for H$\beta$; for HD\,44458, it is 22.3$\pm$1.0\,\kms\ for H$\beta$ and $-5.8\pm1.0$\,\kms\ for the blue RVs. For HD\,44458 and HD\,119682, these orbital solutions must be considered preliminary. For published orbital parameters, references are given in the first column: [1] \citet{naz25}, [2] \citet{naz22}, [3] \citet{rud09}, [4] \citet{naz22Be}, [5] \citet{wan23}, [6] \citet{bjo02}, and [7] \citet{tsu23}. When two references are noted, the first one yields the period and the second one the orbital solution (notably the $K$). The last three columns provide the maximum X-ray luminosity known (after correction for absorption and in the 0.5--10.\,keV range) along with its references: [8] \citet{naz20}, [9] \citet{naz18}, [10] this work, [11] \citet{naz24}, and [12] \citet{naz22evol}.}
\end{table*}

For \vs, the optical spectra indicate strong emission at all times, with EWs of about --50\AA\ for H$\alpha$ (see also Fig. \ref{lineprof}). Line profiles of H$\alpha$ and H$\beta$ appear relatively stable outside of its core, facilitating RV determination. As already hinted in the old UVES dataset, the measured RVs clearly change over time (middle panel of Fig. \ref{lineprof}). A period search \citep{gos01} on the more numerous H$\beta$ velocities (from the double-Gaussian method) led to the detection of a peak at $P=266.7\pm4.0$\,d. When folded with this period, both ESO and BeSS RVs agree well, even though separated by about 16 orbital cycles and from different lines (H$\alpha$, H$\beta$). The RVs display a clear sinusoidal variation, indicating a circular orbit. Fixing the eccentricity to $e=0$, we derived the best-fit orbital solution listed in Table \ref{bin} and shown on the right panel of Fig. \ref{lineprof}. 

HD\,161103 also displays strong emissions, with H$\alpha$ EWs between --30 and --40\AA\ (see also Fig. \ref{l161103}). The H$\beta$ line reveals a change from a double-peaked profile with a stronger blue component in 2020--22 to an equal-peak configuration in 2023 and finally a red-peak-dominated profile in 2025. In addition, the photospheric absorption still seen at high blue and red velocities outside of the main emission appears to be totally filled at some epochs. These profile variations are probably responsible for the long-term trend detected in the RVs (middle panel of Fig. \ref{l161103}). However, a sinusoidal variation clearly appears superimposed on this decreasing trend. To highlight it, a best-fit of the trend was determined for each set of velocities ($M_1$, mirror, double-Gaussian) and then subtracted before a period search was performed. Fixing the eccentricity to $e=0$, we derived the best-fit orbital solution listed in Table \ref{bin} and shown on the right panel of Fig. \ref{l161103}. 

HD\,110432 was the focus of many amateur observations, in addition to several ESO campaigns. Its disk emission is strong but much more variable than for previous targets (see Fig. \ref{l110432}). Indeed, the H$\alpha$ EW goes from --30\AA\ in 2017 to a peak around --50\AA\ in early 2021 and then back to the original value in recent times. The overall line width follows the line strength evolution. Line profiles are complex, with a (skewed) triangular shape or several peaks. For H$\beta$, the profile appears to be dominated by a red peak in 2020 and then by a blue peak in 2023--24. These variations impact velocity determinations and, as for HD\,161103, long-term trends had to be taken out (see the middle panel of Fig. \ref{l110432}). This could only be done for data taken after HJD=2\,458\,900 as only a few observations are available before, which prohibits constraining trends. The period search then revealed a period of about 215\,d, with which the data fold well (see Table \ref{bin} and the right panels of Fig. \ref{l110432}). The best folding is obtained for ESO data taken between mid-2020 and 2022 (the velocity error being larger for amateur data). A continued monitoring would be interesting not only to confirm the orbital solution but also to follow the ongoing disk dissipation.

As for HD\,110432 but with a larger scatter, the emission strength of HD\,44458 reached a peak a few years ago, at the end of 2018, and has declined since then. The H$\alpha$ EW ranges between --20 and --40\AA\ (see Fig. \ref{l44458}). The profile itself appears rather stable, with a double-peaked shape, although the absorption at high red and blue velocities is sometimes filled. A period could be derived from the H$\alpha$ RVs and it provides good folding, but the H$\beta$ velocities show much more scatter (see Table \ref{bin} and the right panels of Fig. \ref{l44458}). Additional monitoring is definitely required to confirm this preliminary orbital solution. 

As mentioned in \citet{naz22evol} and in the previous section, the emission of HD\,119682 was greatly reduced in 2019--21. At that time, for H$\alpha$, the emission basically filled the photospheric absorption (i.e. $EW\sim0$). Without a clear absorption or emission profile, the RV could not be determined. For H$\beta$, the situation is slightly different with a clear absorption seen at that time but a varying absorption+emission profile observed more recently. Unfortunately, RVs determined on so different profiles cannot be compared. For Balmer lines, we therefore decided to evaluate RVs only on the H$\alpha$ emission from the recent spectra, although that limits the number of measurements and the temporal coverage. It may be noted that the amplitude range for the mirror method is small, to accommodate the varying emission strength, but this results in less precise RVs, hence we focus on the double-Gaussian method only in Fig. \ref{l119682}. In parallel, a correlation was done using a small part of the blue spectrum showing metallic lines, which seemed stable over the whole observing range (see previous section). Of course, this could only be done on ESO spectra, not on amateur ones that lack S/N and/or do not cover that blue range. Therefore, this further limits the number of data points and the temporal coverage. The weakness of the blue lines also implied a larger error on the derived RVs. In both cases, the period search suffers from large uncertainties. For example, the observation cadence of a few months of monitoring separated by a year resulted in a set of narrow peaks arranged in a broader one. We tried to improve the data quality by combining all available RVs, taking shifts into account (see the previous section). In the end, a period of about 60\,d is found, yielding a rather convincing folding (see Table \ref{bin} and the right panels of Fig. \ref{l119682}) although the scatter is large (especially in view of the RV amplitude). Continuing the monitoring of HD\,119682 (now that the emission is back), with an enhanced cadence, would thus be extremely useful to confirm this preliminary orbital solution. 

Our targets display periods of 59--322\,d, a lack of strong eccentricity, and small velocity semi-amplitudes $K$ of $\sim$5\,\kms. Considering a wide range of inclinations (40--90$^{\circ}$), the mass function then leads to companions with masses $\sim$1\,M$_{\odot}$. This can be compared to other \gc\ analogs, whose binary properties fill the bottom of Table \ref{bin}\footnote{The case of SAO\,49725 is not shown as its orbital solution remains uncertain, see \citet{naz22}.}. These other \gc\ analogs have $P=76-203$\,d, $e=0$, and $K=4-8$\,\kms. In comparison, most of the newly found binaries have longer periods, but overall display very similar characteristics. This is also in line with the parameters of other spectroscopic solutions linked to Be stars \citep{naz22,lab25}. The presence of such low-mass companions is understood to be the result of a past mass transfer event, with the current Be star being initially the least massive object in the binary. The observed long periods and small eccentricities agree well with predictions of population synthesis models taking into account binary interactions (e.g. \citealt{shao14}). 

\begin{figure}
  \begin{center}
    \includegraphics[width=8cm]{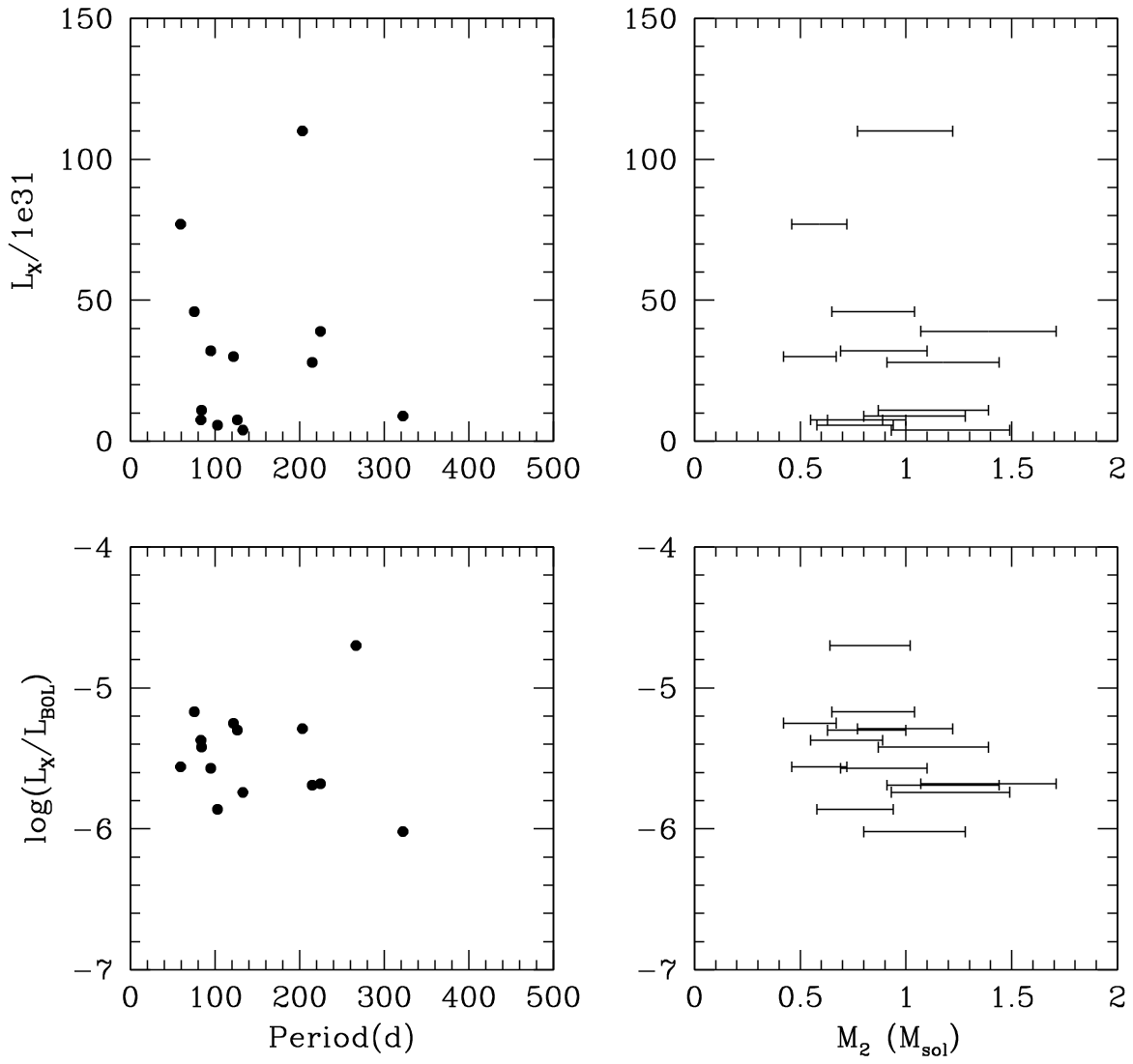}
  \end{center}  
  \caption{Comparison between maximum X-ray luminosities (top) or X-ray to bolometric luminosity ratios (bottom) and the orbital periods (left) or the mass ranges for the companions to the Be stars (right).  \label{compax}}
\end{figure}

\begin{figure*}
  \begin{center}
    \includegraphics[width=6cm]{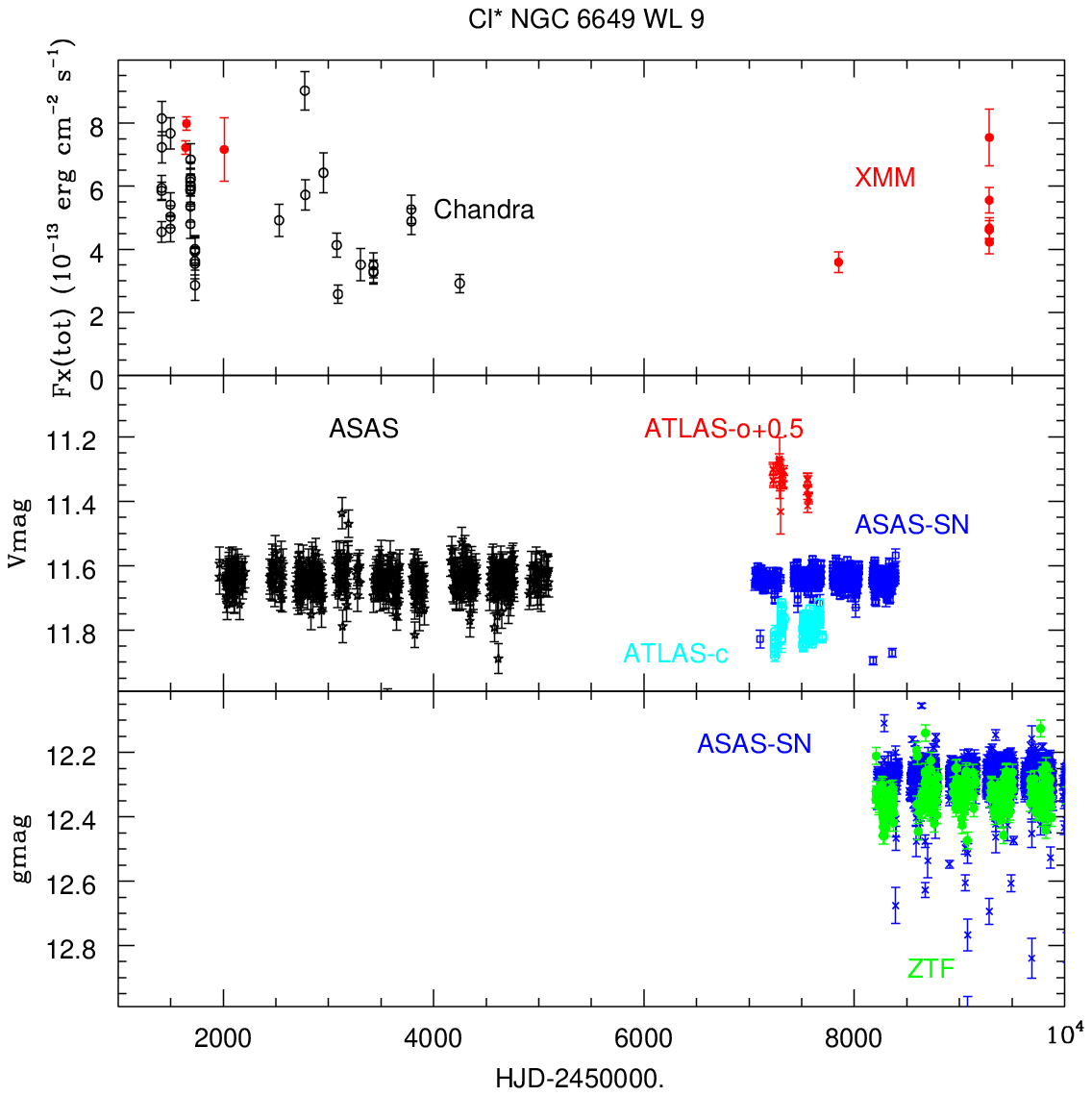}
    \includegraphics[width=6cm]{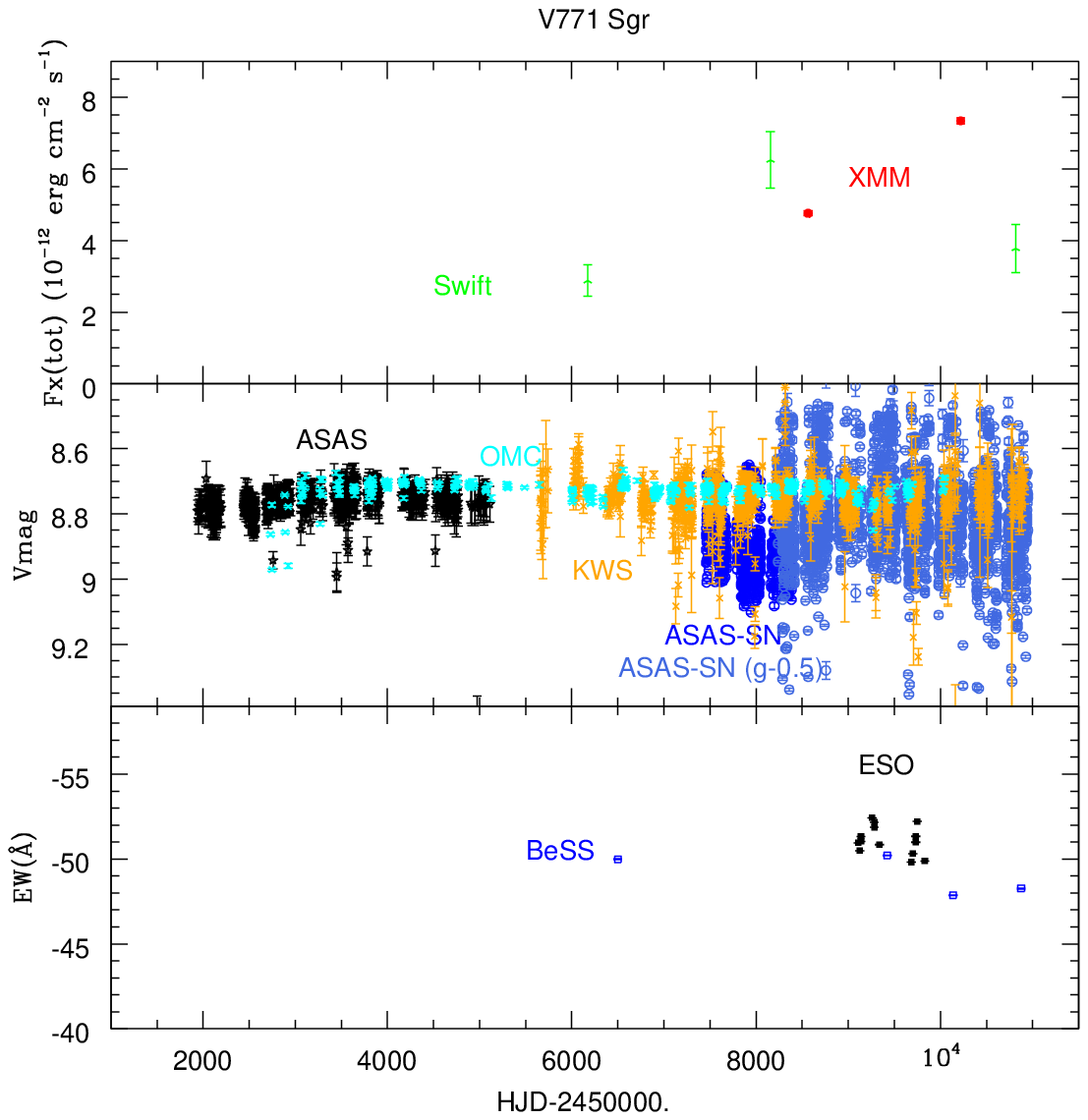}
    \includegraphics[width=6cm]{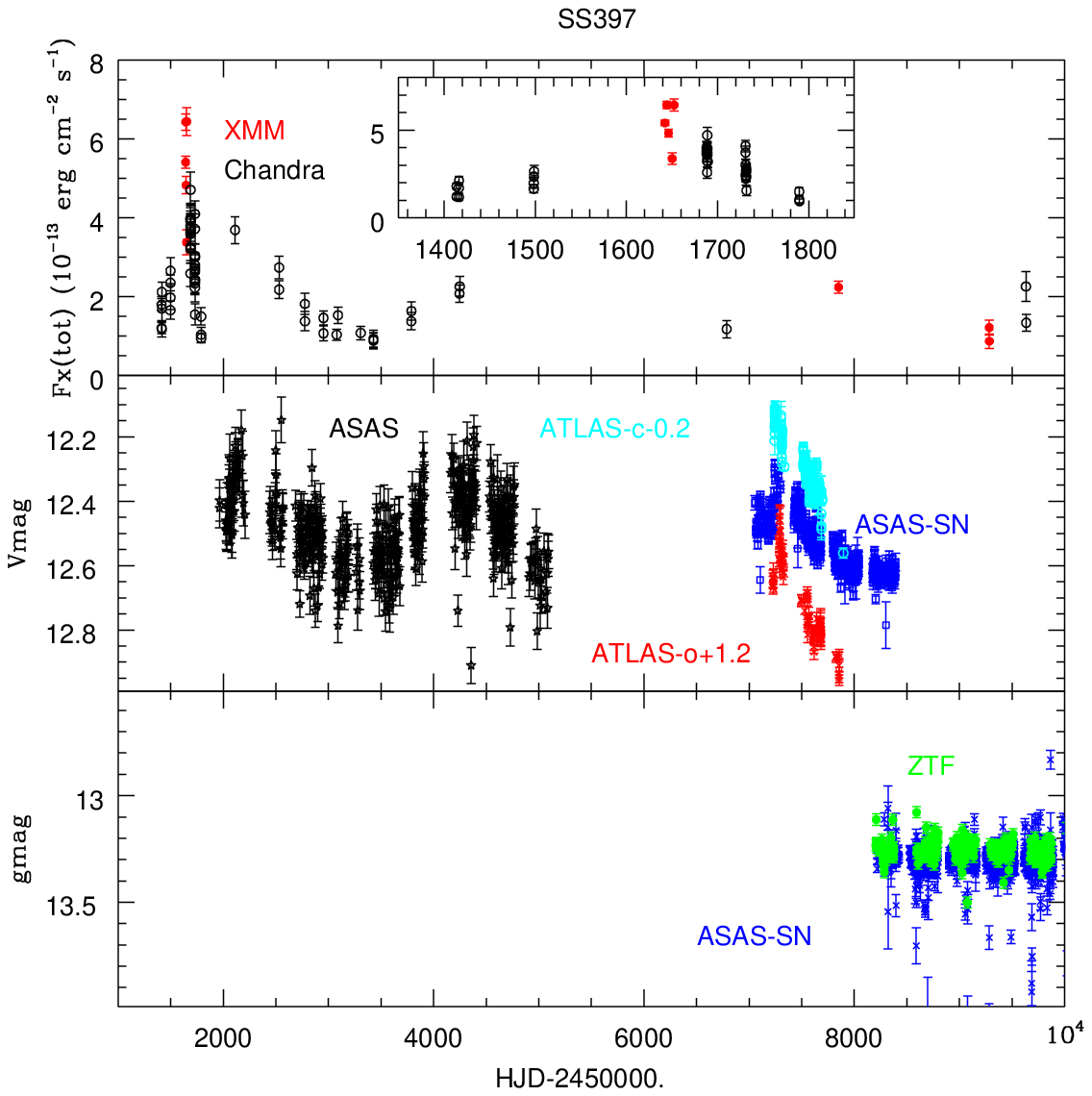}
  \end{center}  
  \caption{Evolution of fluxes over time recorded in X-rays (top panels - \xmm\ in red, \ch\ in black, and \sw\ in green) and in the optical (other panels - different colours correspond to different surveys) for \cl\ (left), \vs\ (middle), and \sgc\ (right). \label{phot}}
\end{figure*}

Since a dozen orbital solutions are now available for \gc\ analogs, it is tempting to examine whether X-ray parameters are linked to orbital properties. Table \ref{bin} provides the maximum X-ray luminosities recorded for these objects, and Fig. \ref{compax} compares them to the orbital period and secondary mass ranges. For X-rays produced by accretion onto a white dwarf, it could be expected that the luminosity will increase if the companion is more massive and closer to the source of material (the Be star). Although Fig. \ref{compax} shows that shorter-period systems display a broader range of X-ray luminosities, with longer-period systems missing large luminosity cases, there is an important exception to this trend: \gc\ itself. In addition, no strong correlation is seen in any of the panels, and this remains true if mass ratios are used instead of secondary masses. Of course, not all sources have been intensively followed in the X-ray range. The full range of X-ray luminosities therefore remains unknown, as some crucial epoch may have been missed. In addition, the accretion luminosity also depends on the accretion rate, and hence on the Be disk content. Unfortunately, simultaneous optical data to estimate, for example, the H$\alpha$ strength at the time of X-ray observations are not always available. All this limits the constraining power of Fig. \ref{compax}. In the future, simultaneous and more continuous monitoring will be required, along with dedicated hydrodynamical simulations.

\subsection{X-ray evolution}
\gc\ analogs are identified by a 'peculiar' high-energy emission, on the basis of several criteria \citep{naz18}. First, the X-ray brightness appears intermediate between those of 'normal' OB stars and of X-ray binaries: $\log[L_X^{ISM cor} {\rm in\, }0.5-10$keV] between 31.6 and 33.2 and/or $\log[L_X/L_{\rm BOL}]$ between --6.2 and --4, while this ratio is $\sim-7$ for normal OB stars and $>-4$ for X-ray binaries. Then, the X-ray spectrum is thermal in nature, but hard ($HR>1.6$, $L_X^{ISM cor}[2-10$keV]$ > 10^{31}$\,erg\,s$^{-1}$, and/or $kT>5$\,keV) with prominent iron lines (at 6.7 and 7.0\,keV), presence of a fluorescence Fe\,K$\alpha$ feature at 6.4\,keV, and strong short-term variability. The last three criteria might be more difficult to assess as detailed short-term light curves can only be extracted on high-quality data of a high-flux source, and the iron complex may be difficult to spot in low-count spectra.

In all X-ray data, \cl\ always appears sufficiently bright and hard to classify it as \gc\ analog. In addition, the combined spectra, despite their limited quality, hint at the presence of the iron complex. In optical, the photometry appears quite stable, with any variation lost in the $\sim$0.1\,mag scatter (left panels of Fig. \ref{phot}) but the total X-ray luminosity of \cl\ varies by a factor of about 3, yielding $\log(L_X/L_{\rm BOL})$ between --6.1 and --5.6. Because of the shortness of the exposures ($\sim10$\,ks), the X-ray 'scatter' could possibly be explained by the large-amplitude short-term variations (see also the case of $\pi$\,Aqr in \citealt{naz19}). 

\begin{figure}
  \begin{center}
    \includegraphics[width=8cm]{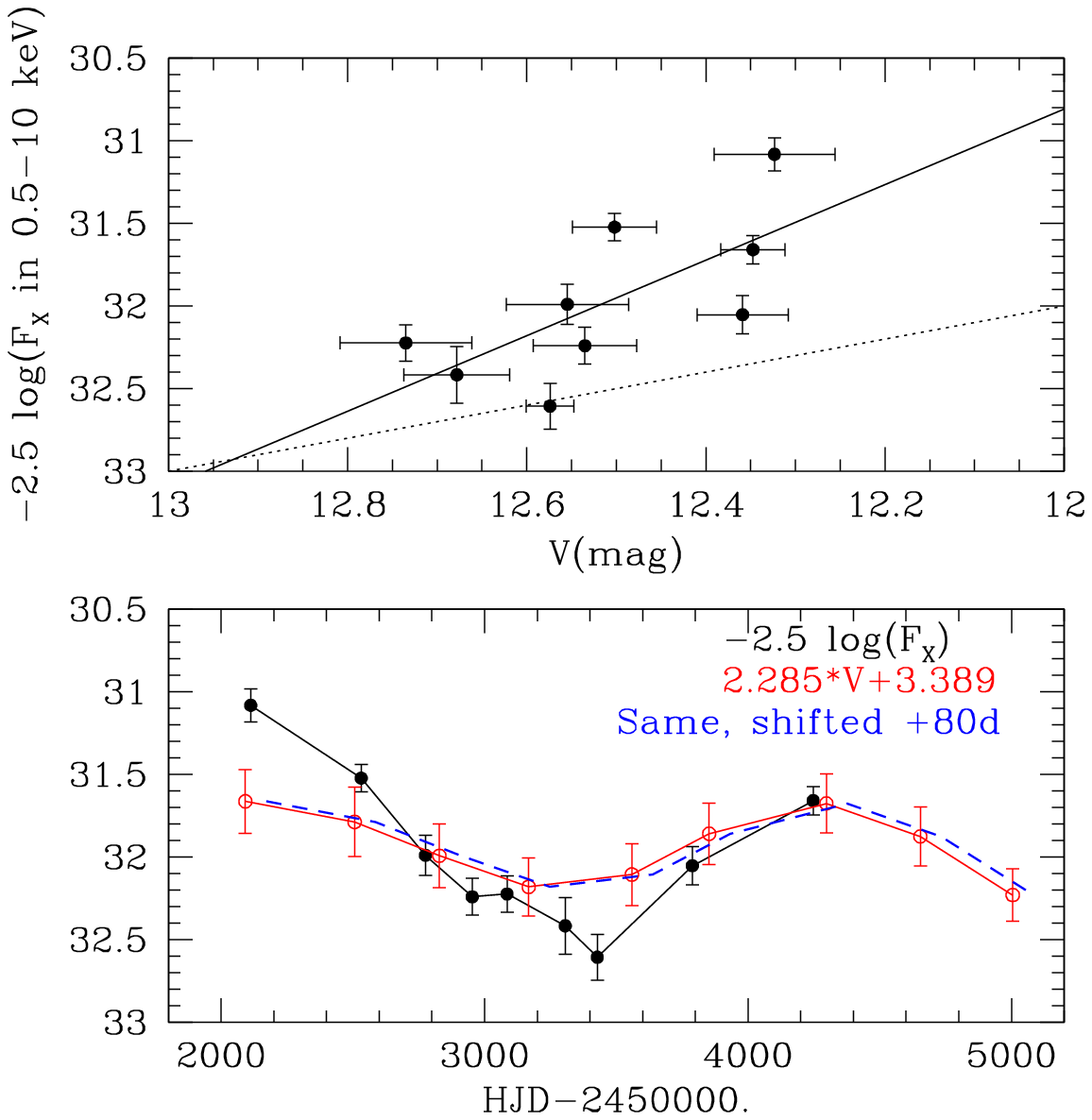}
  \end{center}  
  \caption{{\it Top:} Observed X-ray 'magnitudes' (defined as $-2.5\times \log(F^{obs}_X)$) and mean V-magnitudes from ASAS (error bars indicate dispersions) for \sgc. They appear clearly correlated, but not with a one-by-one relationship (one mag for one mag, shown by the dotted line). Rather, the X-ray changes are larger (the best-fit relation is shown by the solid line). {\it Bottom:} X-ray (black points) and optical curves (red circles). The best-fit correlation is obtained by shifting the optical curve by +80\,d (blue dashed line). \label{correl}}
\end{figure}

As \cl, \vs\ displays a rather stable optical photometry (middle panels of Fig. \ref{phot}). In the X-ray domain, the observed flux in the 0.5--10.\,keV energy band more than doubles (middle-top panel of Fig. \ref{phot}, \citealt{web26}). There is no obvious correlation between these X-ray changes and optical photometry, H$\alpha$ strength, or orbital phase. Older data are scarce, but \vs\ was reported in the {\it ROSAT} faint sources catalogue \citep{vog00} with a PSPC count rate of $(4.99\pm1.45)\times10^{-2}$\,cts\,s$^{-1}$ in the 0.1--2.4\,keV band. This appears in line with the properties recorded in the first \xmm\ spectra. Indeed, folding the modelled spectrum into {\it ROSAT} matrices using WebPIMMS\footnote{https://heasarc.gsfc.nasa.gov/cgi-bin/Tools/w3pimms/w3pimms.pl} yields a {\it ROSAT} count rate of $5.3\times10^{-2}$\,cts\,s$^{-1}$. Although the distance is unknown for \vs, the $\log(L_X/L_{\rm BOL})$ ratio can still be calculated as it is independent of this parameter: values between --4.7 and --5.1 are found. Along with the ratio measured for CQ\,Cir \citep{naz18}, such values appear as the highest among \gc\ analogs. In parallel, while \vs\ clearly remains a hard X-ray source, even the best spectra remain noisy. Therefore, it is difficult to assess the presence of the iron complex. While such lines are always present in \gc\ analogs, hard X-ray spectra of X-ray binaries gathering a Be star and a neutron star do not have them. However, the low mass of the companion of \vs\ excludes a neutron star nature, thereby favouring a \gc\ nature for the system. 

\sgc\ appears to be much more variable, both at X-ray and optical wavelengths (right panels of Fig. \ref{phot}). The ASAS data reveal a sinusoidal variation with a peak-to-peak amplitude of about 0.25\,mag. Two thousand days later, the brightness of \sgc\ decreases by 0.2\,mag in ASAS-SN data (V-band, or 0.4\,mag in the two ATLAS bands, o and c). In the more recent ZTF and ASAS-SN (in g-band) data, the brightness seems to have stabilized. In parallel, the X-ray data display an overall flux variation of a factor of seven (in the 0.5--10.\,keV range). The X-ray light curve starts with a narrow peak corresponding to the highest X-ray luminosity observed in our dataset. Unfortunately, no simultaneous optical data are available. A second peak is followed by a slower decline and then a slow increase, both in line with the sinusoidal change observed in ASAS data. At the time, the observed X-ray flux changes by a factor of about four over slightly less than six years. 

To further assess this apparent correlation, we averaged the X-ray fluxes of neighbouring data points, resulting in nine epochs within $HJD=2452000-5000$ (see the middle panel of Fig. \ref{phot} and Table \ref{journal}). Next we estimated the mean magnitude at these epochs, using ASAS measurements taken within $\pm$10\,d of the X-ray mean observing times. The Pearson correlation coefficient between these V-magnitudes and values $-2.5\times \log(F_X)$ is 70\%, demonstrating a good correlation between the two. The best-fit linear relation was calculated and is shown in Fig. \ref{correl}. Its slope is 2.3, not one, indicating that the relative X-ray flux variations are larger than the optical ones. However, this applies to relative fluxes, not absolute ones. Converting the V-magnitudes into optical fluxes as in \citet{mot15}, using as 'diskless-state' reference the maximum magnitude observed in the ASAS averages, one can then assess the relationship between absolute X-ray and optical 'disk' fluxes. Here, the best-fit linear slope is 0.014, showing that only a small fraction of the disk diagnostic emission (in V-band) is converted to X-rays.

This correlation is reminiscent of the direct correlation between hard X-rays and broad-band optical photometry observed for \gc: over two decades, it underwent a relative change in hard X-ray fluxes by a factor of 1.8 and a change in V-band magnitude of 0.07\,mag \citep{rob02,mot15,rau22}. The changes in \sgc\ are larger and faster, but the optical to X-ray conversion factors were more extreme\footnote{Using only ASM data from 2002 to 2010, the slope was 0.0076 for optical to X-ray fluxes conversion \citep{mot15,rau22}. Considering all available X-ray data (ASM+MAXI), the slopes are 0.0030 for optical to X-ray fluxes conversion and 7.2 for optical to X-ray magnitudes. Note that, for \gc, the reference magnitude is the faintest luminosity state, observed in the 1940s - \sgc\ did not benefit from such a long-term coverage, hence its faintest luminosity state remains somewhat arbitrary.} for \gc. Other \gc\ analogs did not benefit from similar high temporal coverage, which prohibits the detection and analysis of such long-term correlations. 

Since the ASAS data are taken over long time blocks separated by at least 100\,d, we also calculated mean magnitudes for each of these intervals and converted them using the best-fit linear relationship previously mentioned. A linear interpolation then provided the optical values associated with any set of shifted X-ray epochs. The $\chi^2$ was calculated between the X-ray points (defined by the X-ray magnitudes $-2.5\times \log(F_X)$) and these converted optical values. As the first X-ray epoch was taken close to the first ASAS interval, which prevented us from investigating some shifts, we excluded it from the calculation. The minimum $\chi^2$ is formally found for an optical curve shifted by +80\,d. The X-ray thus reacts with a possible delay, of as much as a few months, to optical changes. However, it must be kept in mind that the number of epochs remains limited and the $\chi^2$ minimum is rather flat (allowing 0--150\,d). In comparison, for \gc the delay was constrained to a maximum of one month \citep{mot15}.

The more recent X-ray coverage is more sparse, but apparently the situation seems to have changed. During both the optical brightness decline and its subsequent stabilization, flux doublings are recorded in X-rays. As noted in \cl, the shortness of the X-ray exposures ($\sim10$\,ks) may result in some X-ray scatter. Without better coverage, it is difficult to disentangle both effects and ascertain the (true) long-term variations.

Finally, one may note that the combined spectra, despite their limited quality, clearly hint at the presence of the iron complex in the emission of \sgc. In parallel, the change in the X-ray luminosity leads to $\log(L_X/L_{\rm BOL})$ between --6.8 and --5.9. Sometimes, \sgc\ thus appears too faint to be classified as a \gc\ analog. However, it is remarkable that even when weak, the X-ray emission of \sgc\ remains hard, as demonstrated by its large hardness ratio and its high X-ray luminosity in the hard band. This result does not depend on the details of the spectral fitting, as the CSC aperture photometry also reveals a stable hardness ratio. This is the first case of a \gc\ analog appearing X-ray faint but still hard. In contrast, HD\,45314 no longer fulfilled the \gc\ hardness criterion when becoming as faint as normal OB stars, although some faint hard component remained \citep{rau18}.

\section{Summary and conclusion}
Long-term monitoring of seven \gc\ analogs was conducted at X-ray and/or optical wavelengths. Overall, the new observational results agree well with those previously derived for other \gc\ analogs. This strengthens the statistics, hence our empirical view of these peculiar objects.

High-resolution spectroscopy allowed us to propose the first orbital solutions for HD\,44458, HD\,110432, HD\,119682, HD\,161103, and \vs, although those of HD\,44458 and HD\,119682 remain very preliminary. Thus, half of the \gc\ analogs are now identified as binaries. Most of the remaining ones are much fainter, rendering binary identification more difficult, unfortunately. In all known cases, the derived companion masses are low and the periods are long, in line with the idea that Be stars in general, and \gc\ analogs in particular, correspond to post-mass-transfer systems.

\sgc, \cl, and \vs\ are found to be variable in X-rays. The largest X-ray flux variations are recorded for \sgc, and these X-ray flux changes clearly follow the optical broad-band variability over time. A similar correlation between X-ray and optical emissions was previously found for \gc, HD\,45314, and $\pi$\,Aqr. Thanks to the better temporal coverage of the \sgc\ data, a small delay (a few months) was tentatively detected, although with a large error bar. This delay informs us on the X-ray emission process. Indeed, the optical photometric changes are most probably related to a density change in the (inner) Be disk: if X-rays arise from star-disk interactions, they would react immediately to them, while X-rays born near the companion require some time before the change in the flow of material reaches its surroundings. Another point to emphasise is that \sgc\ displays a very low X-ray luminosity at minimum, while it keeps its typical \gc\ hardness. This had never been observed for \gc\ analogs before. In contrast, the broad-band optical photometry of \cl\ and \vs\ remained stable while their X-ray fluxes changed, but less than for \sgc. This is probably linked to the large short-term variations typical of \gc\ analogs. At all times, these two objects kept their \gc\ character. More generally, \sgc\ (being particularly faint in X-rays) and \vs\ (being particularly bright) seem to represent the extremes in $L_X/L_{\rm BOL}$ ratios among \gc\ analogs. 

To complement a previous study, the short-term photometric behaviours of HD\,161103, \sgc, \cl, and \vs\ were analysed. Their frequency spectra appear similar to those of Be stars, including \gc\ analogs. However, the main signal found in \vs\ corresponds to a high-frequency ($>5$\,d$^{-1}$) periodicity, which is infrequent (but not exceptional) amongst such stars. Until now, only HD\,110432 and $\pi$\,Aqr were found in this category. The next step will be to gather high-cadence spectroscopy for \vs, to better constrain its asteroseismic activity with, for example, mode identification.

\begin{acknowledgements}
The Li\`ege authors acknowledge support from the Fonds National de la Recherche Scientifique (Belgium) and the University of Li\`ege. MAS acknowledges support from \ch\ grant G02-23004X. ADS and CDS were used for preparing this document. This research was based on observations collected at the European Southern Observatory under ESO programmes 105.204D and 109.22V6, plus archival data from programmes 71.C-0367, 085.C-0799, 090.D-0212, 194.C-0833, 0102.C-0699, 113.26KE, 113.26B9. Data are available in their respective archives.
\end{acknowledgements}

\bibliographystyle{aa}
\bibliography{longtermgcas}

\appendix

\section{Additional information}
\begin{figure*}[htb]
  \begin{center}
    \includegraphics[width=6cm]{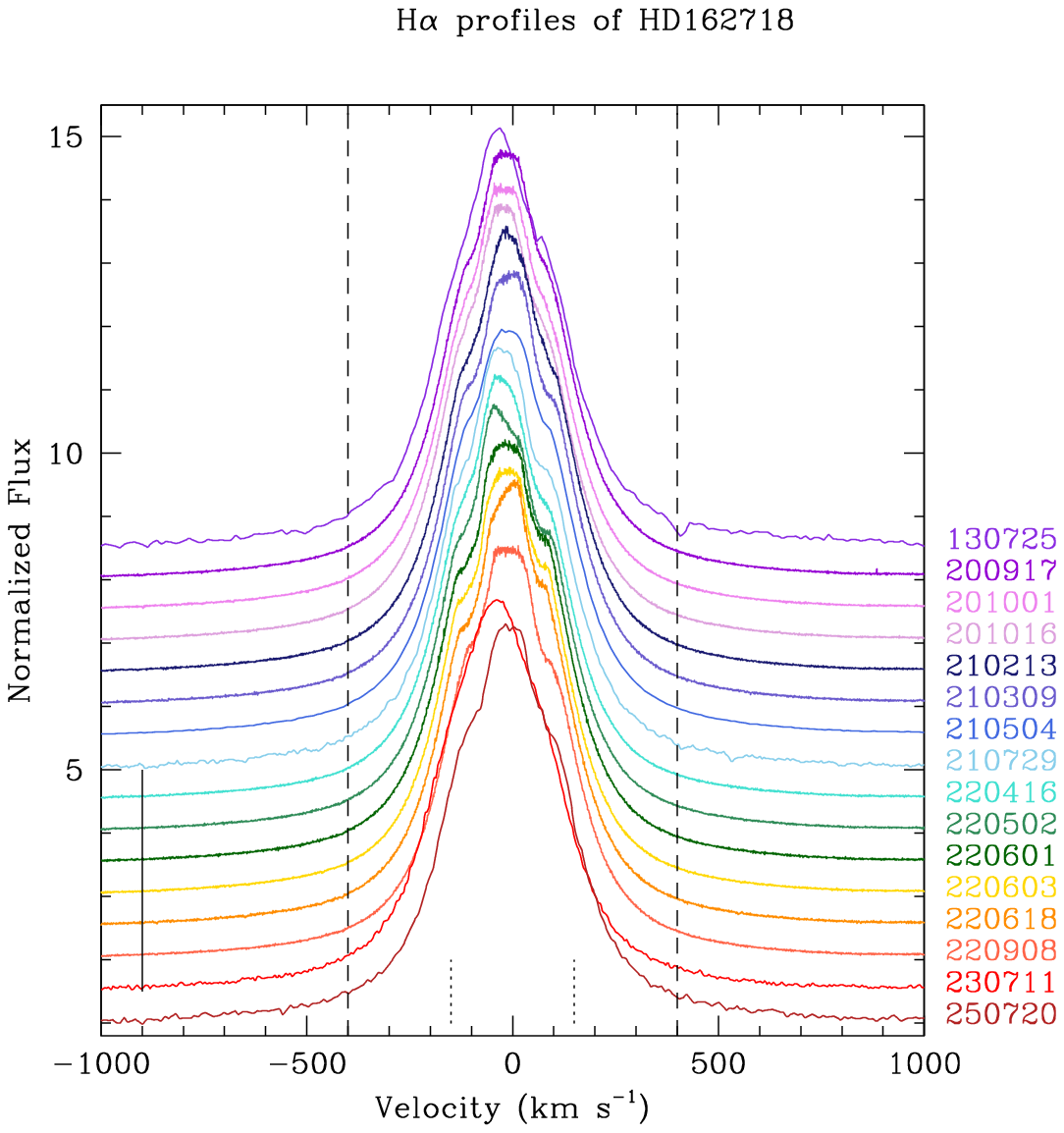}
    \includegraphics[width=6cm]{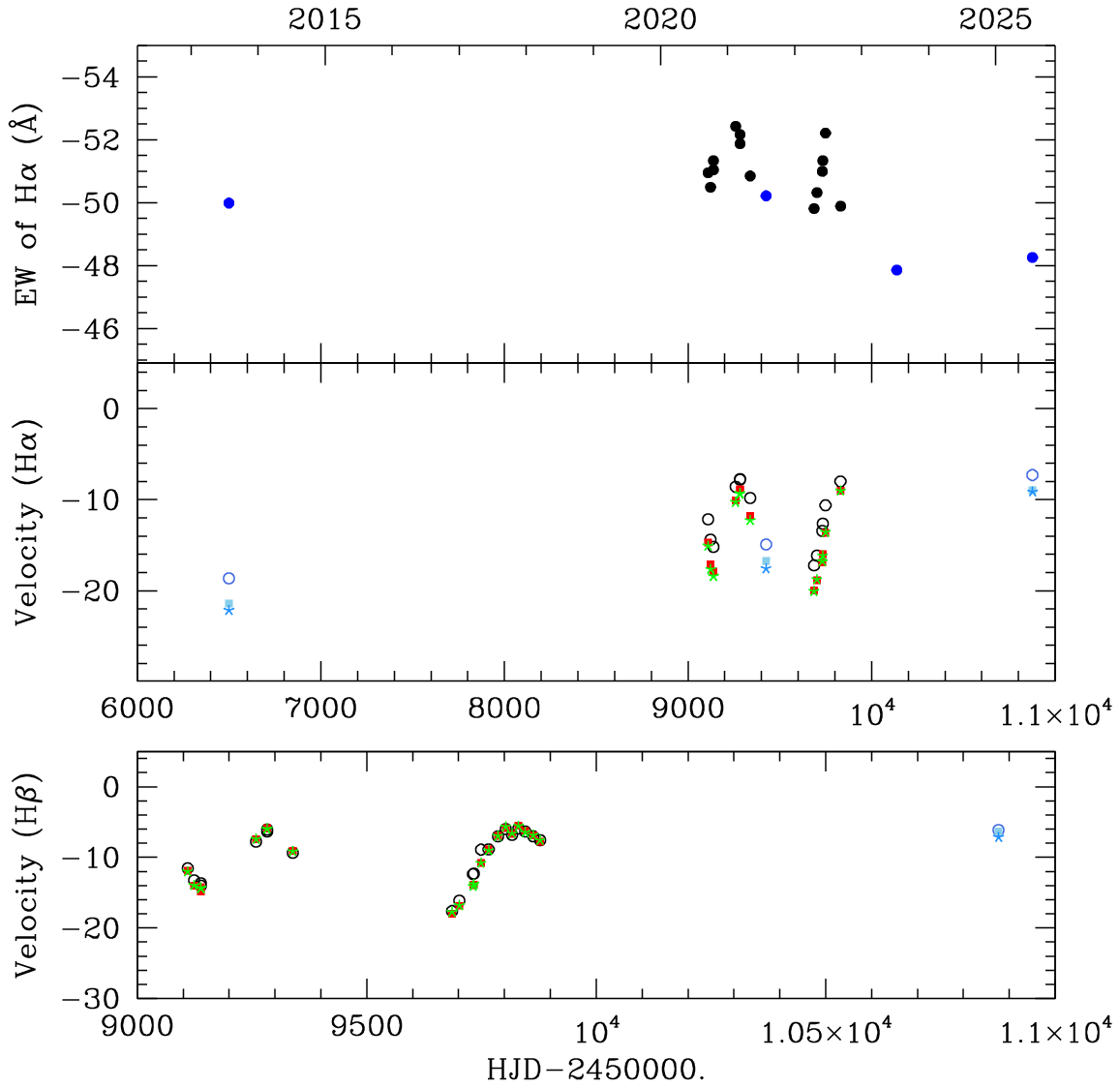}
    \includegraphics[width=6cm]{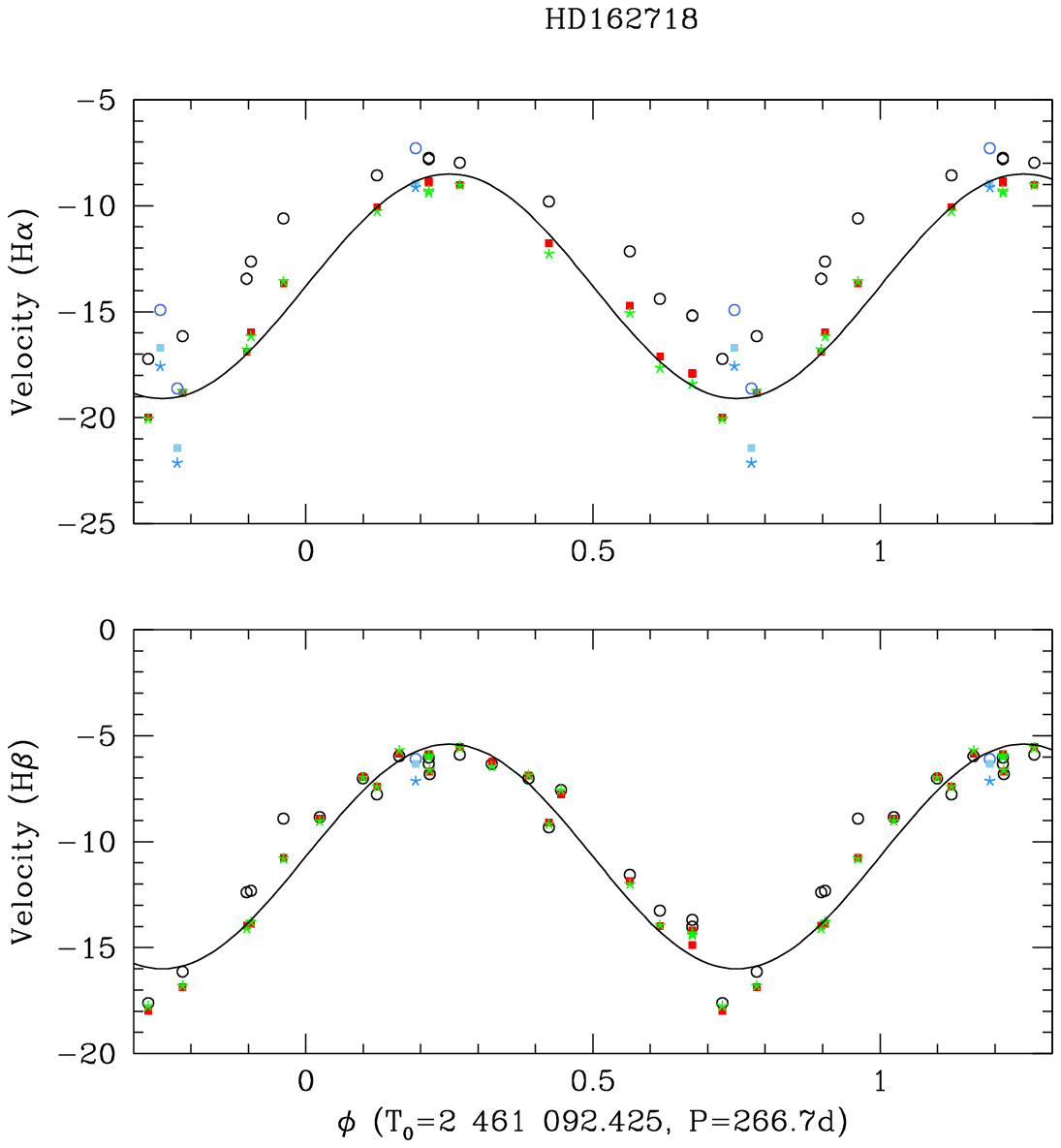}
  \end{center}  
  \caption{{\it Left:} Profile of the H$\alpha$ line in the optical spectra of \vs\ (Table \ref{rv}). The range used for EWs and RVs calculations is delimited by the two vertical dashed lines. Two small dotted lines indicate the separation of the Gaussians for the double Gaussian method while a long solid line on the edge yields the amplitude range considered for the mirror method (see text for details). Dates are given in YYMMDD on the right side. {\it Middle:} Evolution of the EWs of H$\alpha$ and of RVs measured by different methods (black circles for first-order moment, red squares for mirror method, green asterisks for the double-Gaussian method, for H$\alpha$ in the middle and H$\beta$ at the bottom). The blue symbols correspond to the BeSS amateur spectra. {\it Right:} Same RVs but folded considering the 266.7\,d period, with the best-fit orbital solution superimposed. \label{lineprof}}
\end{figure*}

\begin{figure*}
  \begin{center}
    \includegraphics[width=6cm]{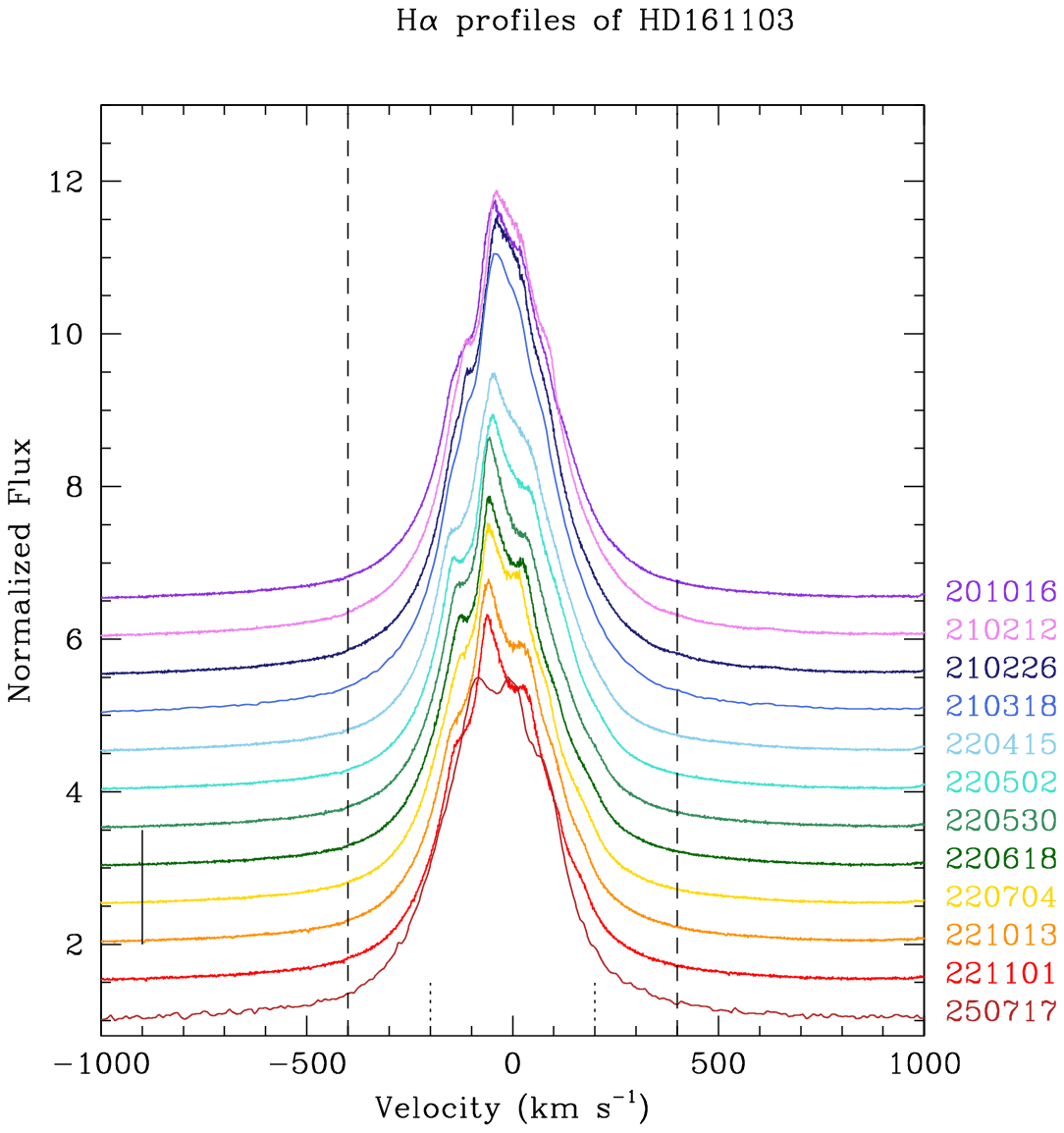}
    \includegraphics[width=6cm]{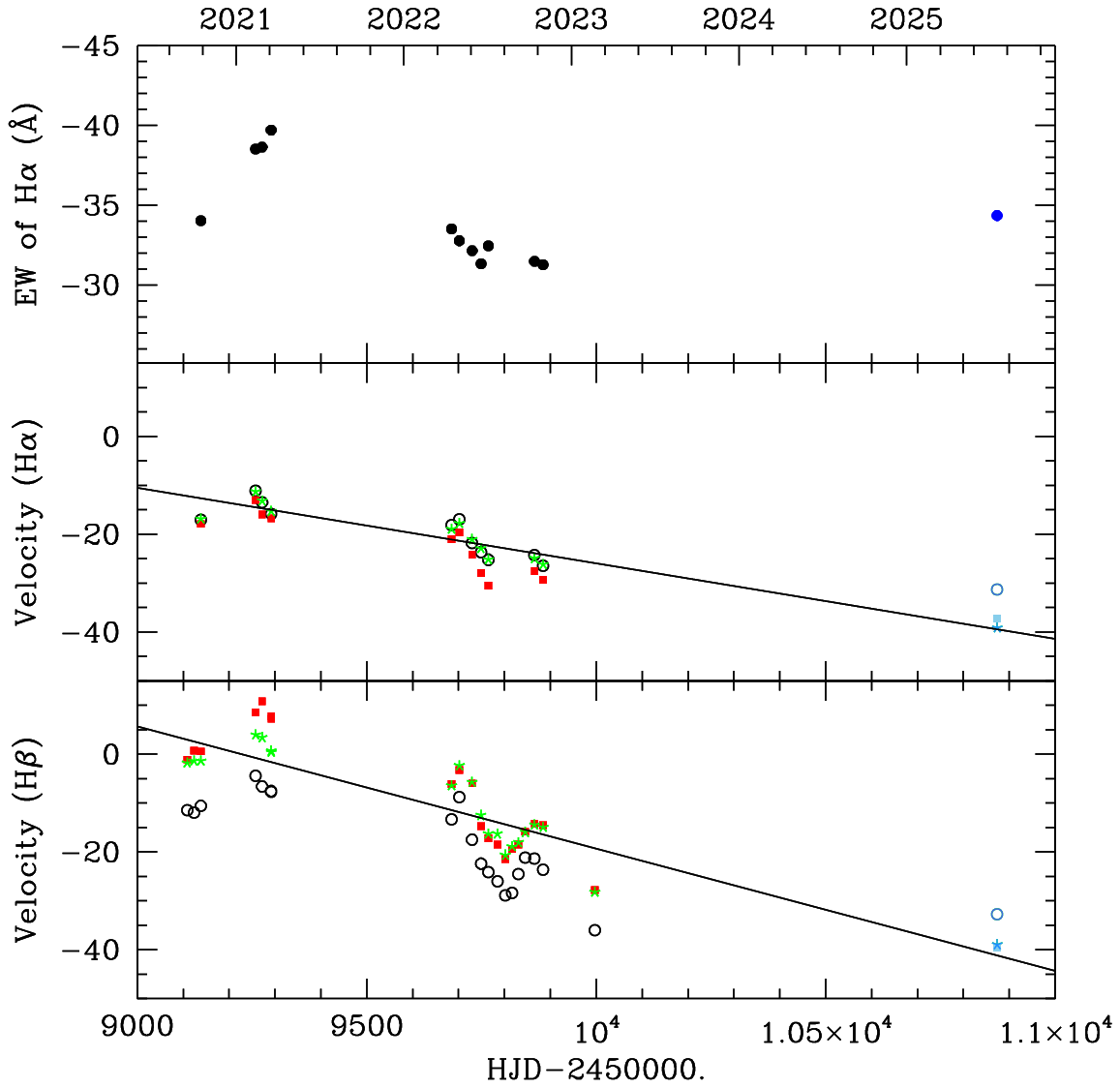}
    \includegraphics[width=6cm]{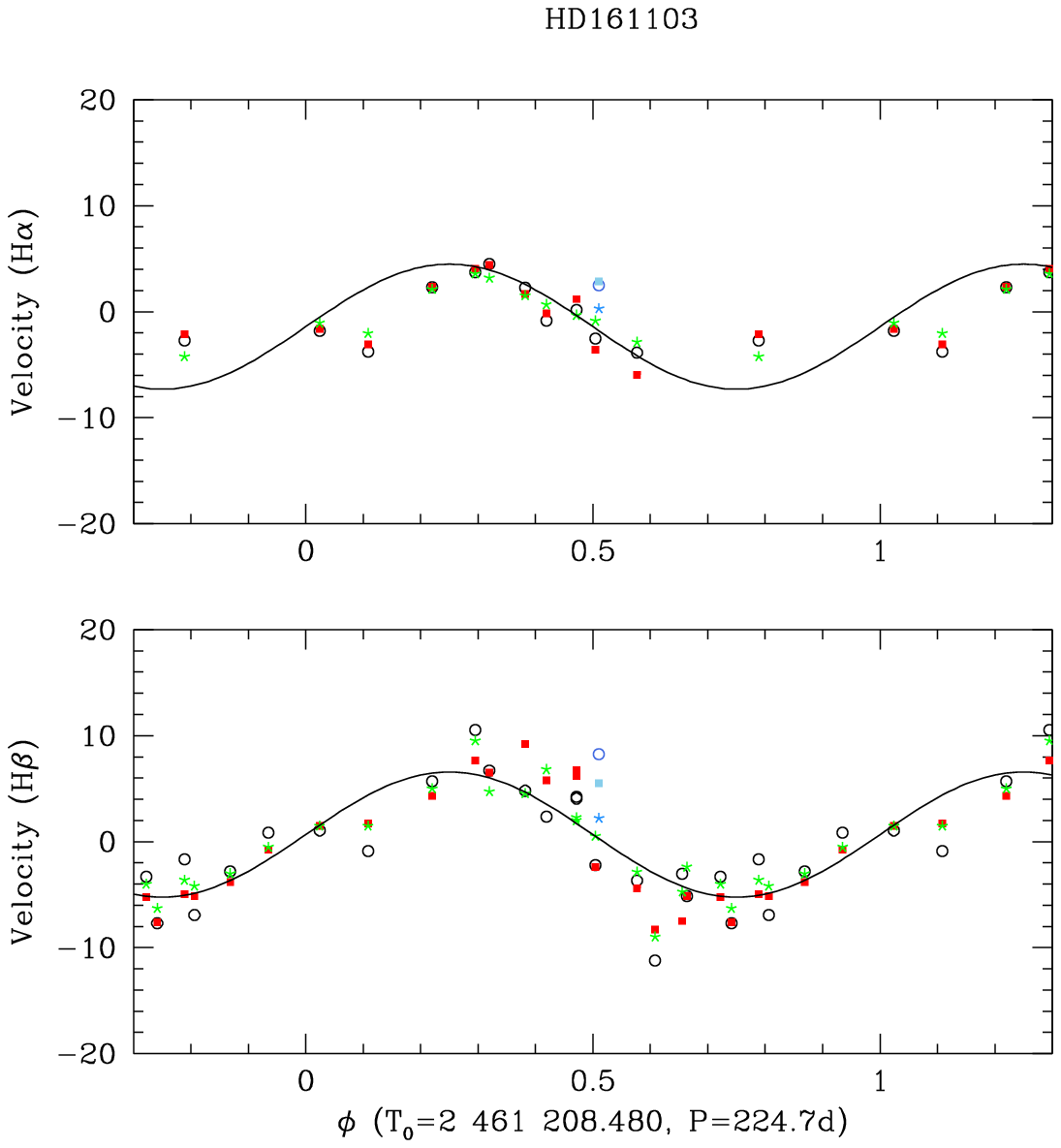}
  \end{center}  
  \caption{Same as Fig. \ref{lineprof} but for HD\,161103. \label{l161103}}
\end{figure*}

\begin{figure*}
  \begin{center}
    \includegraphics[width=6cm]{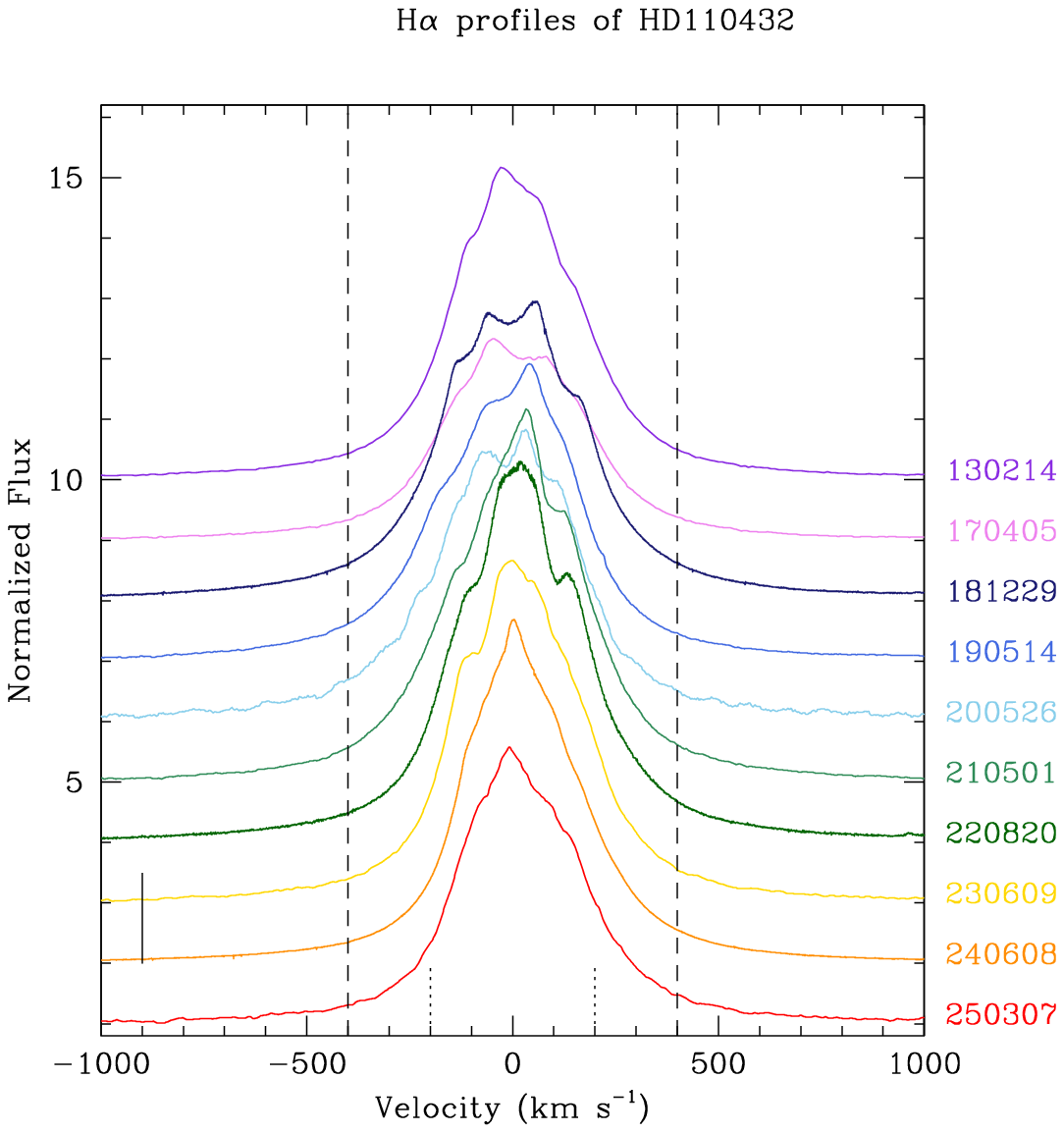}
    \includegraphics[width=6cm]{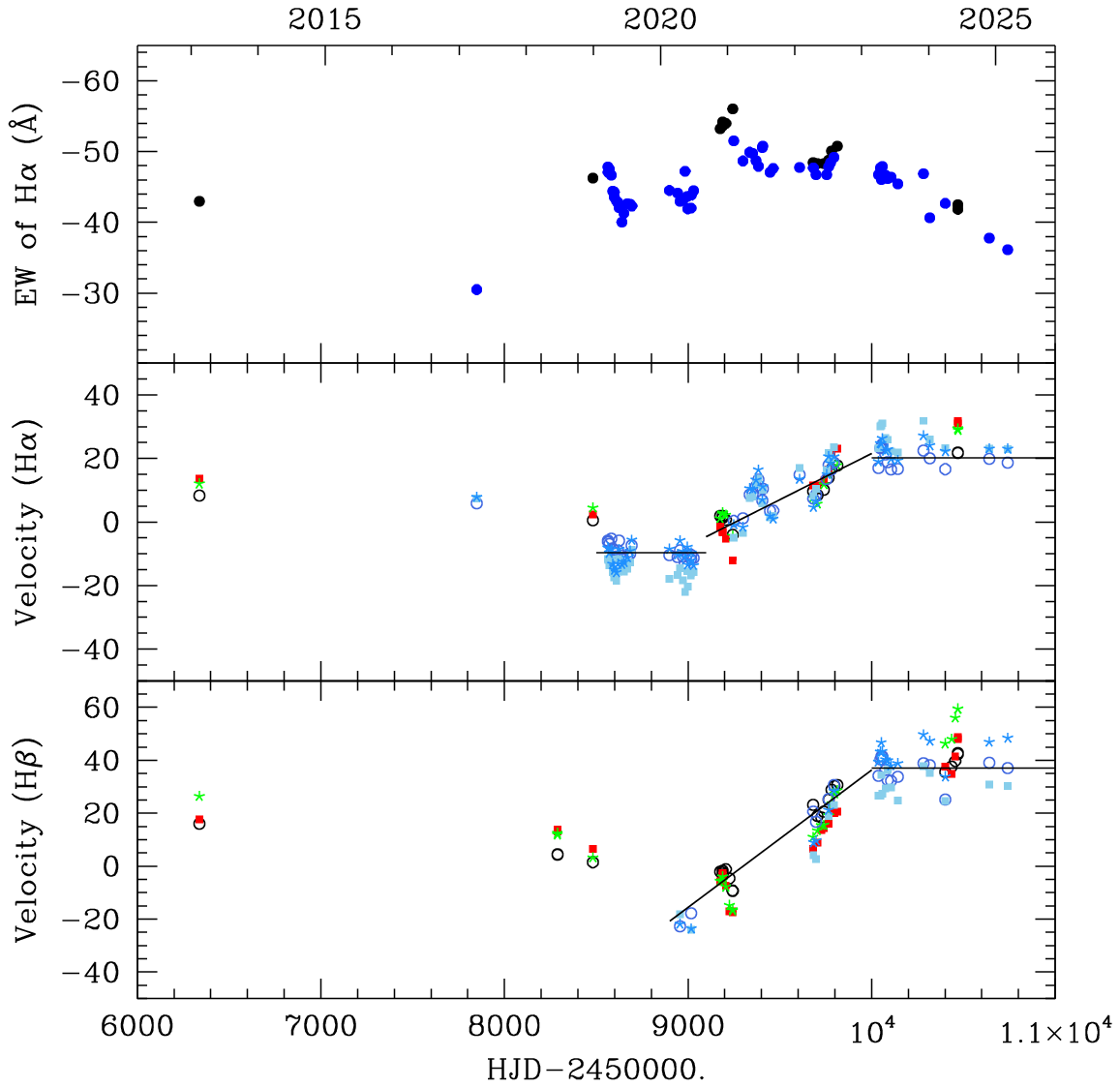}
    \includegraphics[width=6cm]{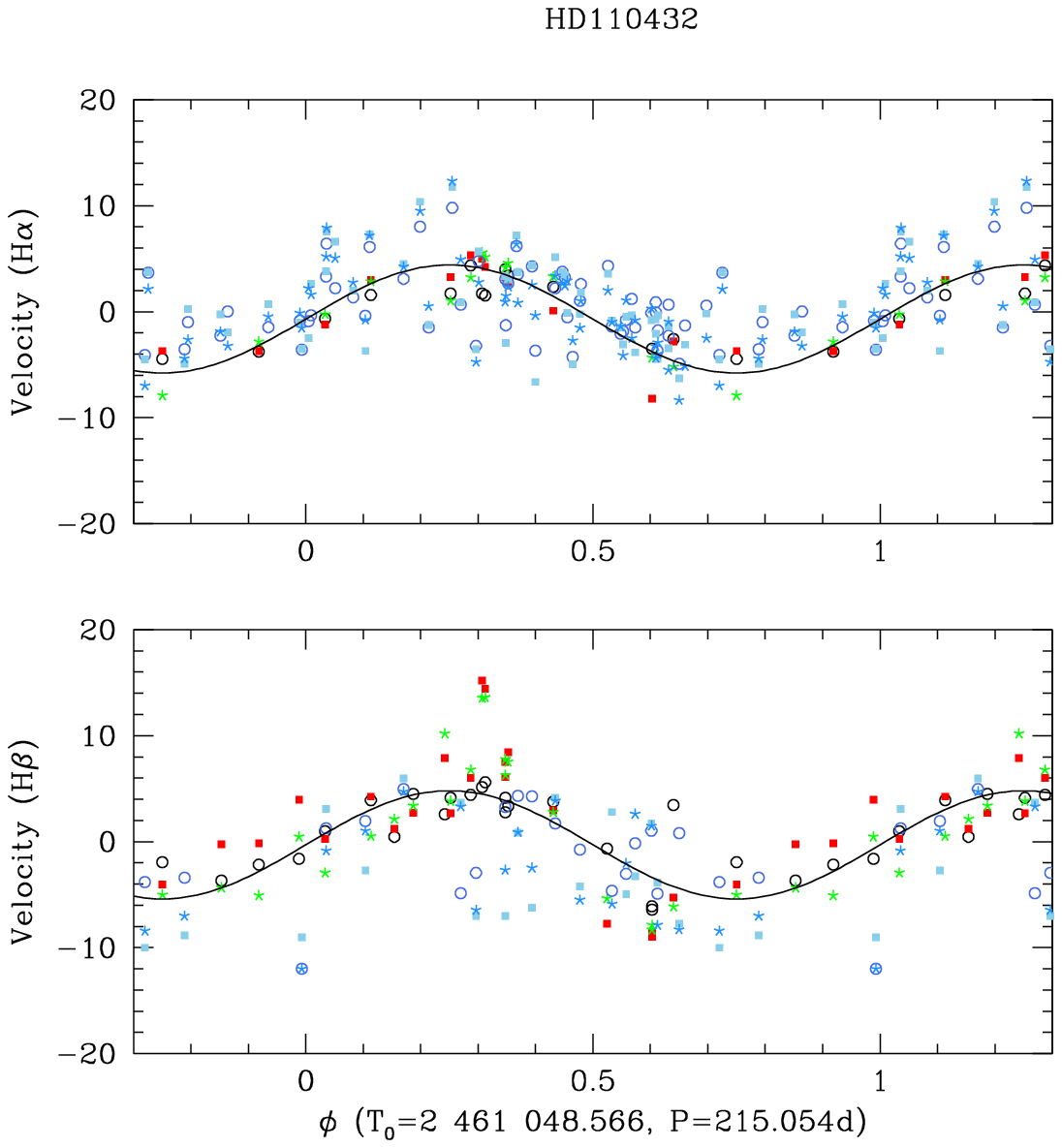}
  \end{center}  
  \caption{Same as Fig. \ref{lineprof} but for HD\,110432. Only one spectrum per year is shown on the left panel. \label{l110432}}
\end{figure*}

\begin{figure*}
  \begin{center}
    \includegraphics[width=6cm]{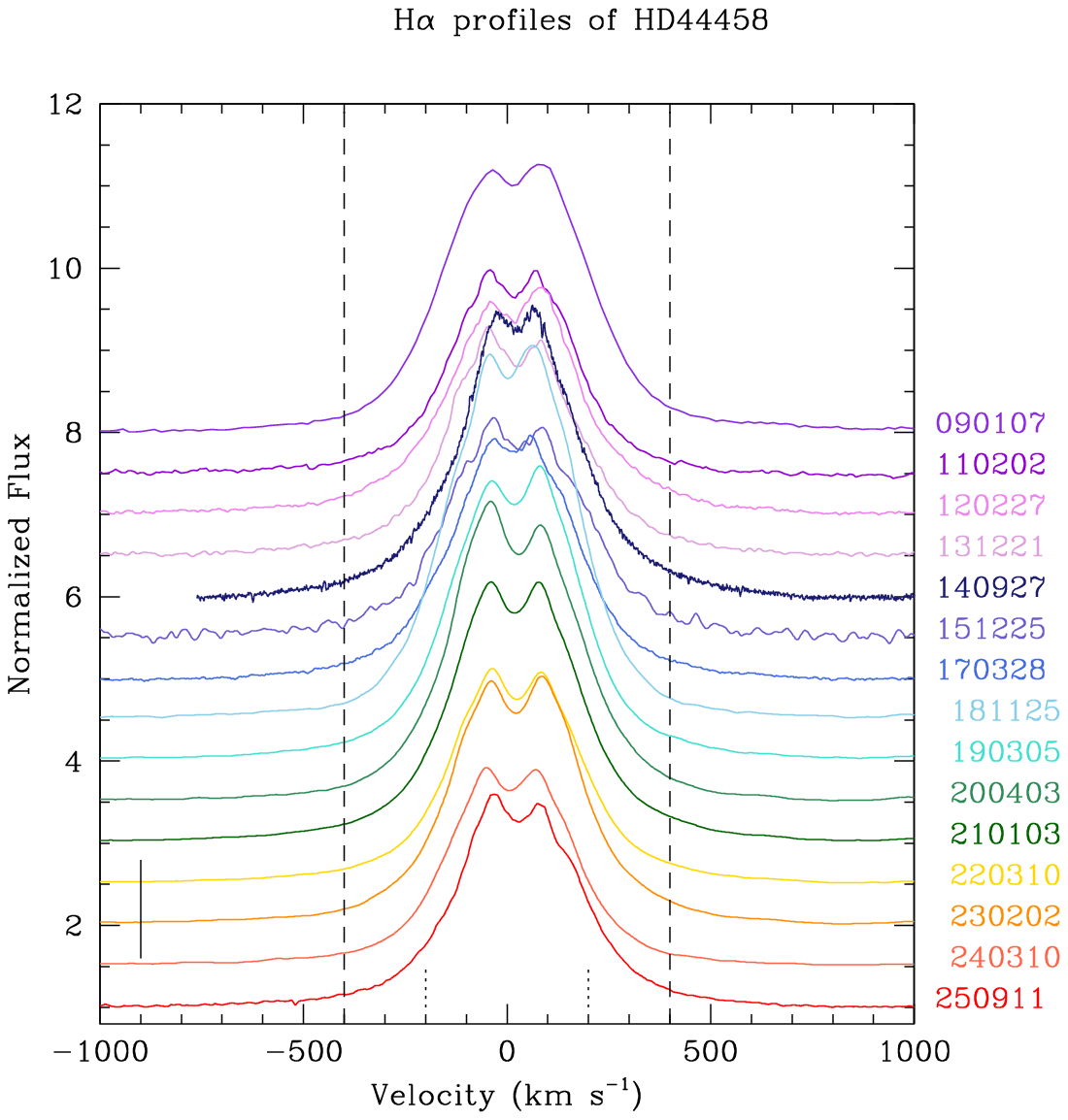}
    \includegraphics[width=6cm]{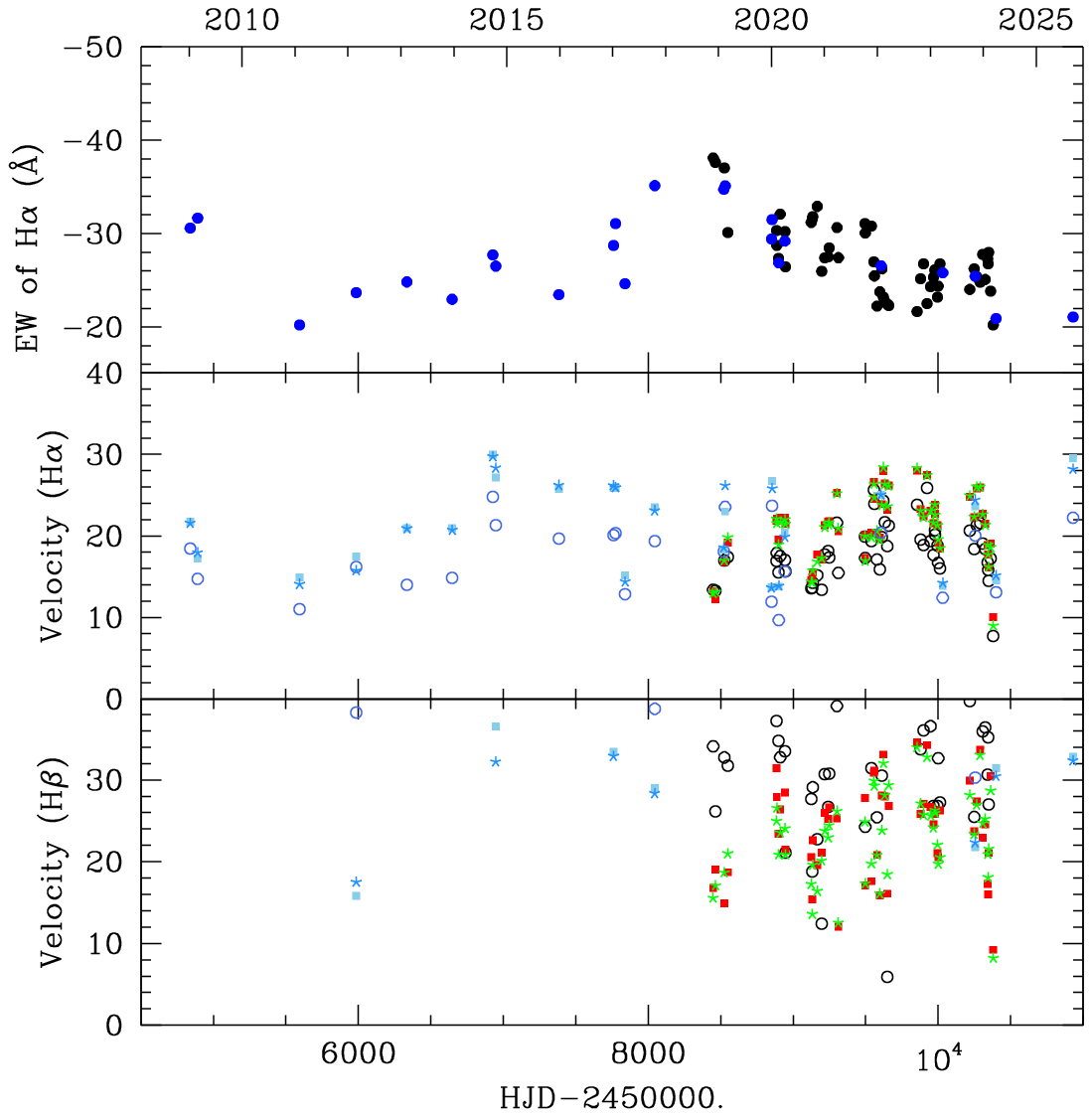}
    \includegraphics[width=6cm]{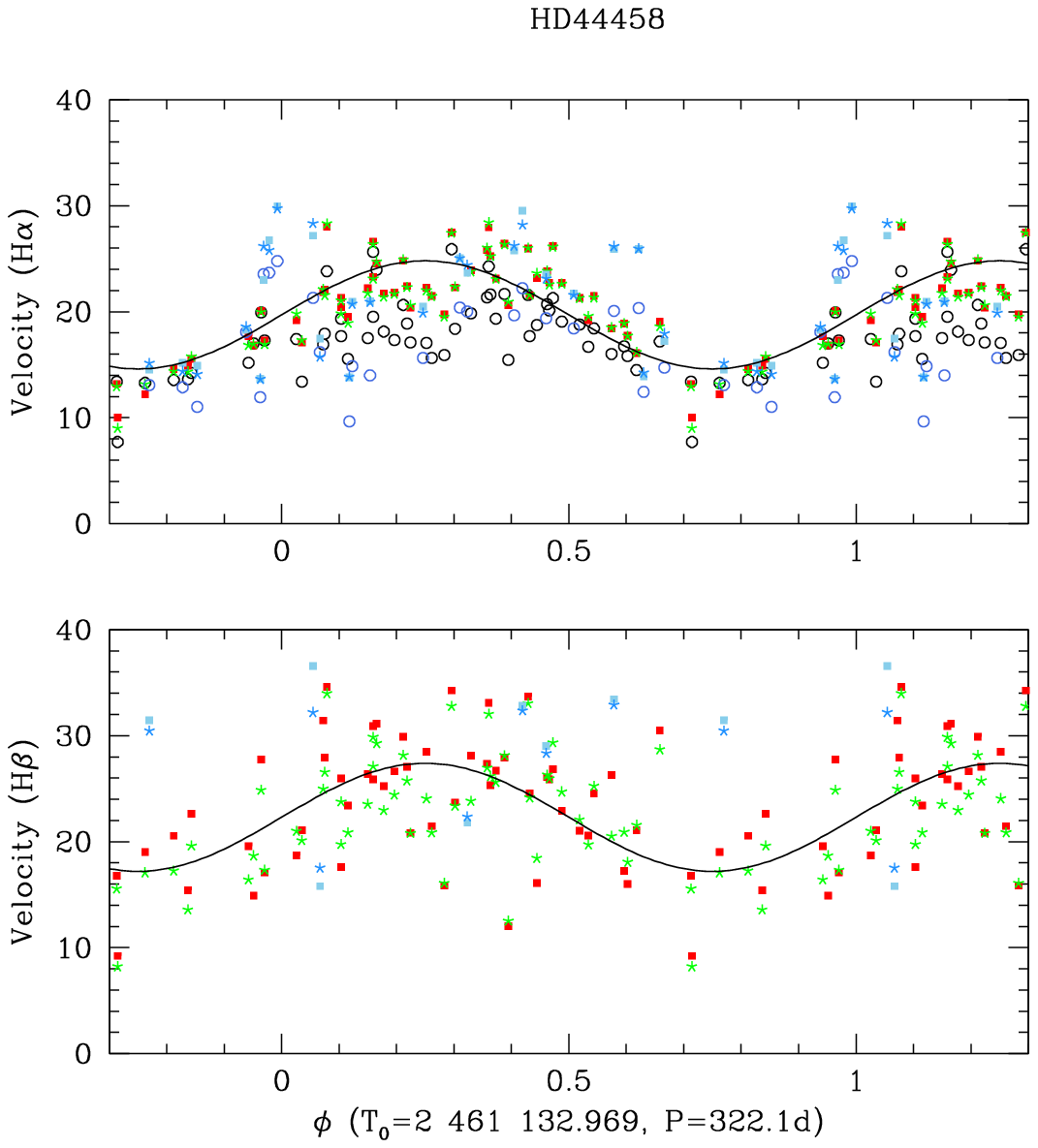}
  \end{center}  
  \caption{Same as Fig. \ref{lineprof} but for HD\,44458. Only one spectrum per year is shown on the left panel. \label{l44458}}
\end{figure*}

\begin{figure*}
  \begin{center}
    \includegraphics[width=6cm]{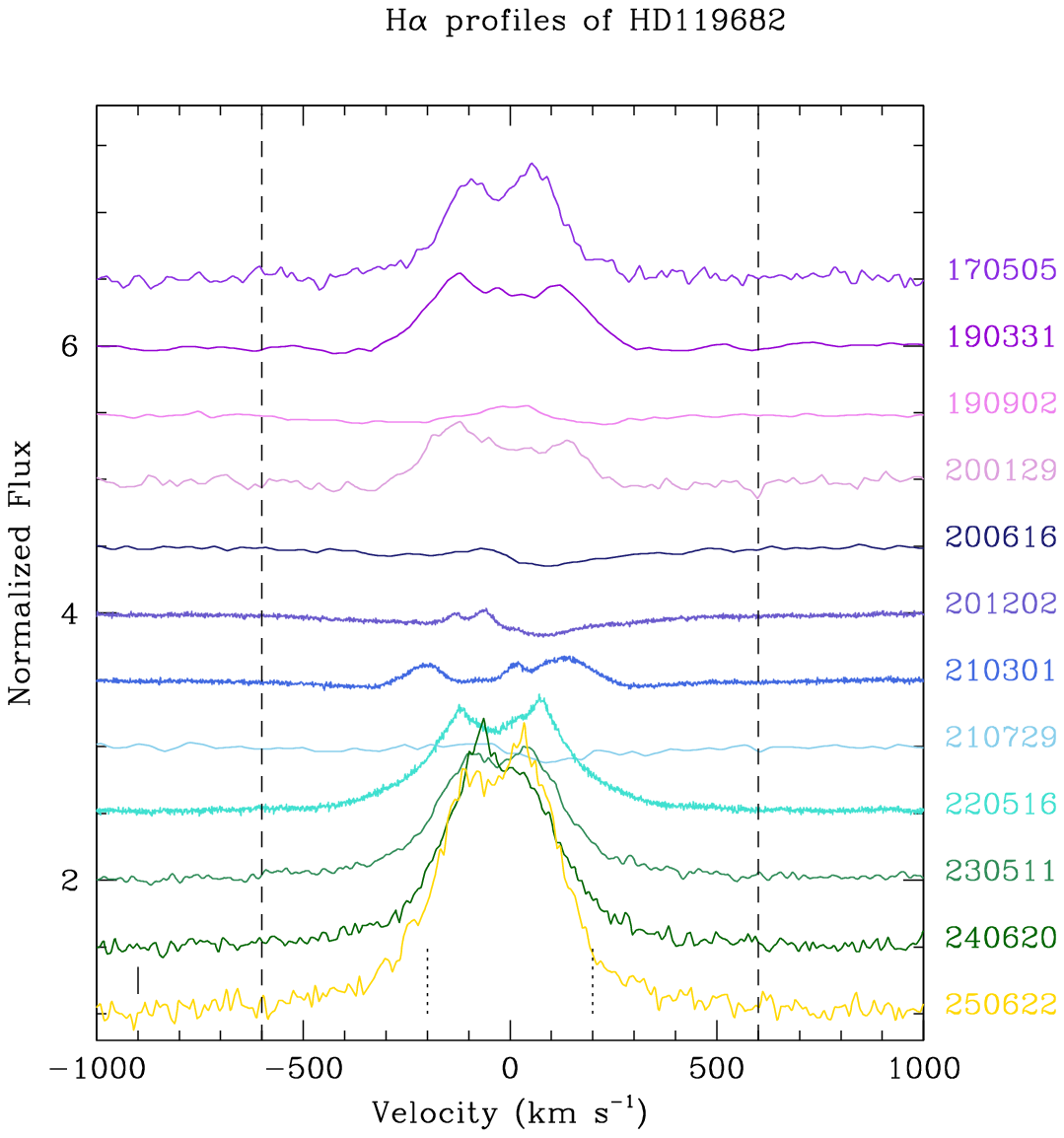}
    \includegraphics[width=6cm]{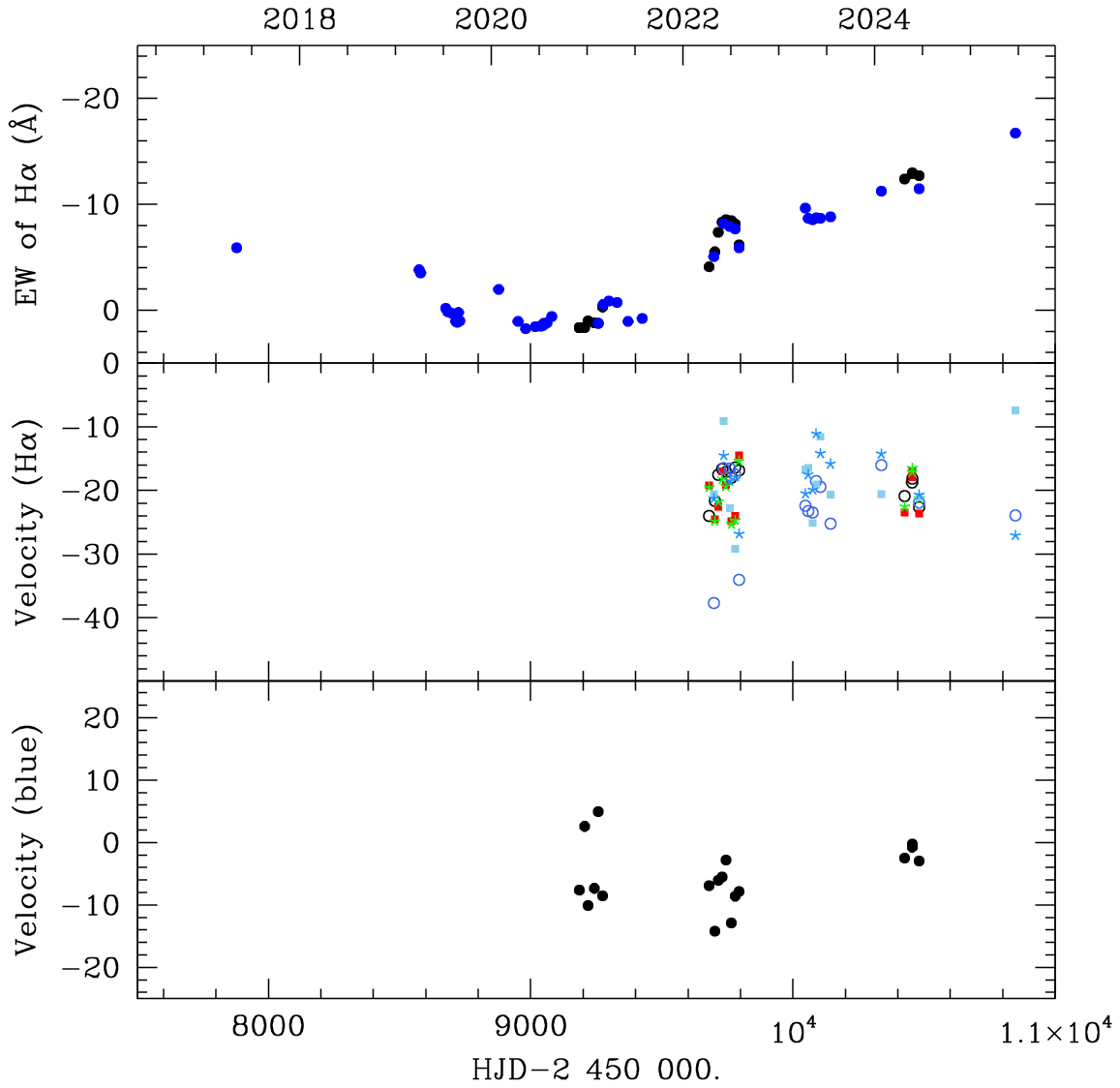}
    \includegraphics[width=6cm]{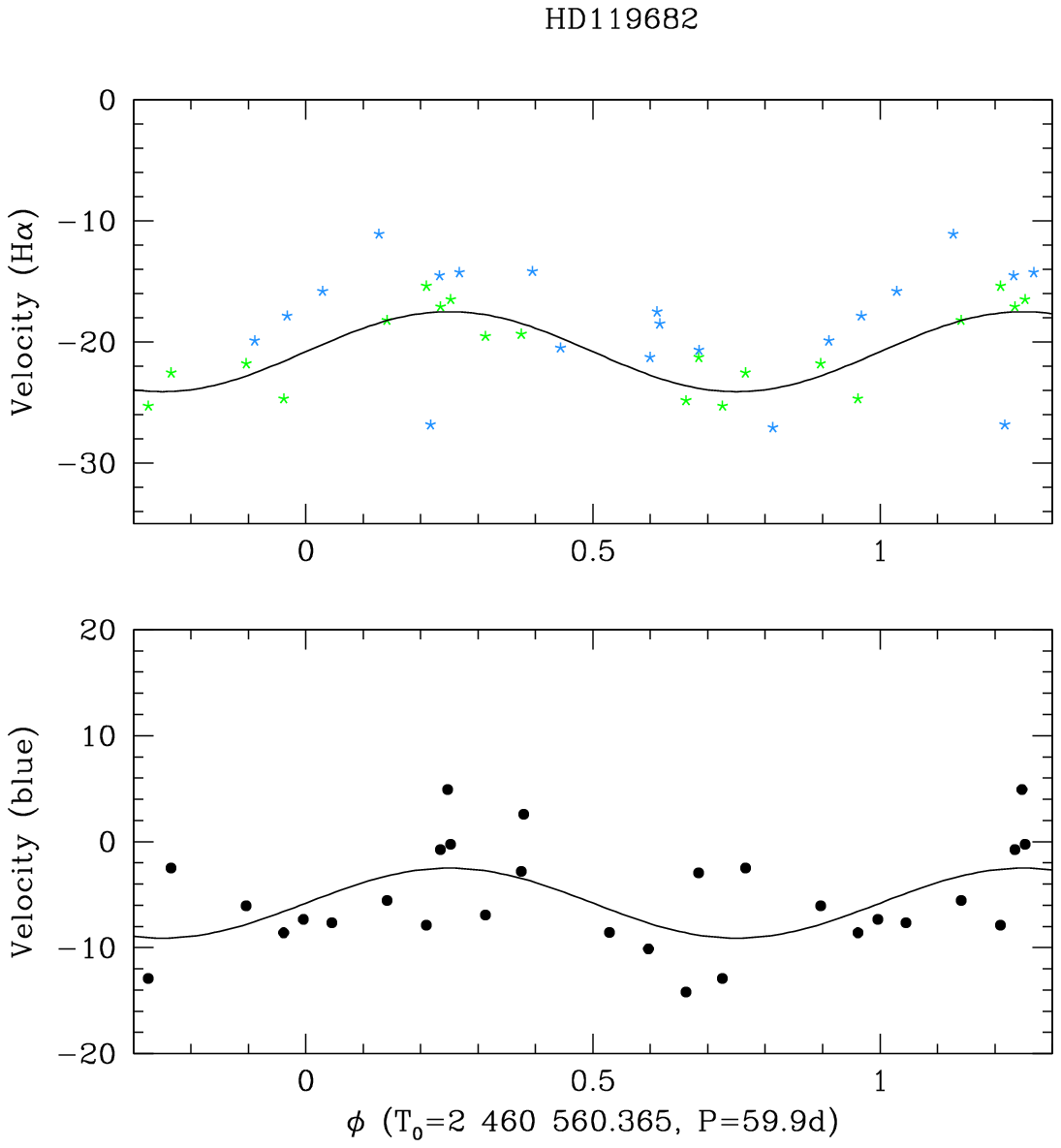}
  \end{center}  
  \caption{Same as Fig. \ref{lineprof} but for HD\,119682, with the H$\beta$ velocities replaced by those from a correlation using blue metallic lines. Only a limited number of representative spectra are shown on the left panel. \label{l119682}}
\end{figure*}

\begin{figure*}
  \begin{center}
    \includegraphics[width=6cm]{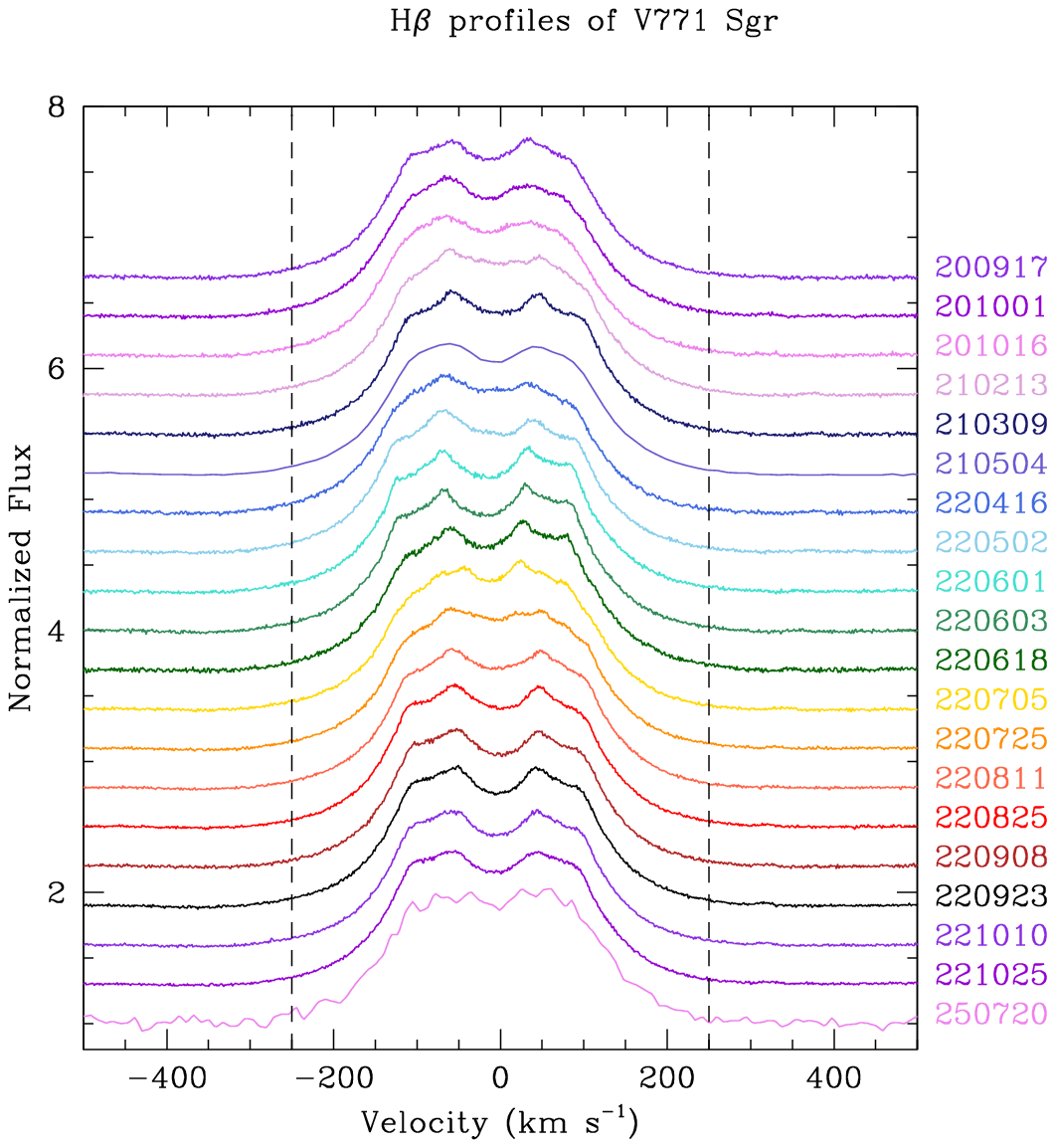}
    \includegraphics[width=6cm]{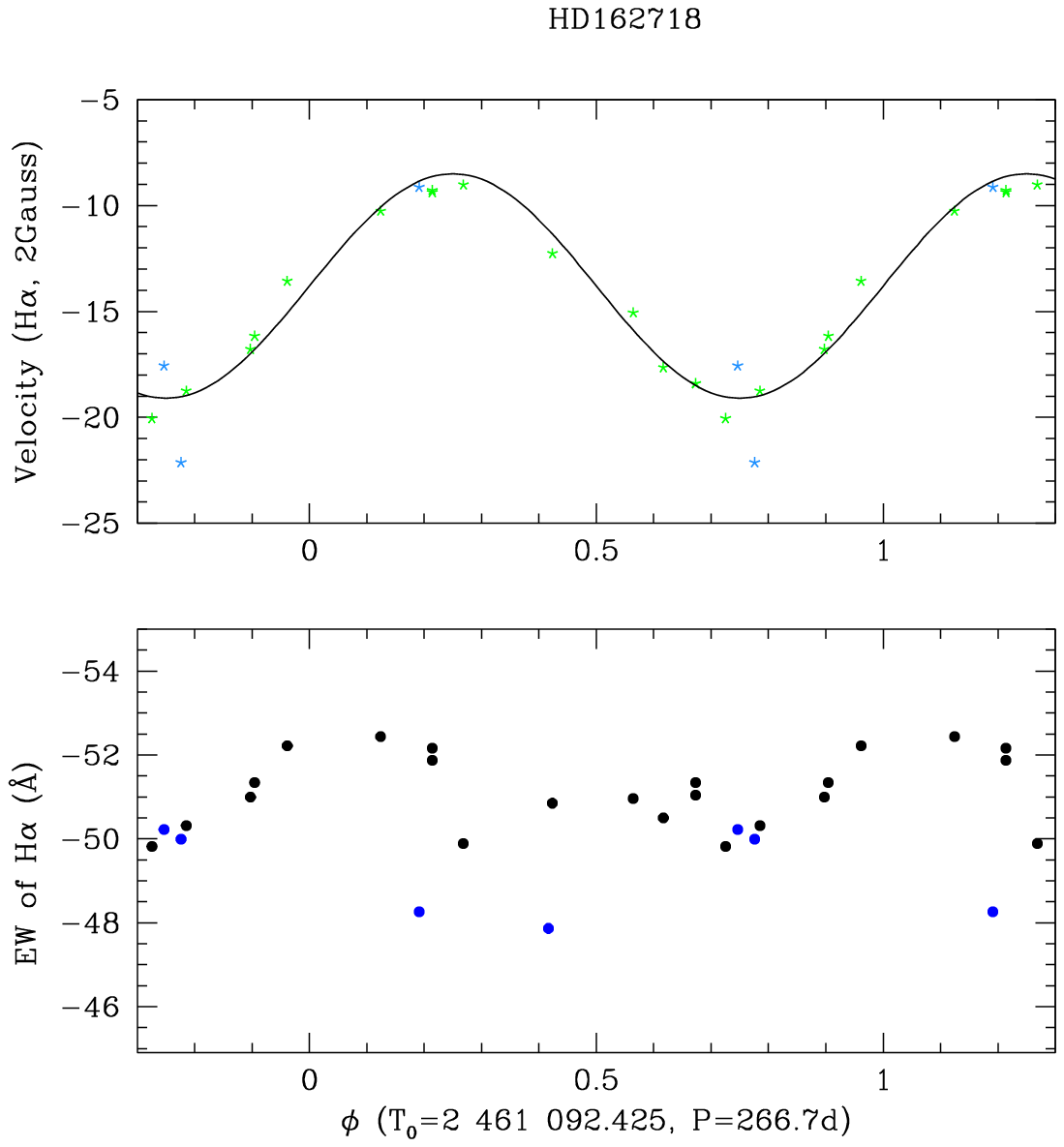}
    \includegraphics[width=6cm]{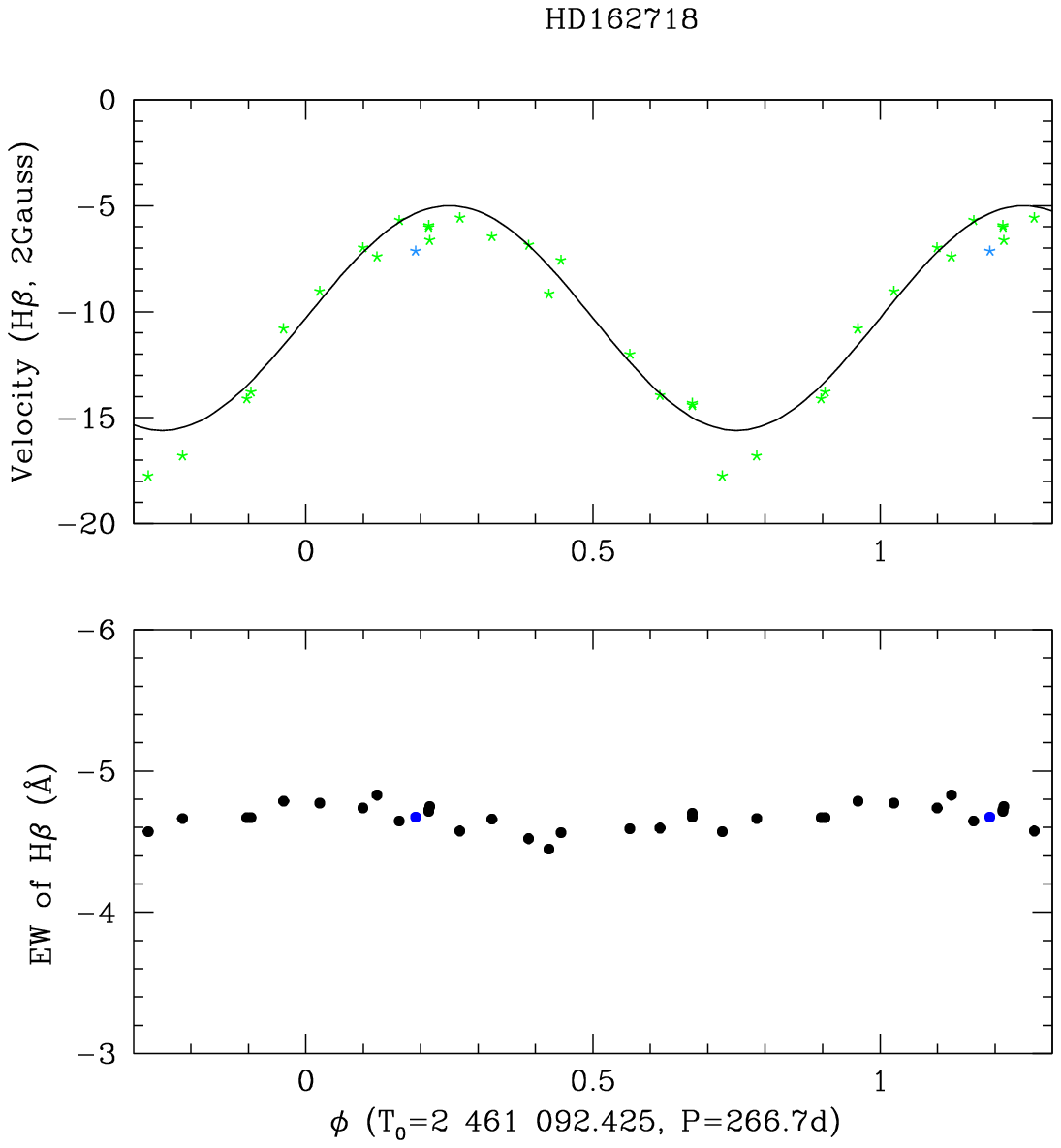}
  \end{center}  
  \caption{{\it Left:} Profiles of the H$\beta$ line in the optical spectra of \vs\ (H$\alpha$ are shown on Fig. \ref{lineprof}). {\it Middle and Right:} RVs (from the double-Gaussian method) and EWs of H$\alpha$ and H$\beta$ folded with the best-fit ephemerides (see Tables \ref{rv} and \ref{bin}). Symbols as in Fig. \ref{lineprof}. \label{v771new}}
\end{figure*}

\begin{figure*}
  \begin{center}
    \includegraphics[width=6cm]{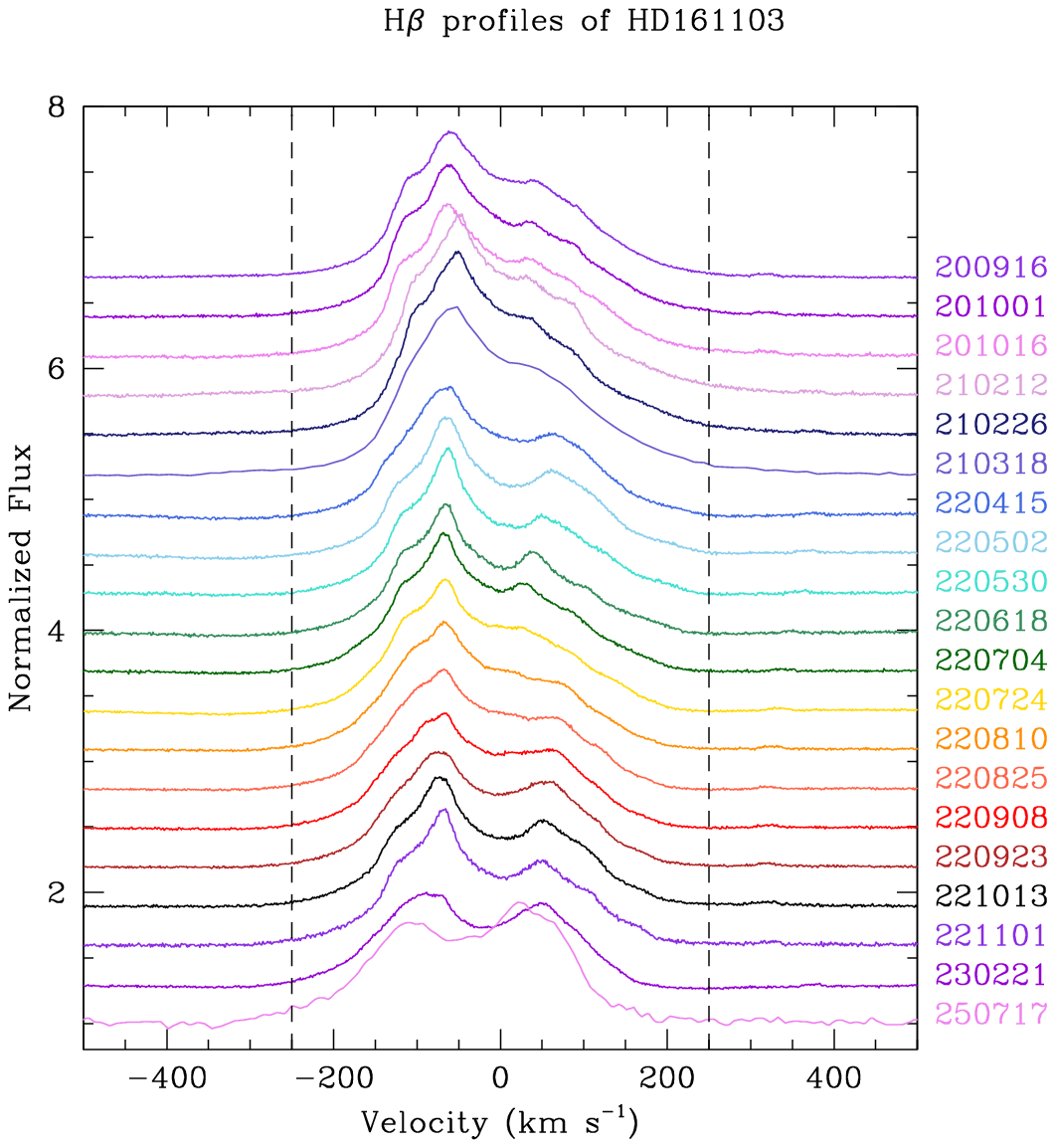}
    \includegraphics[width=6cm]{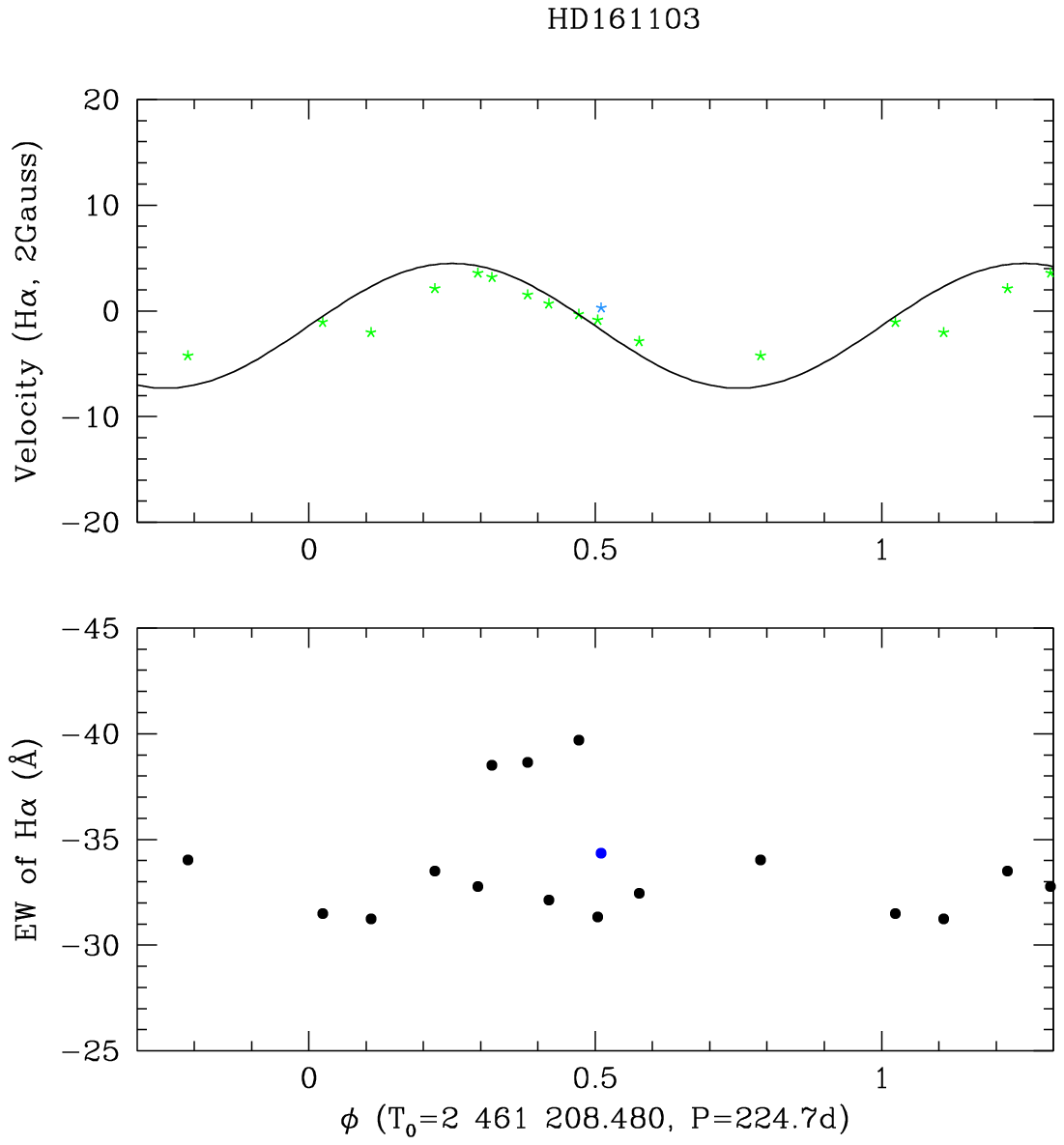}
    \includegraphics[width=6cm]{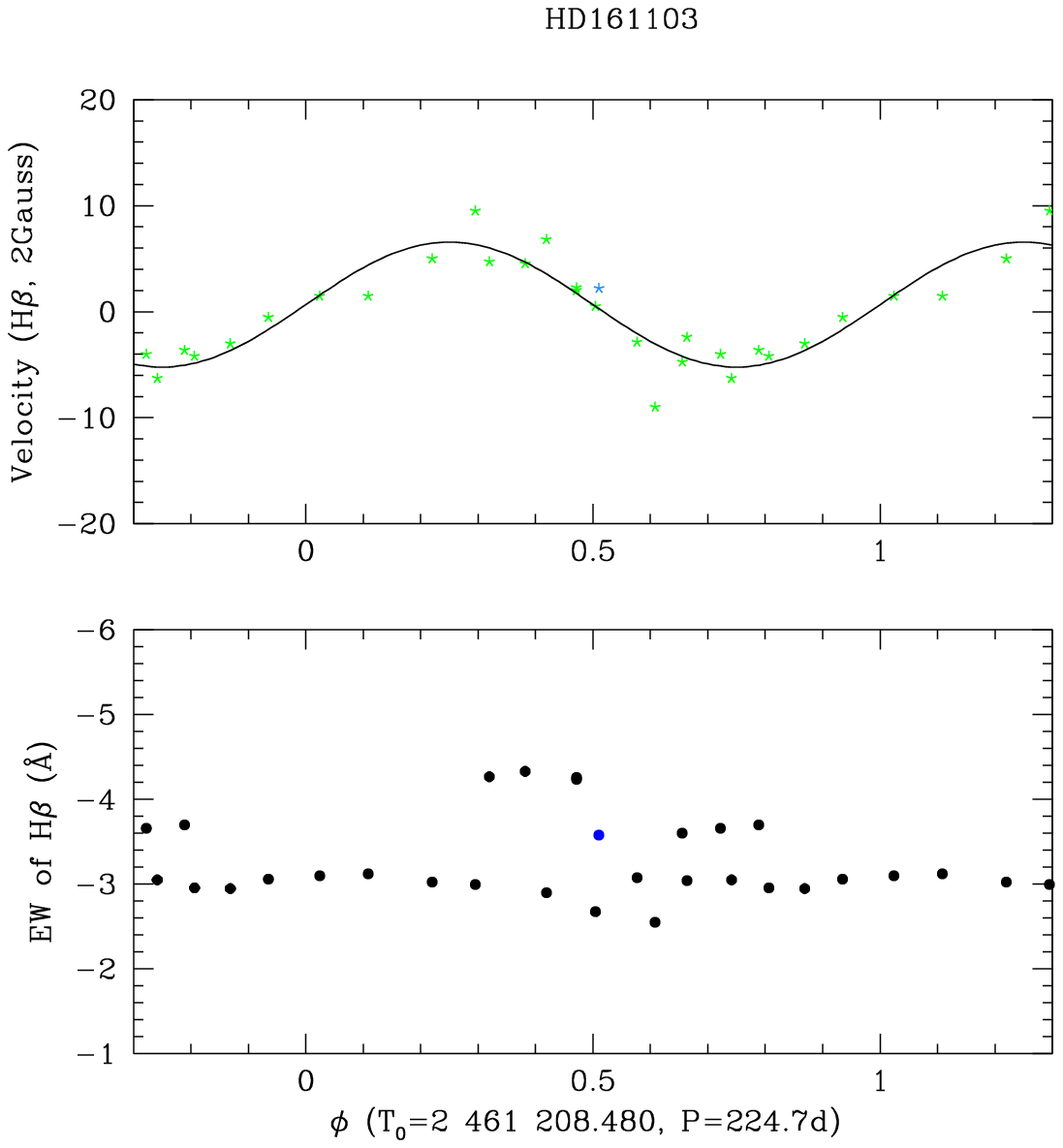}
  \end{center}  
  \caption{Same as Fig. \ref{v771new} but for HD\,161103. \label{161new}}
\end{figure*}

\begin{figure*}
  \begin{center}
    \includegraphics[width=6cm]{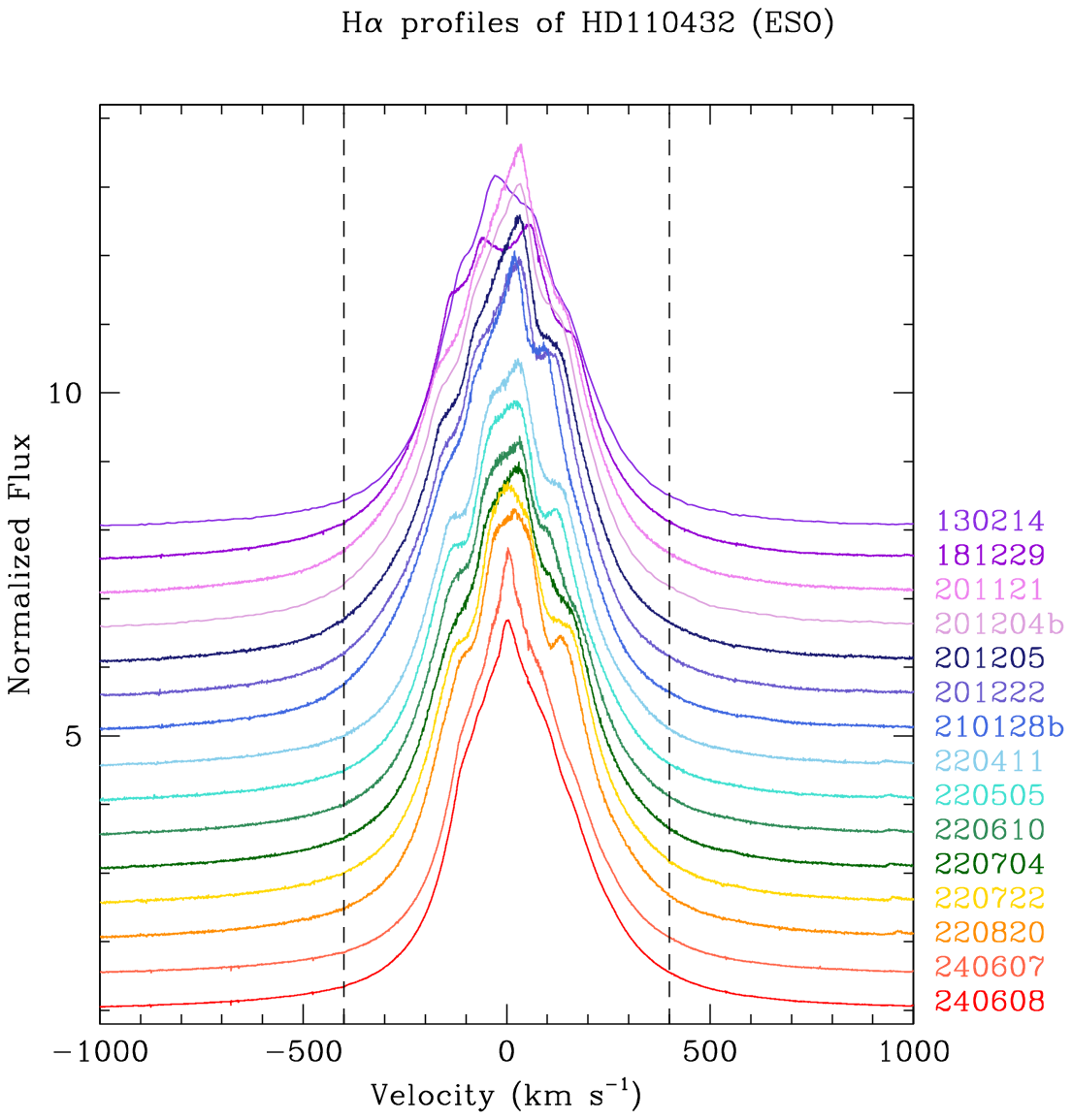}
    \includegraphics[width=6cm]{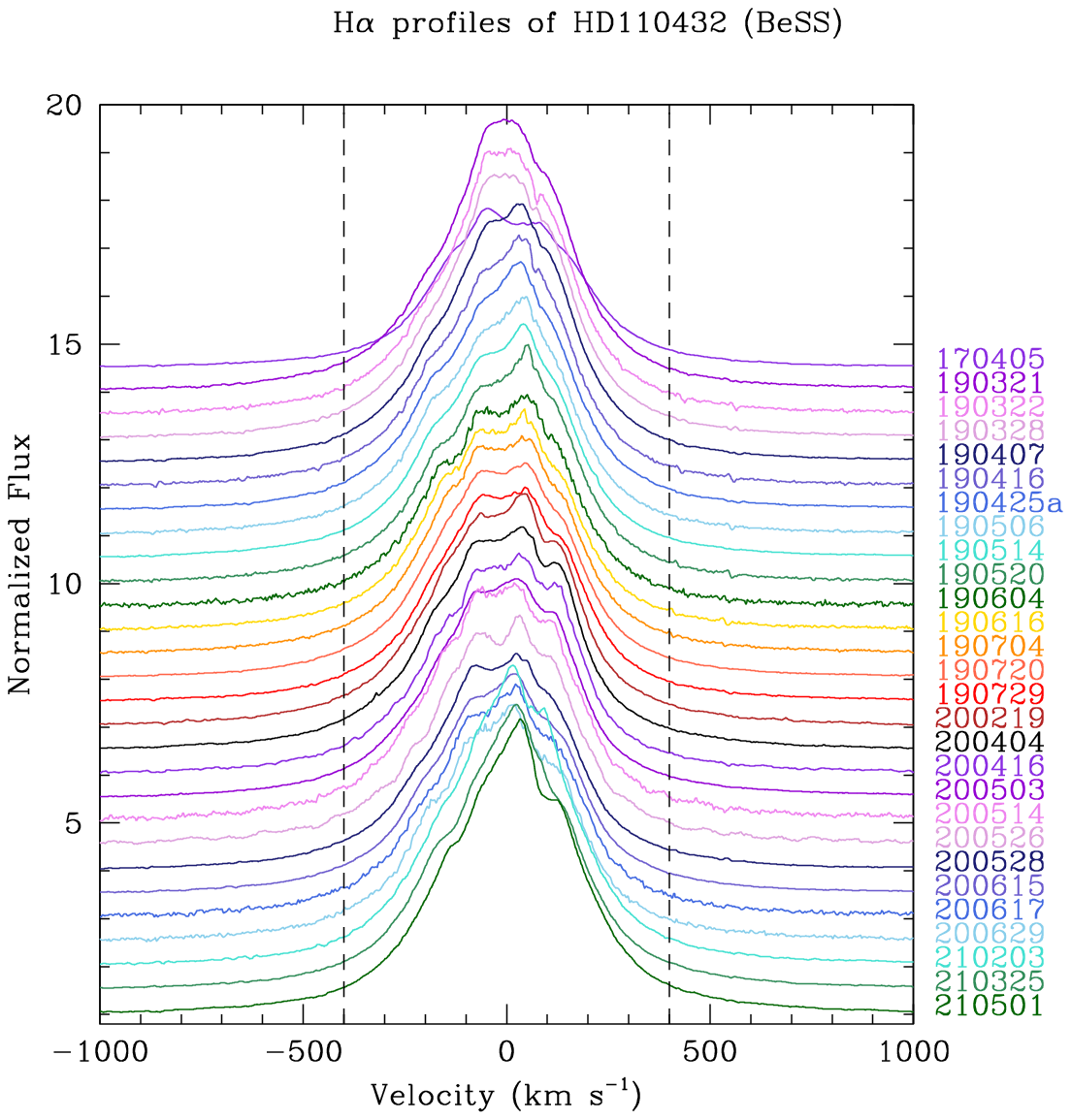}
    \includegraphics[width=6cm]{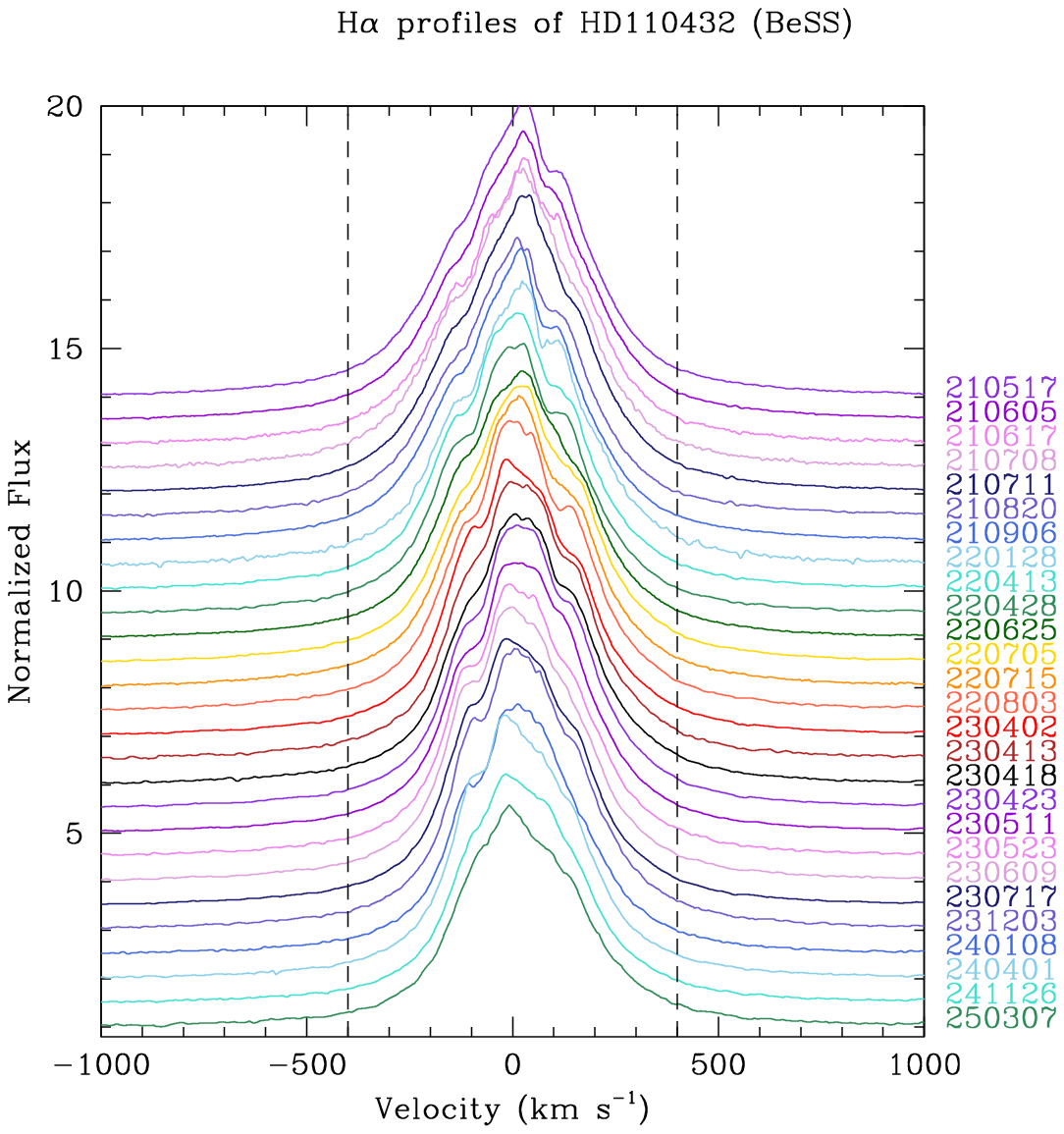}
    \includegraphics[width=4.5cm]{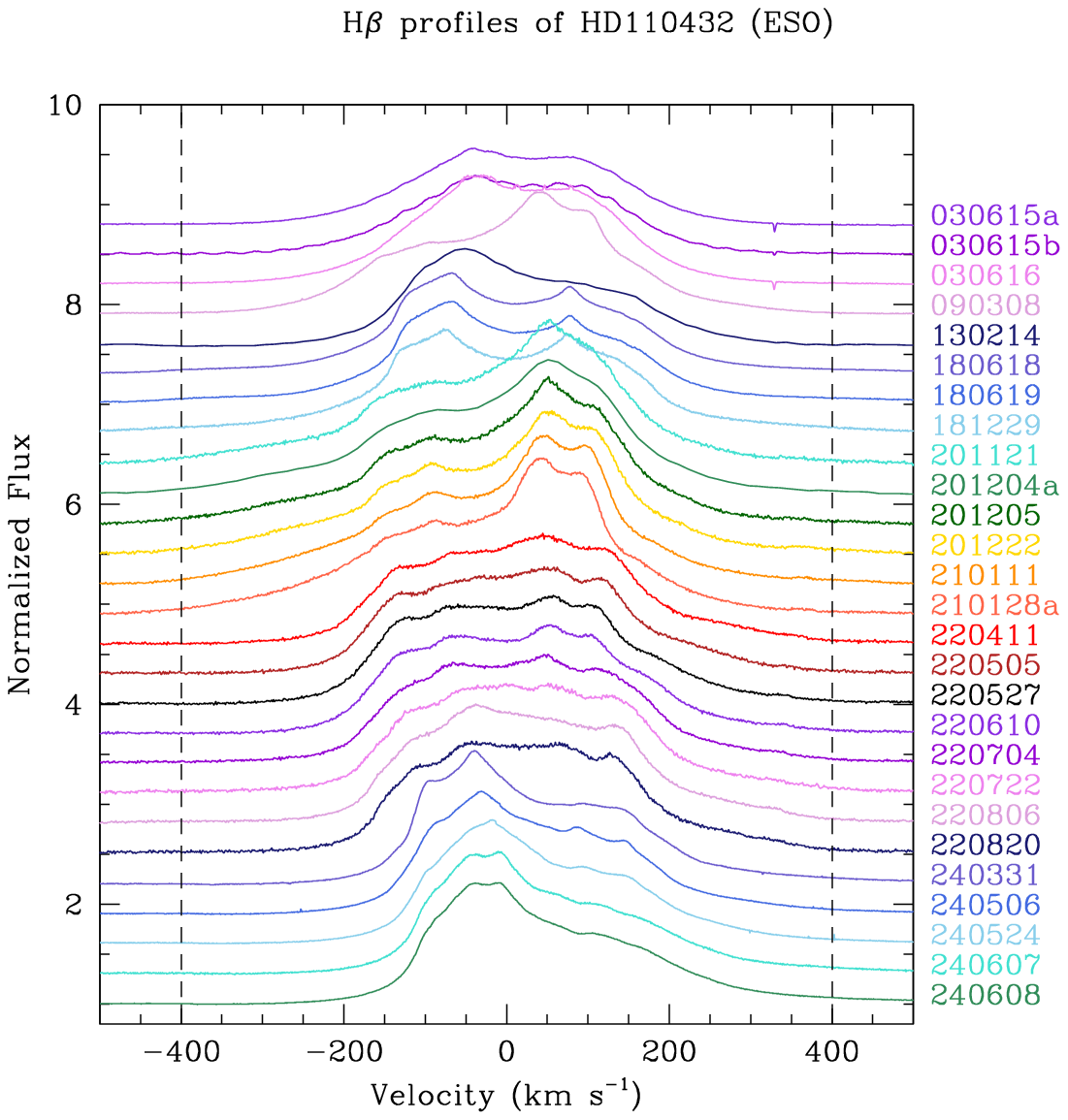}
    \includegraphics[width=4.5cm]{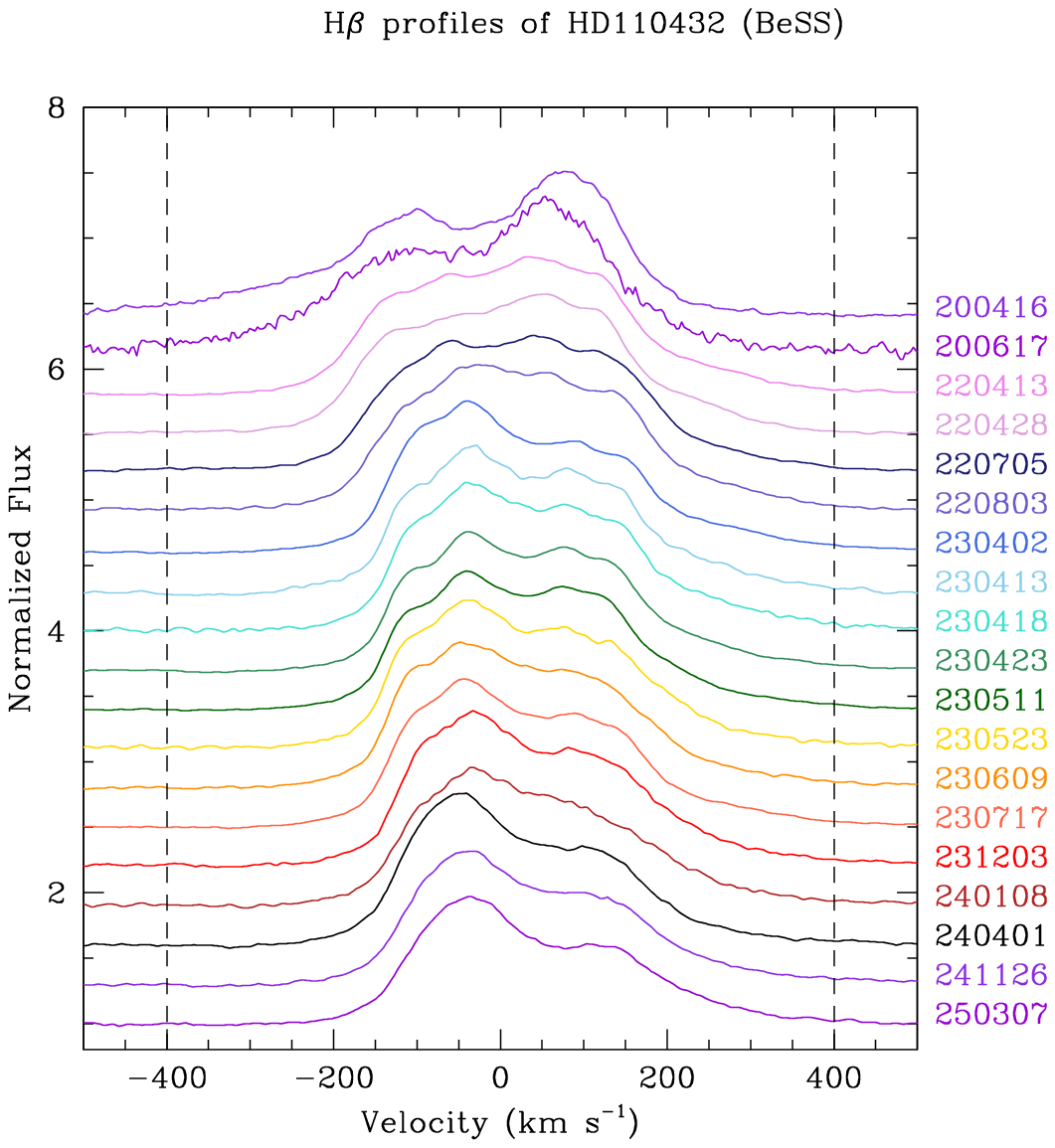}
    \includegraphics[width=4.5cm]{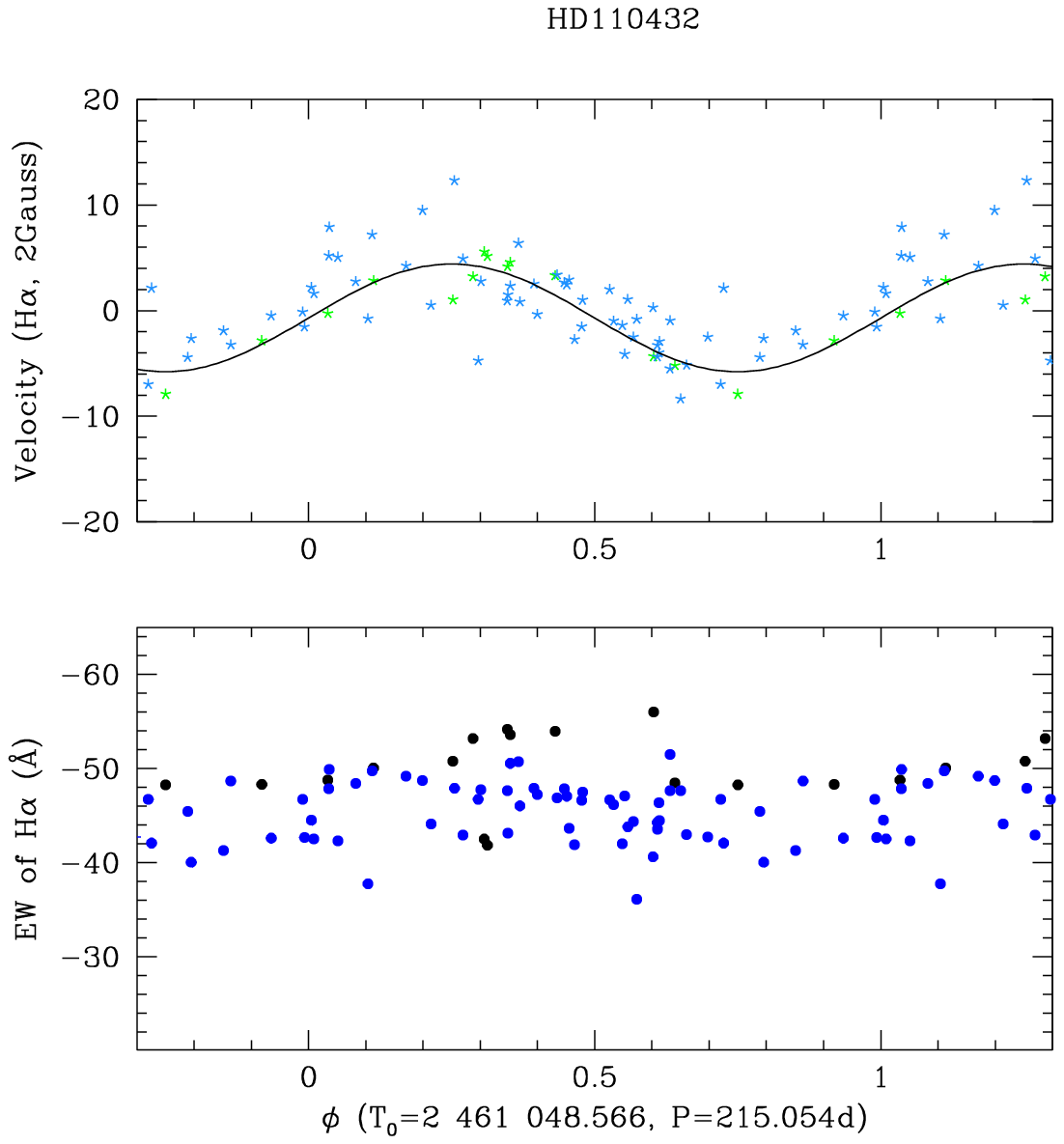}
    \includegraphics[width=4.5cm]{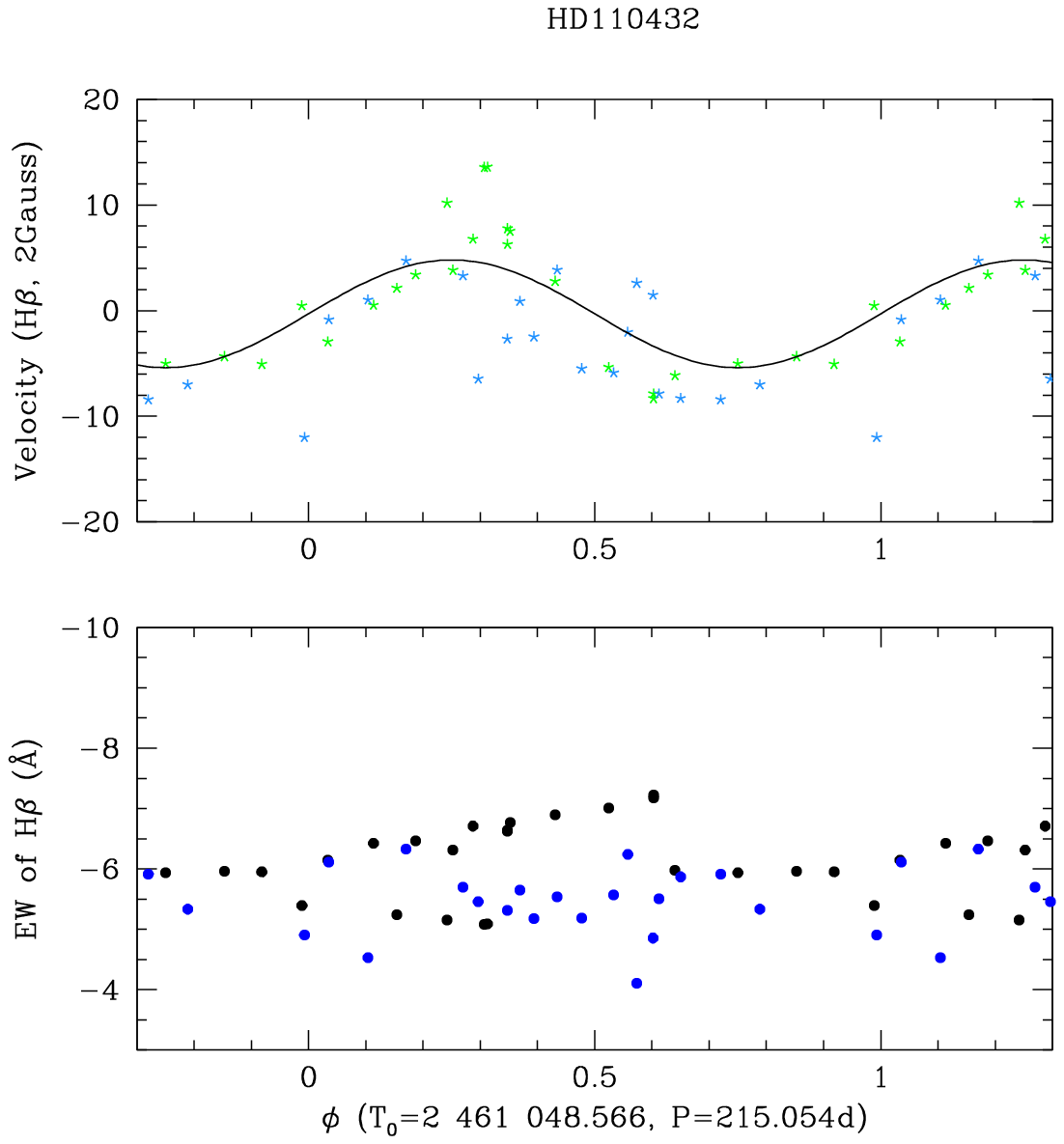}
  \end{center}  
  \caption{{\it Top:} Profiles of the H$\alpha$ line in the whole set of optical spectra of HD\,110432. {\it Bottom left:} Same for H$\beta$. {\it Bottom right:} RVs (from the double-Gaussian method) and EWs of H$\alpha$ and H$\beta$ folded with the best-fit ephemerides (see Tables  \ref{rv110432} and \ref{bin}).  \label{110new}}
\end{figure*}

\begin{figure*}
  \begin{center}
    \includegraphics[width=4.5cm]{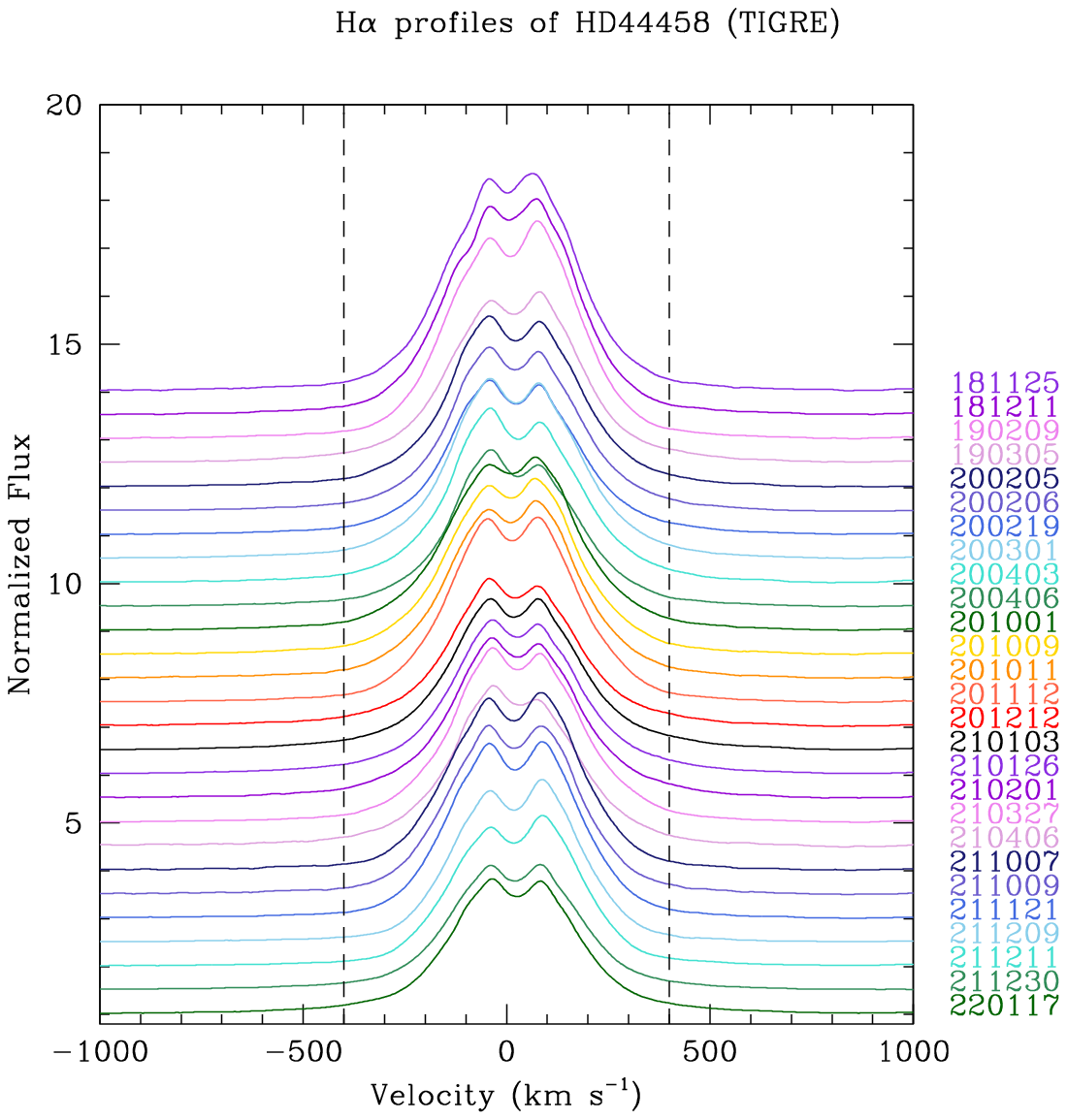}
    \includegraphics[width=4.5cm]{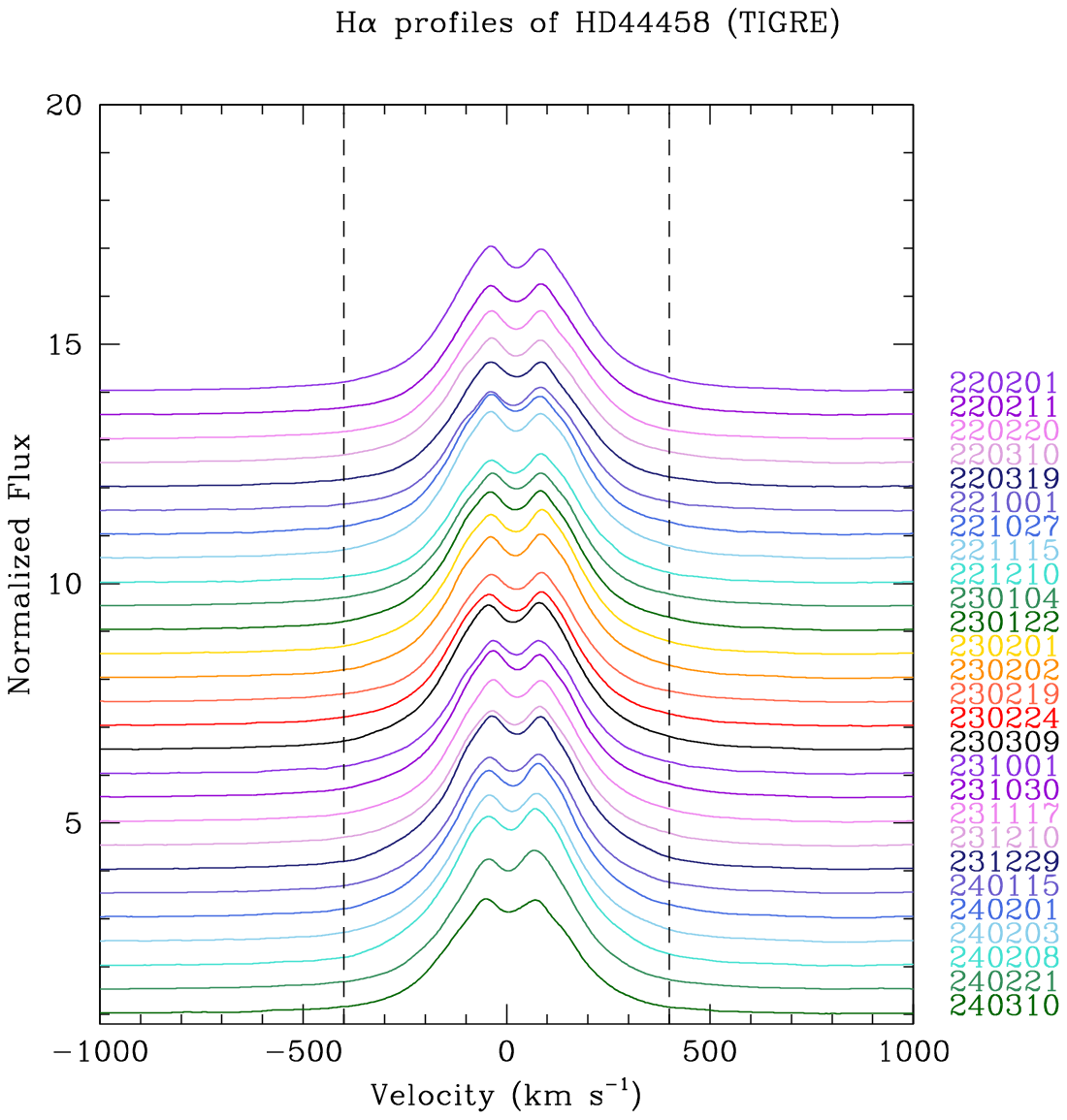}
    \includegraphics[width=4.5cm]{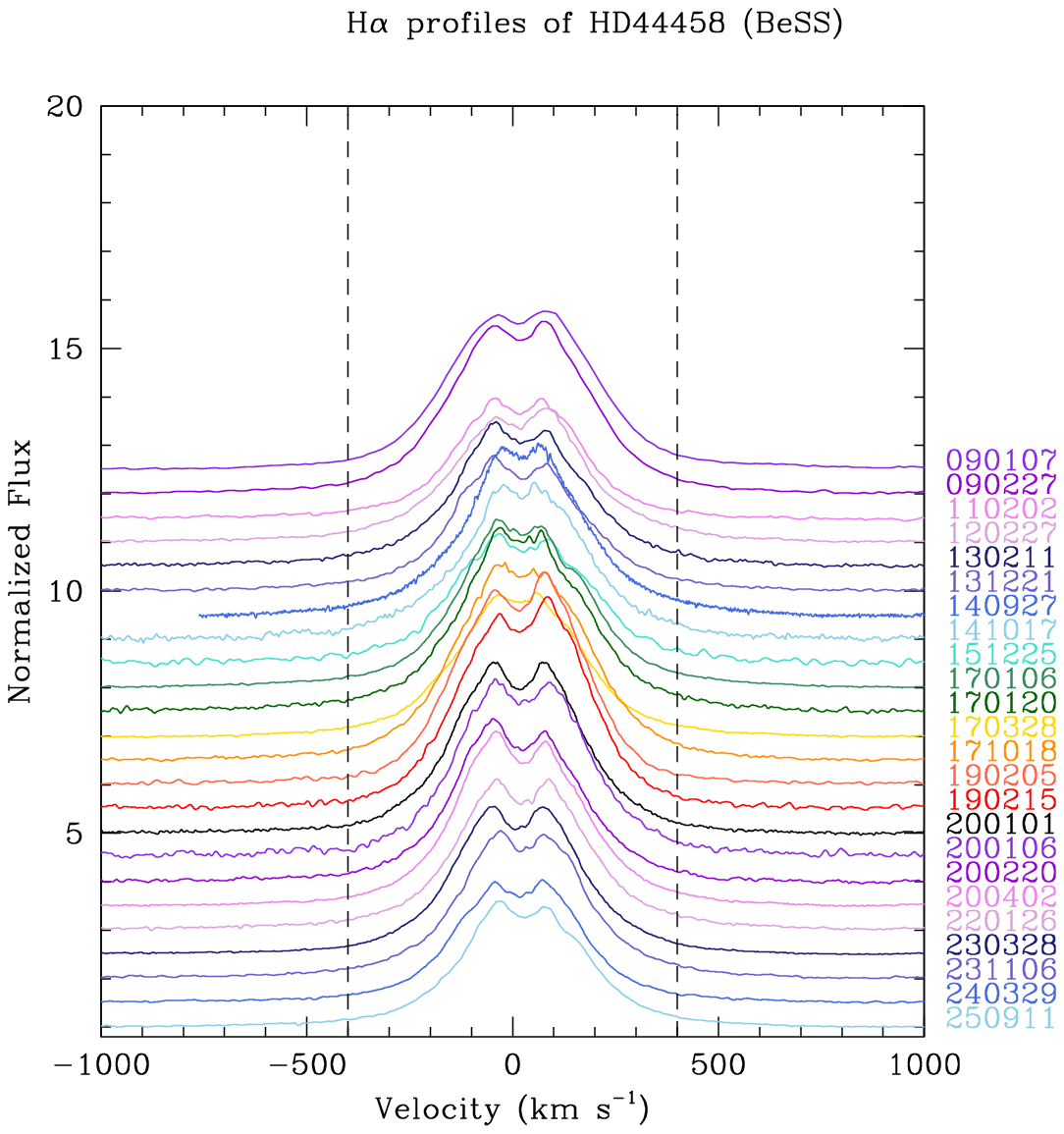}
    \includegraphics[width=4.5cm]{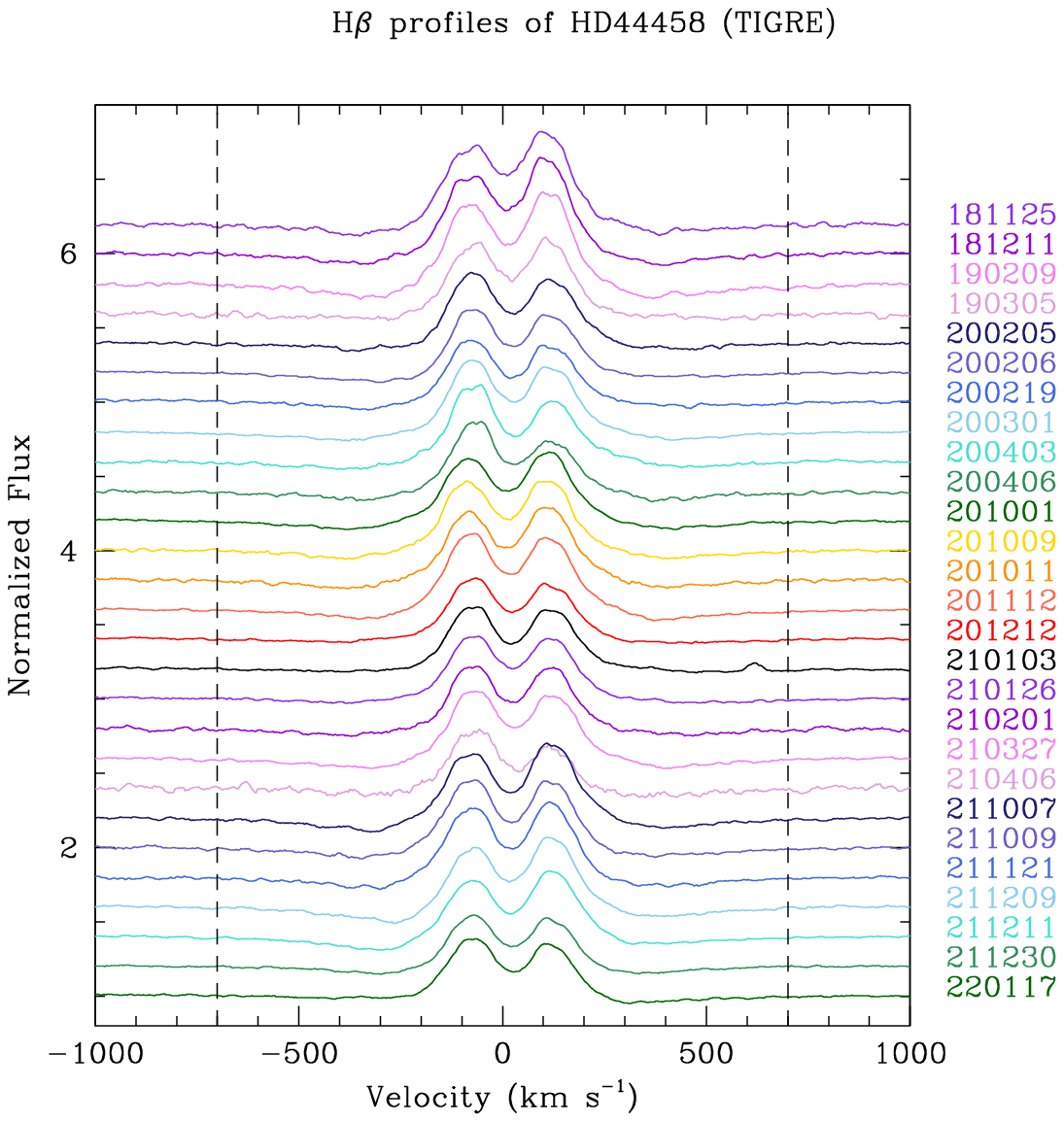}
    \includegraphics[width=4.5cm]{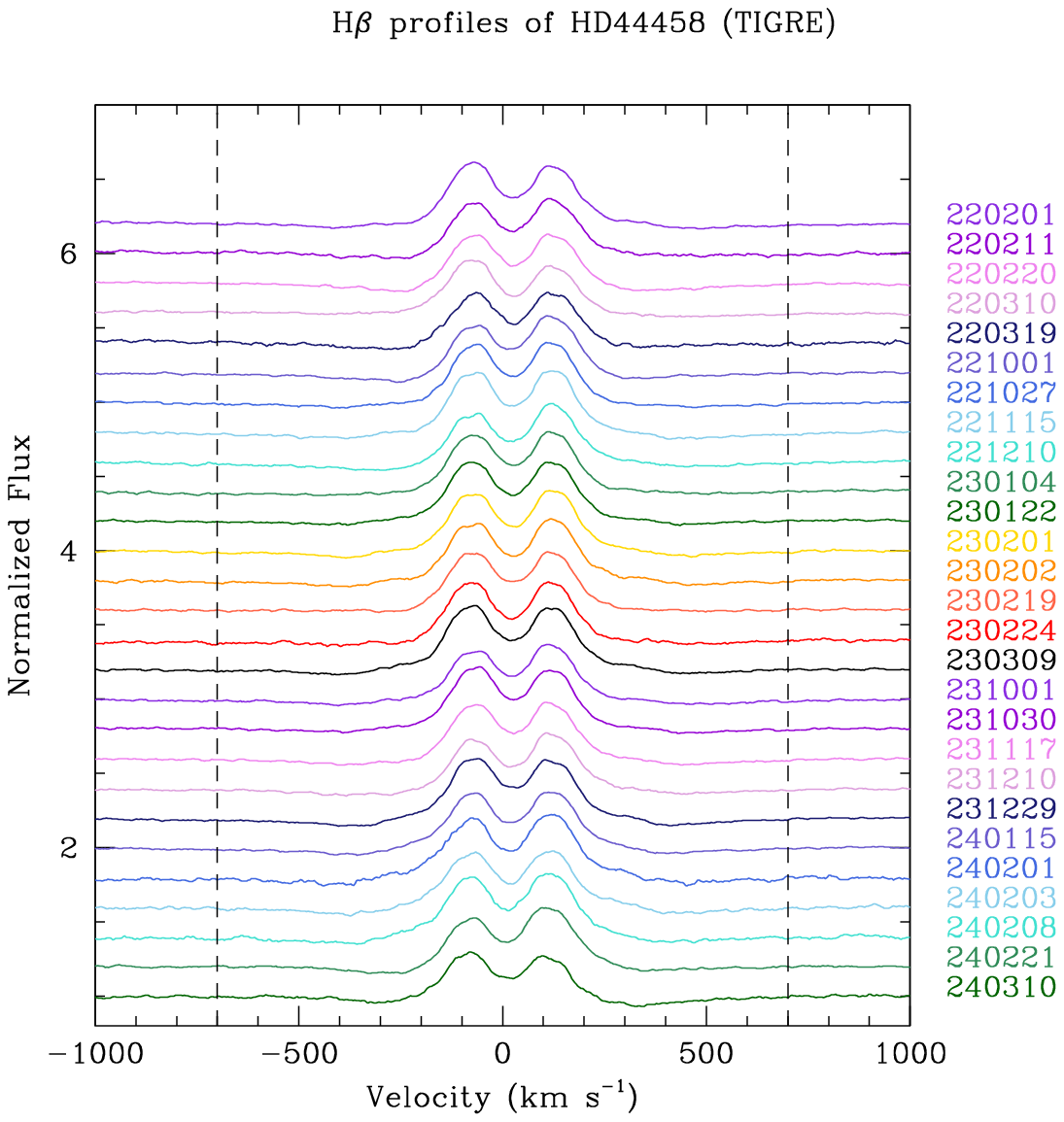}
    \includegraphics[width=4.5cm]{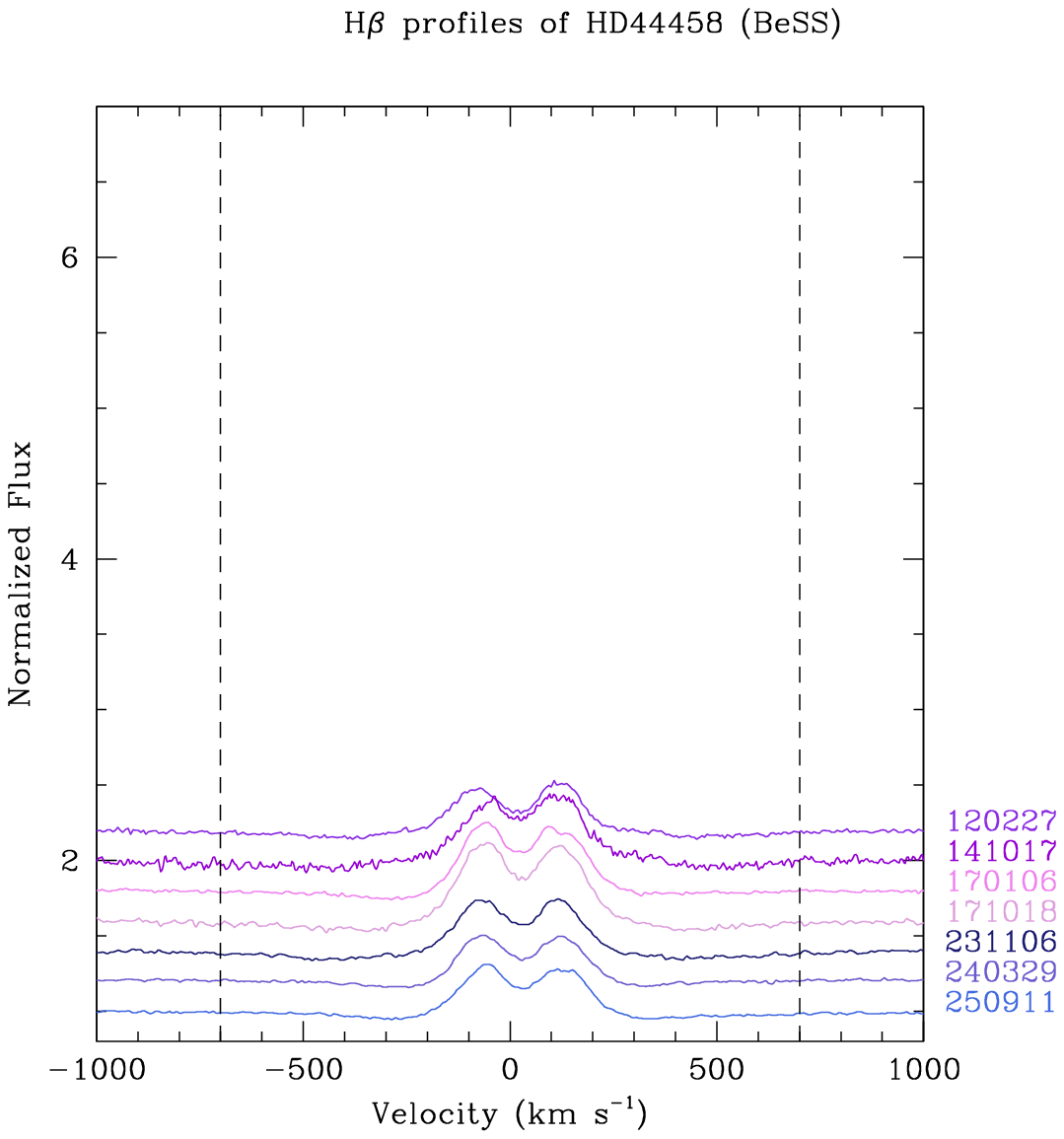}
    \includegraphics[width=4.5cm]{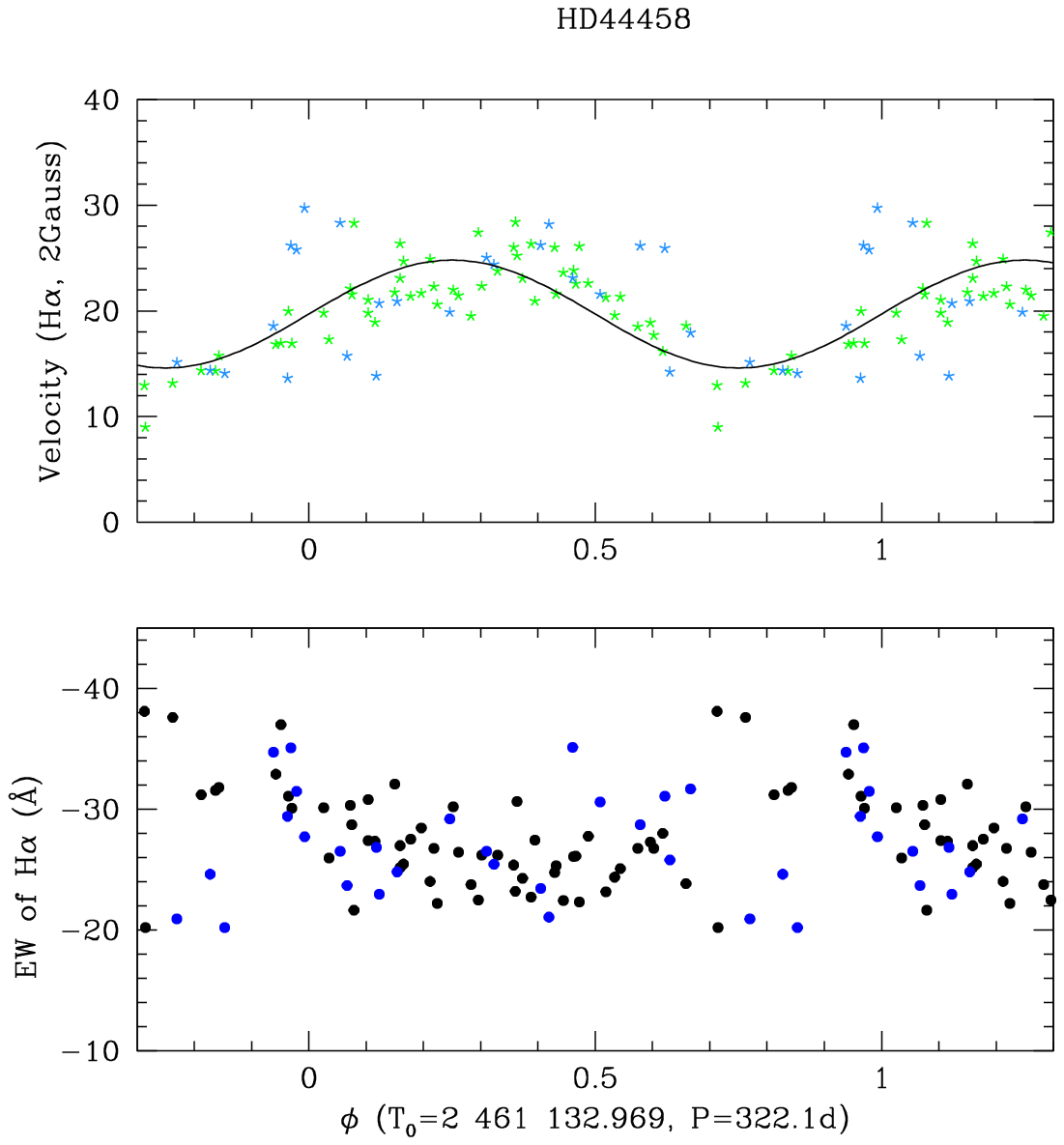}
    \includegraphics[width=4.5cm]{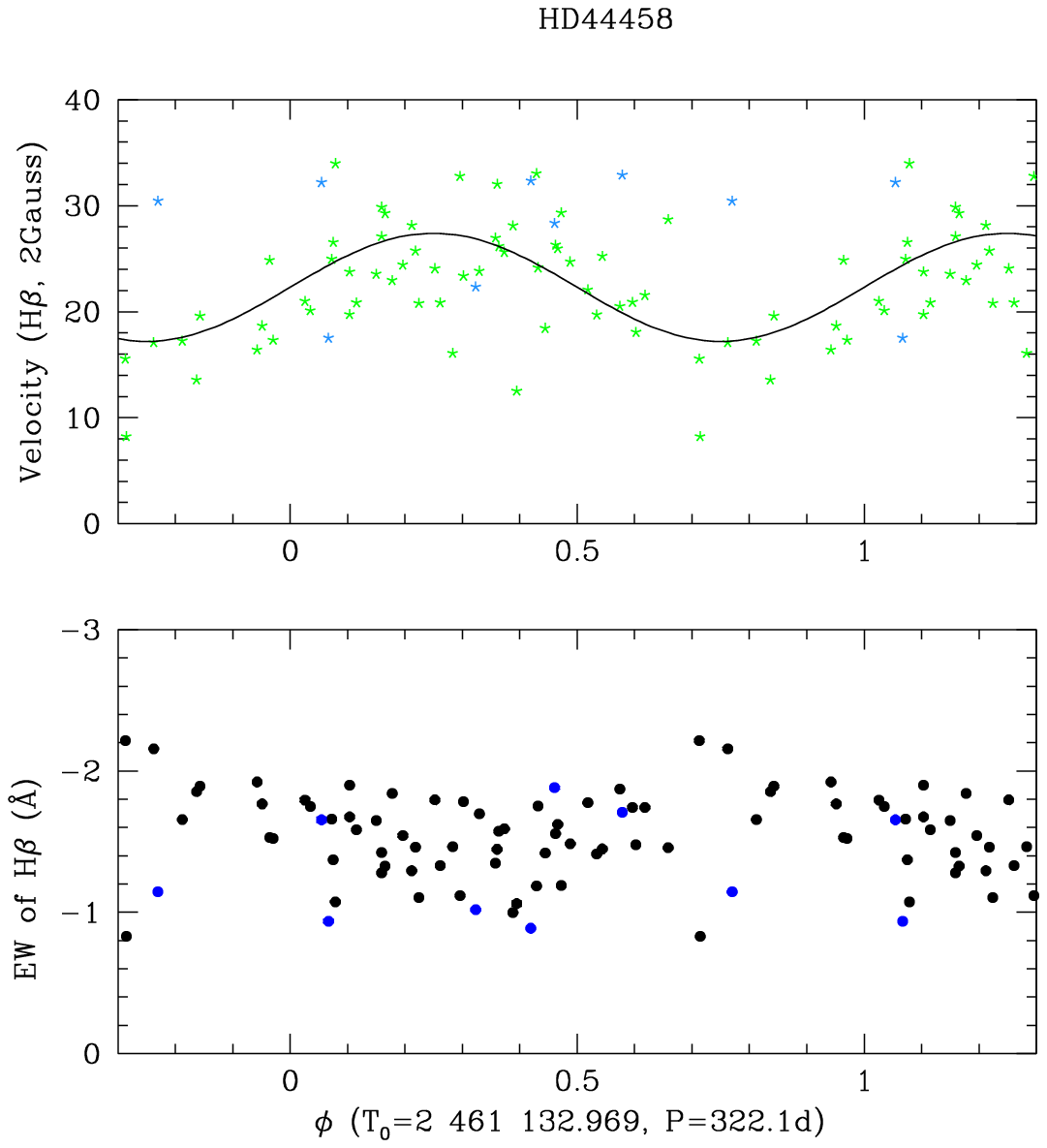}
  \end{center}  
  \caption{Same as Fig. \ref{110new} but for HD\,44458 (but see Table \ref{rv44458} for individual RV and EW values). \label{44new}}
\end{figure*}

\begin{figure*}
  \begin{center}
    \includegraphics[width=6cm]{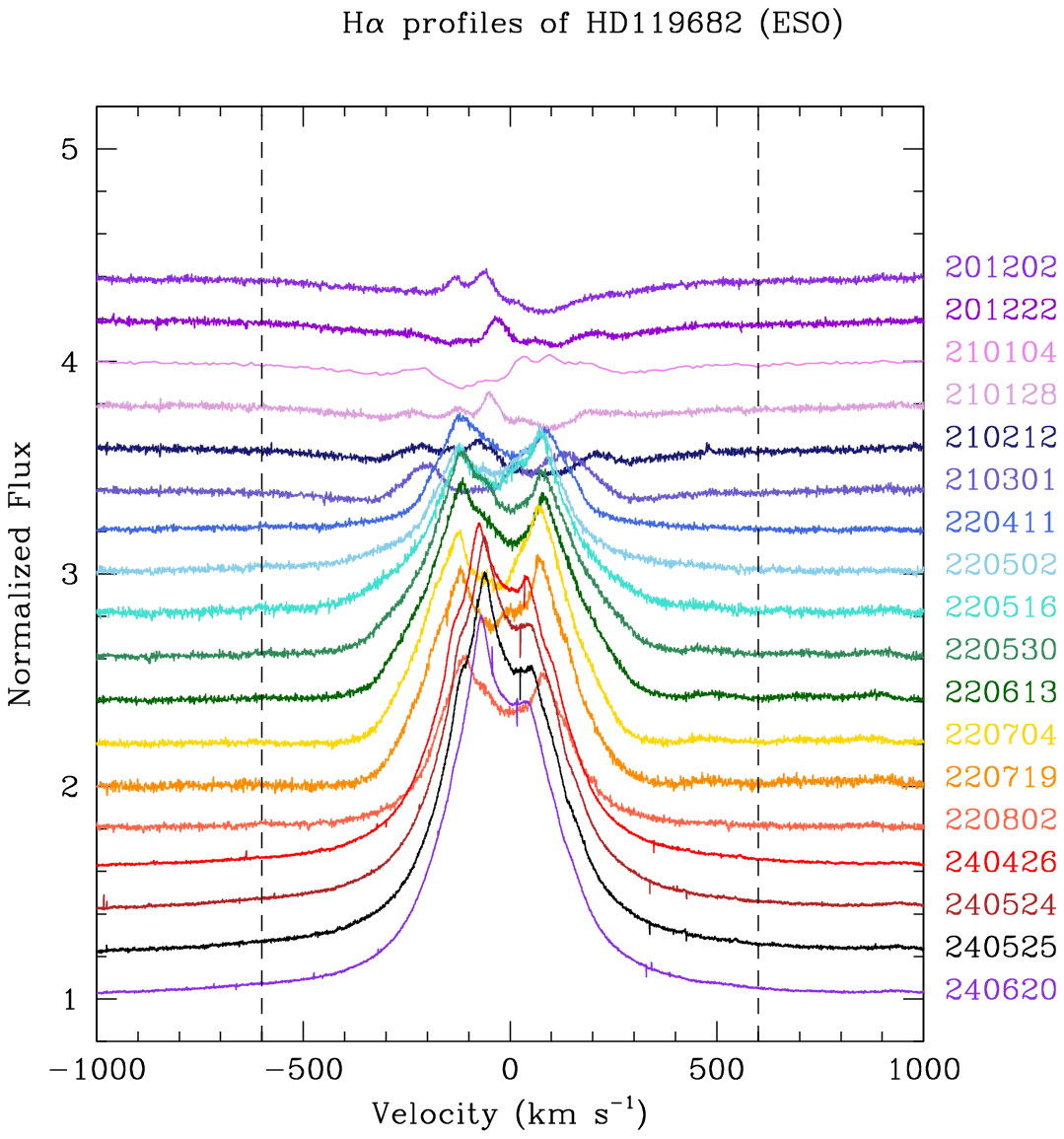}
    \includegraphics[width=6cm]{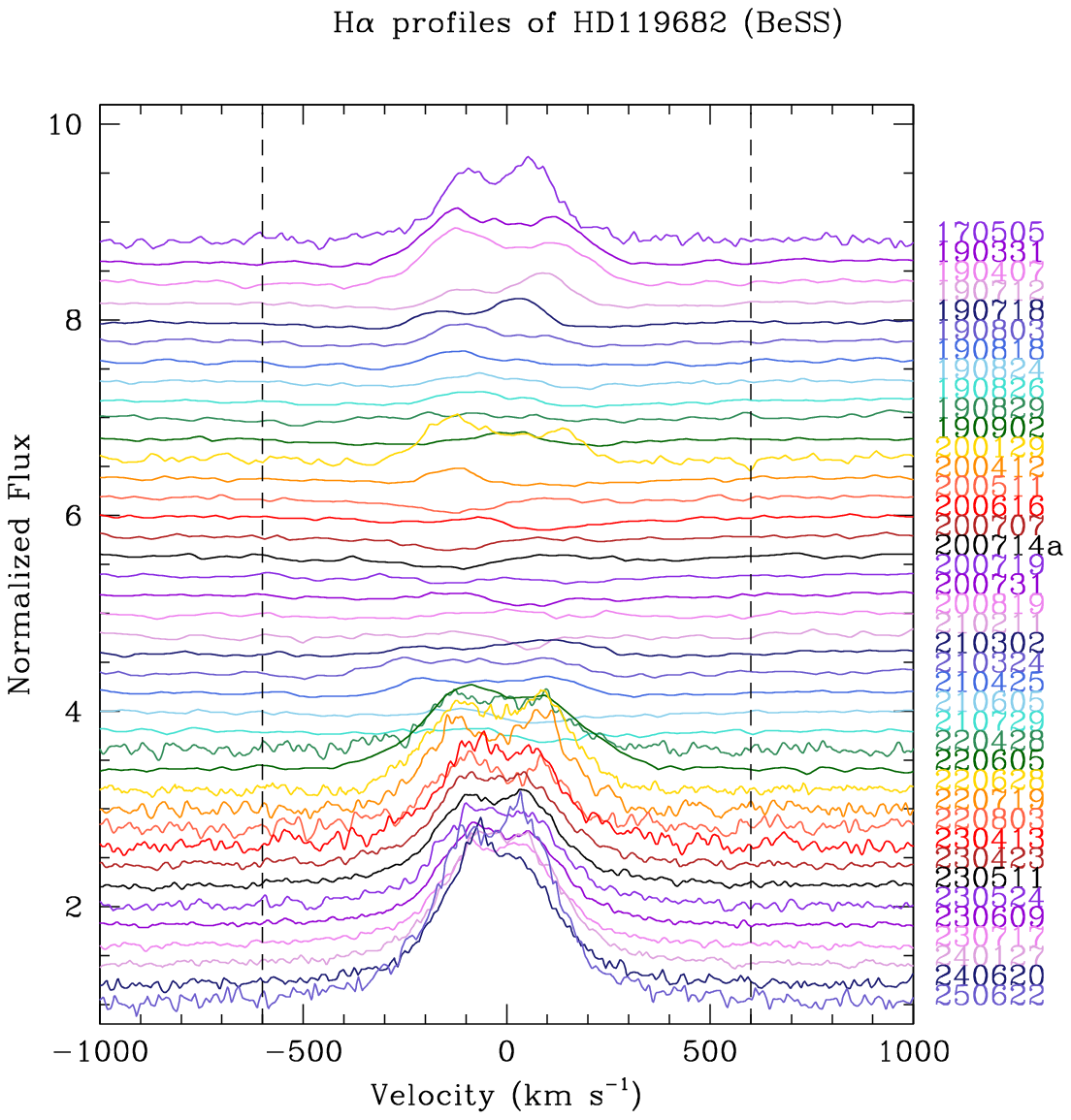}
    \includegraphics[width=6cm]{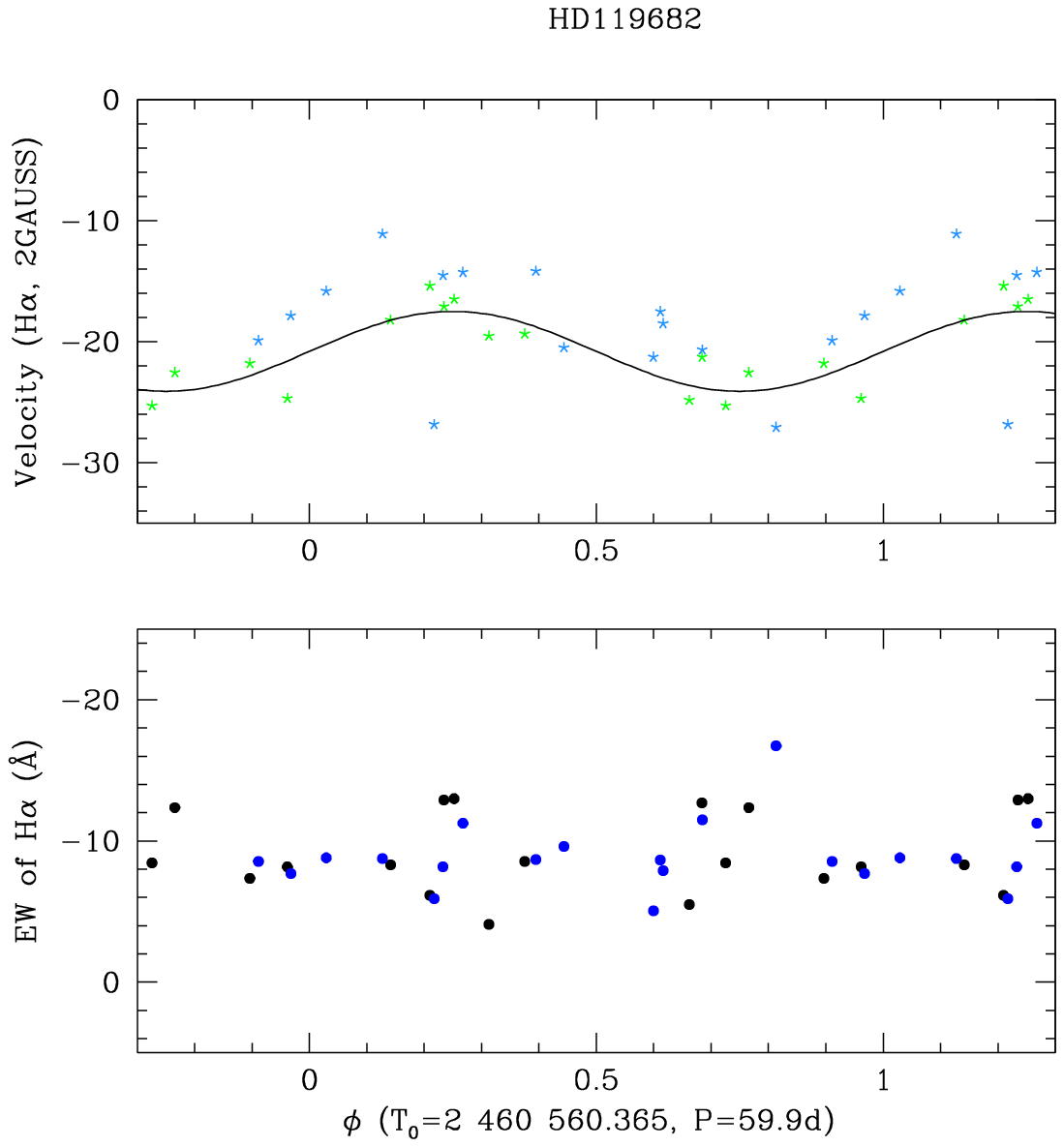}
  \end{center}  
  \caption{{\it Left and Middle:} Profiles of the H$\alpha$ line in the whole set of optical spectra of HD\,119682. {\it Right:} RVs (from the double-Gaussian method) and EWs of H$\alpha$ folded with the best-fit ephemerides (see Tables \ref{rv119682} and \ref{bin}). \label{119new}}
\end{figure*}

Tables \ref{rv} to \ref{rv119682} yield the measured EWs and RVs, while Figures \ref{lineprof} to \ref{l119682} display the (typical) H$\alpha$ line profiles and the RV evolution with time or orbital phase. In addition, Table \ref{journal} provides the results of the X-ray spectral fitting. Finally, Figures \ref{v771new} to \ref{119new} provide the whole sets of H$\alpha$ and H$\beta$ line profiles and phase-folded RVs and EWs. With the possible exception of \vs, the EWs do not show obvious phase-related variations. Also, even when RV trends and EW changes were spotted, de-trended RVs taken at the same phase agree well.

\begin{table*}
  \begin{center}
  \scriptsize
  \caption{EWs and RVs of H$\alpha$ and H$\beta$ measured for \vs\ and HD\,161103.
 \label{rv}}
  \begin{tabular}{lcccc|lcccc}
    \hline
    \multicolumn{5}{c|}{\vs} & \multicolumn{5}{c}{HD\,161103}\\
mid-HJD & \multicolumn{2}{c}{H$\alpha$} & \multicolumn{2}{c|}{H$\beta$} & mid-HJD & \multicolumn{2}{c}{H$\alpha$} & \multicolumn{2}{c}{H$\beta$}\\
$-2\,450\,000$ & EW(\AA) & RV(\kms) & EW(\AA) & RV(\kms) & $-2\,450\,000$ & EW(\AA) & RV(\kms) & EW(\AA) & RV(\kms) \\
    \hline
  6499.389$^b$ & -50.0 & -22.1 &          &            &9108.618 &	   &	         &   -3.6 &  -1.8 (-4.7)\\     
  9109.590 &   -51.0 & -15.1  &     -4.59 &  -12.0     &9123.559 &	   &		 &   -3.7 &  -1.4 (-4.0)\\     
  9123.602 &   -50.5 & -17.6  &     -4.60 &  -13.9     &9138.546 &  -34.0 & -16.9 (-4.2)&   -3.7 &  -1.4 (-3.6)\\     
  9138.574 &   -51.3 & -18.4  &     -4.70 &  -14.4     &9257.848 &  -38.5 & -11.3 ( 3.2)&   -4.3 &   4.0 ( 4.7)\\     
  9258.870 &   -52.4 & -10.3  &     -4.83 &   -7.4     &9271.856 &  -38.6 & -13.2 ( 1.5)&   -4.3 &   3.4 ( 4.5)\\     
  9282.837 &   -52.2 &  -9.4  &     -4.71 &   -5.9     &9291.894$^x$ &        &             &   -4.3 &   0.4 ( 2.0)\\ 
9338.711$^x$&  -50.9 & -12.3  &     -4.45 &   -9.2     &9291.897$^x$ &  -39.7 & -15.3 (-0.3)&   -4.2 &   0.7 ( 2.3)\\ 
  9424.783$^b$ & -50.2 & -17.6 &          &            &9684.883 &  -33.5 & -18.9 ( 2.1)&   -3.0 &  -6.4 ( 5.0)\\     
  9685.825 &   -49.8 & -20.1  &     -4.57 &  -17.7     &9701.778 &  -32.8 & -17.7 ( 3.6)&   -3.0 &  -2.3 ( 9.5)\\     
  9701.792 &   -50.3 & -18.8  &     -4.66 &  -16.8     &9729.580 &  -32.1 & -21.1 ( 0.7)&   -2.9 &  -5.7 ( 6.8)\\     
  9731.673 &   -51.0 & -16.8  &     -4.67 &  -14.1     &9748.762 &  -31.3 & -22.9 (-0.9)&   -2.7 & -12.5 ( 0.5)\\     
  9733.625 &   -51.3 & -16.2  &     -4.67 &  -13.8     &9765.008 &  -32.5 & -25.2 (-2.9)&   -3.1 & -16.3 (-2.9)\\     
  9748.776 &   -52.2 & -13.6  &     -4.79 &  -10.8     &9784.553 & 	   &		 &   -3.0 & -16.3 (-2.4)\\     
  9765.520 &         &        &     -4.77 &   -9.0     &9802.002 & 	   &		 &   -3.0 & -20.6 (-6.3)\\     
  9785.631 &         &        &     -4.74 &   -7.0     &9816.595 & 	   &		 &   -3.0 & -18.9 (-4.2)\\     
  9802.528 &         &        &     -4.65 &   -5.7     &9830.620 & 	   &		 &   -2.9 & -18.1 (-3.0)\\     
  9816.609 &         &        &     -4.75 &   -6.6     &9845.510 & 	   &		 &   -3.1 & -16.0 (-0.5)\\     
  9830.669 &   -49.9 & -9.0   &     -4.58 &   -5.6     &9865.574 &  -31.5 & -24.9 (-1.1)&   -3.1 & -14.5 ( 1.5)\\     
  9845.529 &         &        &     -4.66 &   -6.4     &9884.538 &  -31.3 & -26.2 (-2.0)&   -3.1 & -14.9 ( 1.5)\\     
  9862.582 &         &        &     -4.52 &   -6.9     &9996.856 & 	   & 	    	 &   -2.5 & -28.2 (-9.0)\\     
  9877.511 &         &        &     -4.56 &   -7.6    &10873.762$^b$ &  -34.3 & -39.2 ( 0.3)&   -3.6 & -38.9 ( 2.2)\\ 
 10136.797$^b$ & -47.9 &       &       &        \\
 10876.745$^b$ & -48.3 &  -9.1 & -4.67 & -7.1   \\
\hline
  \end{tabular}
  \end{center}
\vspace*{-0.5cm}  
\tablefoot{The superscripts $^x$ and $^b$ identify the Xshooter and BeSS spectra, respectively, all other values correspond to UVES spectra. The RV values correspond to those derived using the double Gaussian method; typical RV error is about 1\,\kms\ for professional data and about 5\,\kms\ for amateur data. Saturation explains the missing H$\alpha$ values for ESO spectra, and a problem of wavelength calibration prohibited the derivation of a correct RV in the penultimate BeSS spectrum. For \vs, as in \citet{naz22}, the velocity range for calculations on H$\alpha$ is $\pm$400\,\kms\ (Fig. \ref{lineprof}), the amplitude range for the mirror method is 1.5--5, and the two Gaussians are centered on $\pm$150\,\kms\ with a 10\,\kms\ width (values are $\pm$250\,\kms, 1.2--1.6, $\pm$150\,\kms, and 10\,\kms\ for H$\beta$). For HD\,161103, the velocity range for calculations on H$\alpha$ is $\pm$400\,\kms, the amplitude range for the mirror method is 2--3.5, and the two Gaussians are centered on $\pm$200\,\kms\ with a 10\,\kms\ width (values are $\pm$250\,\kms, 1.15--1.4, $\pm$150\,\kms, and 10\,\kms\ for H$\beta$) - the values are slightly different from \citet{naz22}, explaining small changes in RV or EW values. The velocities provided between parentheses are corrected for the long-term trend (Fig. \ref{l161103}).
 }
\end{table*}

\begin{table*}
  \begin{center}
  \scriptsize
  \caption{EWs and RVs of H$\alpha$ and H$\beta$ measured for HD\,110432.
 \label{rv110432}}
  \begin{tabular}{lcccc|lcccc}
    \hline
mid-HJD & \multicolumn{2}{c}{H$\alpha$} & \multicolumn{2}{c|}{H$\beta$}& mid-HJD & \multicolumn{2}{c}{H$\alpha$} & \multicolumn{2}{c}{H$\beta$}\\
$-2\,450\,000$ & EW(\AA) & RV(\kms) & EW(\AA) & RV(\kms) & $-2\,450\,000$ & EW(\AA) & RV(\kms) & EW(\AA) & RV(\kms)\\
\hline
2805.543 &       &               &  -3.7 &   8.2               &9298.922$^b$ & -48.7 &  -1.6 (-3.2) \\                           
2806.462 &       &               &  -3.8 &   8.3               &9335.943$^b$ & -49.9 &  10.6 ( 7.9) \\                           
2807.463 &       &               &  -5.2 &   6.9               &9351.973$^b$ & -49.8 &  10.4 ( 7.2) \\                           
4898.702 &       &               &  -5.6 &  -3.5               &9370.950$^b$ & -48.7 &  13.2 ( 9.5) \\                           
6337.897$^x$ & -43.0 &  12.0         &  -4.1 &  26.5           &9382.973$^b$ & -47.9 &  16.4 (12.3) \\                           
7849.076$^b$ & -30.5 &   7.9         &       &                 &9403.929$^b$ & -50.6 &   7.1 ( 2.4) \\                           
8288.497 &       &               &  -5.1 &  11.9               &9407.003$^b$ & -50.8 &  11.2 ( 6.4) \\                           
8288.519 &       &               &  -5.3 &  12.5               &9446.959$^b$ & -47.1 &   1.9 (-4.1) \\                           
8481.822$^e$ & -46.3 &   4.5         &  -5.9 &   3.2           &9463.936$^b$ & -47.6 &   1.0 (-5.5) \\                           
8563.961$^b$ & -47.9 &  -8.1 ( 2.7)  &       &                 &9607.975$^b$ & -47.7 &  13.5 ( 2.8) \\                           
8564.950$^b$ & -47.1 &  -8.3 ( 2.5)  &       &                 &9680.903 & -48.5 &   7.7 (-5.2)  &  -6.0 &  11.0 (-6.1) \\       
8570.915$^b$ & -47.5 &  -9.7 ( 1.0)  &       &                 &9682.988$^b$ & -47.7 &   4.6 (-8.4) &  -5.9 &   8.9 (-8.3) \\    
8581.044$^b$ & -46.7 &  -8.7 ( 2.0)  &       &                 &9698.022$^b$ & -46.7 &   6.4 (-7.0) &  -5.9 &   9.7 (-8.4) \\    
8590.007$^b$ & -44.4 & -13.2 (-2.5)  &       &                 &9704.528 & -48.3 &   5.7 (-7.9)  &  -5.9 &  13.5 (-5.0) \\       
8598.973$^b$ & -44.3 & -14.0 (-3.3)  &       &                 &9726.636 &       &               &  -6.0 &  15.4 (-4.3) \\       
8598.987$^b$ & -43.6 & -15.1 (-4.4)  &       &                 &9740.674 & -48.3 &  11.8 (-2.8)  &  -5.9 &  15.5 (-5.1) \\       
8609.988$^b$ & -43.0 & -15.9 (-5.1)  &       &                 &9756.005$^b$ & -46.7 &  15.0 (-0.1) \\ 				 
8617.987$^b$ & -42.8 & -13.2 (-2.5)  &       &                 &9765.489 & -48.8 &  15.2 (-0.2)  &  -6.1 &  19.1 (-2.9) \\       
8623.933$^b$ & -42.1 &  -8.6 ( 2.1)  &       &                 &9765.886$^b$ & -47.9 &  20.6 ( 5.2) &  -6.1 &  21.2 (-0.8) \\    
8638.898$^b$ & -40.0 & -13.4 (-2.6)  &       &                 &9775.928$^b$ & -48.4 &  18.5 ( 2.8) \\ 				 
8650.992$^b$ & -41.3 & -12.6 (-1.9)  &       &                 &9782.599 & -50.1 &  18.8 ( 2.9)  &  -6.4 &  23.6 ( 0.5) \\       
8668.895$^b$ & -42.6 & -11.2 (-0.5)  &       &                 &9794.892$^b$ & -49.2 &  20.5 ( 4.2) &  -6.3 &  28.5 ( 4.7) \\    
8684.968$^b$ & -42.5 &  -9.1 ( 1.6)  &       &                 &9798.490 &       &               &  -6.5 &  27.4 ( 3.4) \\       
8693.947$^b$ & -42.3 &  -5.7 ( 5.1)  &       &                 &9812.494 & -50.8 &  17.8 ( 1.0)  &  -6.3 &  28.6 ( 3.8) \\       
8898.999$^b$ & -44.5 &  -8.5 ( 2.2)  &       &                 &10037.023$^b$ & -46.8 &  19.1 (-4.7) &  -5.5 &  39.4 (-6.5) \\   
8944.022$^b$ & -44.1 & -10.2 ( 0.5)  &       &                 &10047.953$^b$ & -47.7 &  24.7 ( 0.9) &  -5.3 &  43.2 (-2.7) \\   
8956.054$^b$ & -43.0 &  -5.8 ( 4.9)  &  -5.7 & -21.9 ( 3.3)    &10052.721$^b$ & -46.0 &  24.6 ( 0.8) &  -5.6 &  46.7 ( 0.9) \\   
8972.980$^b$ & -43.1 &  -9.2 ( 1.5)  &       &                 &10057.972$^b$ & -47.9 &  26.3 ( 2.5) &  -5.2 &  43.4 (-2.5) \\   
8984.002$^b$ & -47.2 & -11.1 (-0.3)  &       &                 &10075.939$^b$ & -46.6 &  22.3 (-1.5) &  -5.2 &  40.3 (-5.5) \\   
8995.921$^b$ & -43.7 &  -7.8 ( 2.9)  &       &                 &10087.996$^b$ & -46.2 &  22.8 (-1.0) &  -5.6 &  40.0 (-5.9) \\   
8998.002$^b$ & -41.9 & -13.4 (-2.7)  &       &                 &10105.046$^b$ & -46.4 &  19.8 (-4.0) &  -5.5 &  38.0 (-7.9) \\   
9015.908$^b$ & -42.0 & -12.1 (-1.4)  &       &                 &10142.901$^b$ & -45.4 &  19.4 (-4.4) &  -5.3 &  38.8 (-7.0) \\   
9017.996$^b$ & -43.8 &  -9.7 ( 1.1)  &  -6.2 & -23.6 (-2.0)    &10281.811$^b$ & -46.9 &  27.2 ( 3.4) &  -5.5 &  49.7 ( 3.9) \\   
9029.927$^b$ & -44.5 & -13.6 (-2.9)  &       &                 &10317.768$^b$ & -40.6 &  24.1 ( 0.3) &  -4.9 &  47.3 ( 1.5) \\   
9174.824 & -53.2 &   1.2 ( 3.2)  &  -6.7 &  -5.6 ( 6.8)        &10400.901$^e$ &       &               &  -5.4 &  46.3 ( 0.5) \\  
9187.840$^x$ &       &               &  -6.6 &  -5.4 ( 6.3)    &10401.854$^b$ & -42.7 &  22.3 (-1.5) &  -4.9 &  33.8 (-12.0) \\  
9187.842$^x$ & -54.2 &   2.5 ( 4.2)  &  -6.6 &  -3.9 ( 7.8)    &10436.573$^e$ &       &               &  -5.2 &  48.0 ( 2.1) \\  
9188.827 & -53.6 &   2.9 ( 4.6)  &  -6.8 &  -4.1 ( 7.5)        &10455.458$^e$ &       &               &  -5.2 &  56.0 (10.2) \\  
9205.744 & -54.0 &   2.2 ( 3.3)  &  -6.9 &  -7.8 ( 2.8)        &10469.465$^e$ & -42.5 &  29.4 ( 5.6)  &  -5.1 &  59.4 (13.6) \\  
9225.856 &       &               &  -7.0 & -14.8 (-5.4)        &10470.592$^e$ & -41.8 &  29.0 ( 5.2)  &  -5.1 &  59.4 (13.6) \\  
9242.689 &       &               &  -7.2 & -16.3 (-7.9)        &10640.808$^b$ & -37.8 &  23.0 (-0.8) &  -4.5 &  46.9 ( 1.0) \\   
9242.710 & -56.0 &  -4.4 (-4.3)  &  -7.2 & -16.8 (-8.3)        &10741.811$^b$ & -36.1 &  23.0 (-0.8) &  -4.1 &  48.4 ( 2.6) \\   
9248.994$^b$ & -51.5 &  -0.8 (-1.0) \\ 
\hline
  \end{tabular}
  \end{center}
\vspace*{-0.5cm}  
\tablefoot{The superscripts $^e$, $^x$ and $^b$ identify the Espresso, Xshooter and BeSS spectra, respectively, all other values correspond to UVES spectra. The RV values correspond to those derived using the double Gaussian method; typical RV error is about 1\,\kms\ for professional data and about 5\,\kms\ for amateur data. Saturation or insufficient data coverage explains the missing H$\alpha$ values for ESO spectra. The velocities provided between parentheses are corrected for the long-term trend (Fig. \ref{l110432}). For H$\alpha$, the velocity range for calculations is $\pm$400\,\kms, the amplitude range for the mirror method is 2--3.5, and the two Gaussians are centered on $\pm$200\,\kms\ with a 10\,\kms\ width (values are $\pm$400\,\kms, 1.2--1.6, $\pm$200\,\kms, and 10\,\kms\ for H$\beta$) - the H$\alpha$ values are slightly different from \citet{naz22}, explaining small changes in RV or EW values.}
\end{table*}

\begin{table*}
  \begin{center}
  \scriptsize
  \caption{EWs and RVs of H$\alpha$ and H$\beta$ measured for HD\,44458.
 \label{rv44458}}
  \begin{tabular}{lcccc|lcccc}
    \hline
mid-HJD & \multicolumn{2}{c}{H$\alpha$} & \multicolumn{2}{c|}{H$\beta$}& mid-HJD & \multicolumn{2}{c}{H$\alpha$} & \multicolumn{2}{c}{H$\beta$}\\
$-2\,450\,000$ & EW(\AA) & RV(\kms) & EW(\AA) & RV(\kms) & $-2\,450\,000$ & EW(\AA) & RV(\kms) & EW(\AA) & RV(\kms)\\
\hline
 4839.375$^b$ & -30.6 &  21.6 &       &              &9494.977 & -31.1 &  20.0 & -1.53 &  24.9 \\     
 4890.333$^b$ & -31.7 &  17.9 &       &              &9496.997 & -30.1 &  16.9 & -1.52 &  17.3 \\     
 5594.521$^b$ & -20.2 &  14.1 &       &              &9539.872 & -30.8 &  19.8 & -1.68 &  19.7 \\     
 5985.405$^b$ & -23.7 &  15.7 & -0.94 &  17.5        &9557.795 & -27.0 &  26.4 & -1.28 &  29.9 \\     
 6335.477$^b$ & -24.8 &  20.9 &       &              &9559.826 & -25.5 &  24.7 & -1.33 &  29.3 \\     
 6647.535$^b$ & -23.0 &  20.7 &       &              &9578.764 & -22.2 &  20.6 & -1.10 &  20.8 \\     
 6927.675$^b$ & -27.7 &  29.7 &       &              &9597.699 & -23.8 &  19.5 & -1.47 &  16.1 \\     
 6947.627$^b$ & -26.5 &  28.3 & -1.65 &  32.2        &9606.389$^b$ & -26.5 &  25.0 &  \\              
 7382.460$^b$ & -23.5 &  26.2 &       &              &9612.653 & -26.2 &  23.8 & -1.70 &  23.8 \\     
 7760.498$^b$ & -28.7 &  26.2 & -1.71 &  32.9        &9622.655 & -23.2 &  28.4 & -1.45 &  32.0 \\     
 7774.363$^b$ & -31.1 &  25.9 &       &              &9631.588 & -22.7 &  26.3 & -1.00 &  28.1 \\     
 7840.653$^b$ & -24.6 &  14.4 &       &              &9649.593 & -22.4 &  23.6 & -1.42 &  18.4 \\     
 8044.656$^b$ & -35.1 &  23.1 & -1.88 &  28.3        &9658.629 & -22.3 &  26.1 & -1.19 &  29.3 \\     
 8447.873 & -38.1 &  13.0 & -2.22 &  15.6            &9853.957 & -21.7 &  28.3 & -1.07 &  34.0 \\     
 8463.801 & -37.6 &  13.2 & -2.16 &  17.1            &9879.912 & -25.2 &  23.1 & -1.42 &  27.1 \\     
 8520.386$^b$ & -34.7 &  18.6 &       &              &9898.886 & -26.8 &  22.3 & -1.46 &  25.8 \\     
 8524.664 & -37.0 &  17.0 & -1.77 &  18.7            &9923.860 & -22.5 &  27.4 & -1.12 &  32.8 \\     
 8530.302$^b$ & -35.1 &  26.2 &       &              &9948.770 & -24.3 &  23.1 & -1.59 &  25.6 \\     
 8548.618 & -30.1 &  19.8 & -1.79 &  21.0            &9967.685 & -25.3 &  21.6 & -1.75 &  24.2 \\     
 8850.489$^b$ & -29.4 &  13.6 &       &              &9977.562 & -26.1 &  23.8 & -1.56 &  26.3 \\     
 8855.445$^b$ & -31.5 &  25.8 &       &              &9978.679 & -26.1 &  22.6 & -1.62 &  26.0 \\     
 8885.714 & -30.3 &  22.1 & -1.66 &  25.0            &9995.566 & -23.2 &  21.3 & -1.78 &  22.1 \\     
 8886.624 & -28.7 &  21.6 & -1.37 &  26.6           &10000.662 & -24.4 &  19.6 & -1.41 &  19.7 \\     
 8899.632 & -27.3 &  18.9 & -1.58 &  20.9           &10013.570 & -26.8 &  18.5 & -1.87 &  20.5 \\     
 8900.358$^b$ & -26.9 &  13.9 &       &             &10031.659$^b$ & -25.8 &  14.2 &  \\              
 8910.661 & -32.1 &  21.7 & -1.65 &  23.5           &10218.911 & -24.0 &  24.9 & -1.29 &  28.2 \\     
 8941.627$^b$ & -29.2 &  19.9 &       &             &10247.878 & -26.2 &  22.4 & -1.78 &  23.4 \\     
 8943.622 & -30.2 &  22.0 & -1.80 &  24.1           &10254.804$^b$ & -25.4 &  24.4 & -1.02 &  22.3 \\ 
 8946.585 & -26.4 &  21.5 & -1.33 &  20.9           &10265.872 & -25.4 &  26.0 & -1.35 &  27.0 \\     
 9123.953 & -31.2 &  14.4 & -1.66 &  17.2           &10288.806 & -24.8 &  26.0 & -1.19 &  33.0 \\     
 9131.869 & -31.6 &  14.3 & -1.85 &  13.6           &10307.793 & -27.8 &  22.6 & -1.49 &  24.7 \\     
 9133.879 & -31.8 &  15.8 & -1.89 &  19.6           &10325.749 & -25.1 &  21.3 & -1.45 &  25.2 \\     
 9165.846 & -32.9 &  16.8 & -1.92 &  16.4           &10342.744 & -27.3 &  18.9 & -1.74 &  20.9 \\     
 9195.791 & -26.0 &  17.3 & -1.75 &  20.1           &10344.687 & -26.8 &  17.7 & -1.48 &  18.1 \\     
 9217.819 & -27.4 &  21.1 & -1.90 &  23.8           &10349.664 & -28.0 &  16.2 & -1.74 &  21.6 \\     
 9241.688 & -27.5 &  21.4 & -1.84 &  23.0           &10362.695 & -23.9 &  18.6 & -1.46 &  28.7 \\     
 9247.647 & -28.5 &  21.7 & -1.54 &  24.4           &10380.706 & -20.2 &   9.0 & -0.83 &   8.2 \\     
 9301.588 & -30.6 &  25.2 & -1.57 &  26.2           &10398.589$^b$ & -20.9 &  15.1 & -1.14 &  30.5 \\ 
 9311.583 & -27.4 &  20.9 & -1.06 &  12.5           &10929.871$^b$ & -21.1 &  28.2 & -0.89 &  32.4 \\ 
\hline
  \end{tabular}
  \end{center}
\vspace*{-0.5cm}  
\tablefoot{The superscript $^b$ identifies the BeSS spectra, all other values correspond to TIGRE spectra. The RV values correspond to those derived using the double Gaussian method; typical RV error is about 1\,\kms\ for professional data and about 5\,\kms\ for amateur data. For H$\alpha$, the velocity range for calculations is $\pm$400\,\kms\ (Fig. \ref{l44458}), the amplitude range for the mirror method is 1.6--2.8, and the two Gaussians are centered on $\pm$200\,\kms\ with a 15\,\kms\ width (values are $\pm$700\,\kms, 1.07--1.15, $\pm$175\,\kms, and 15\,\kms\ for H$\beta$) - the values are slightly different from \citet{naz22} for H$\beta$, explaining small changes in RV or EW values.}
\end{table*}

\begin{table*}
  \begin{center}
  \scriptsize
  \caption{EWs and RVs of H$\alpha$ and RVs from blue metallic lines derived for HD\,119682.
 \label{rv119682}}
  \begin{tabular}{lc|lccc|lccc}
    \hline
mid-HJD & H$\alpha$ & mid-HJD & \multicolumn{2}{c}{H$\alpha$} & blue & mid-HJD & \multicolumn{2}{c}{H$\alpha$} & blue\\
$-2\,450\,000$ & EW(\AA) & $-2\,450\,000$ & EW(\AA) & RV(\kms) & RV(\kms) & $-2\,450\,000$ & EW(\AA) & RV(\kms) & RV(\kms)\\
\hline
 7879.037$^b$ &  -5.9 &  9185.837 &   1.6 &       & -7.6        & 9765.495 &  -8.4 & -18.0 & -12.9 \\     
 8573.977$^b$ &  -3.8 &  9205.842 &   1.6 &       &  2.6        & 9779.608 &  -8.2 & -16.4 &  -8.6 \\     
 8580.956$^b$ &  -3.5 &  9218.850$^x$ &   1.0 &       & -10.1   & 9779.975$^b$ &  -7.7 & -17.5 &       \\ 
 8676.913$^b$ &  -0.2 &  9242.753 &   1.2 &       &  -7.3       & 9794.496 &  -6.1 & -16.8 &  -7.9 \\     
 8682.954$^b$ &   0.1 &  9256.984$^b$ &   1.2 &       &         & 9794.928$^b$ &  -5.9 & -34.0 &       \\ 
 8698.991$^b$ &   0.3 &  9257.801 &   1.2 &       &   4.9       &10048.001$^b$ &  -9.6 & -22.4 &       \\ 
 8713.963$^b$ &   1.1 &  9274.662 &  -0.3 &       &  -8.6       &10058.073$^b$ &  -8.7 & -23.2 &       \\ 
 8719.949$^b$ &   1.2 &  9275.999$^b$ &  -0.5 &       &         &10075.988$^b$ &  -8.6 & -23.5 &       \\ 
 8721.940$^b$ &   1.0 &  9297.953$^b$ &  -0.9 &       &         &10088.953$^b$ &  -8.7 & -18.5 &       \\ 
 8724.925$^b$ &   0.2 &  9329.994$^b$ &  -0.7 &       &         &10104.953$^b$ &  -8.7 & -19.4 &       \\ 
 8728.935$^b$ &   1.0 &  9371.045$^b$ &   1.1 &       &         &10142.950$^b$ &  -8.8 & -25.2 &       \\ 
 8878.205$^b$ &  -2.0 &  9424.977$^b$ &   0.8 &       &         &10336.867$^b$ & -11.2 & -16.0 &       \\ 
 8952.069$^b$ &   1.1 &  9680.897 &  -4.1 & -24.0 &  -6.9       &10426.561$^e$ & -12.4 & -20.9 &  -2.5 \\ 
 8981.055$^b$ &   1.7 &  9698.049$^b$ &  -5.1 & -37.7 &         &10454.653$^e$ & -12.9 & -18.8 &  -0.8 \\ 
 9017.003$^b$ &   1.6 &  9701.804 &  -5.5 & -21.7 & -14.2       &10455.715$^e$ & -13.0 & -18.2 &  -0.3 \\ 
 9037.950$^b$ &   1.5 &  9715.832 &  -7.4 & -17.5 &  -6.1       &10481.579$^e$ & -12.7 & -22.6 &  -2.9 \\ 
 9044.996$^b$ &   1.5 &  9730.496 &  -8.3 & -16.6 &  -5.5       &10481.596$^b$ & -11.5 & -21.9 &       \\ 
 9049.936$^b$ &   1.3 &  9735.984$^b$ &  -8.2 & -16.5 &         &10848.594$^b$ & -16.7 & -23.9 &       \\ 
 9061.954$^b$ &   1.2 &  9744.505 &  -8.6 & -17.0 &  -2.8 \\    
 9080.974$^b$ &   0.6 &  9758.953$^b$ &  -7.9 & -16.6 &       \\
\hline
  \end{tabular}
  \end{center}
\vspace*{-0.5cm}  
\tablefoot{The superscripts $^x$, $^e$, and $^b$ identify the Xshooter, Espresso, and BeSS spectra, respectively, all other values correspond to UVES spectra. For BeSS data, the lower SNR and lower resolution data were not considered (see \citealt{naz22evol} for details). For H$\alpha$, the RV values correspond to those derived using the double Gaussian method when the profile showed clear emission; the other RVs come from a cross-correlation in the 4233--4324\AA\ range. Typical RV error is about 1\,\kms\ for H$\alpha$ and about 5\,\kms\ for correlation and amateur data. For H$\alpha$ (Fig. \ref{l119682}), the velocity range for calculations is $\pm$600\,\kms, the amplitude range for the mirror method is 1.15--1.35, and the two Gaussians are centered on $\pm$200\,\kms\ with a 10\,\kms\ width.}
\end{table*}

\begin{table*}
  \begin{center}
  \scriptsize
  \caption{Journal of the X-ray observations and results of the spectral fits.
 \label{journal}}
  \begin{tabular}{lcccccccccc}
    \hline
Facility & ObsID & mid-HJD & $N_{\rm H}^{ISM}$ & $N_{\rm H}^{add}$ & $kT$ & $norm$ & $\chi^2$(dof) & $F_{tot}^{obs}$ &  $F_{tot}^{unabs}$ & $HR$ \\
& & $-2\,450\,000$ & \multicolumn{2}{c}{($10^{22}$\,cm$^{-2}$)} & (keV) & (cm$^{-5}$) & &  \multicolumn{2}{c}{(erg\,cm$^{-2}$\,s$^{-1}$)} & \\
    \hline
\multicolumn{6}{l}{\vs}\\
\sw  & 43742/9         &  6172.753 &  0.47 & 0.07$\pm$0.21 & 20.0 (fixed) & (1.86$\pm$0.32)e-3 & 7.4(5)     & (2.88$\pm$0.44)e-12 & (3.29$\pm$0.50)e-12 & 3.30$\pm$0.84 \\
\sw  & 10533           &  8155.541 &  0.47 & 0.81$\pm$0.34 & 20.0 (fixed) & (4.48$\pm$0.61)e-3 & 4.5(10)    & (6.25$\pm$0.79)e-12 & (6.63$\pm$0.84)e-12 & 7.27$\pm$1.35 \\
\xmm & 0840910501/3531 &  8565.045 &  0.47 & 0.13$\pm$0.02 & 19.9$\pm$2.5 & (3.10$\pm$0.05)e-3 & 611.2(514) & (4.76$\pm$0.07)e-12 & (5.36$\pm$0.08)e-12 & 3.59$\pm$0.07 \\
\xmm & 0886090801/4361 & 10219.964 &  0.47 & 0.15$\pm$0.01 & 36.1$\pm$3.9 & (5.00$\pm$0.09)e-3 & 738.3(622) & (7.34$\pm$0.08)e-12 & (8.16$\pm$0.09)e-12 & 4.13$\pm$0.06 \\
\sw  & 89907           & 10817.698 &  0.47 & 0.41$\pm$0.62 & 20.0 (fixed) & (2.58$\pm$0.74)e-3 & 5.1(4)     & (3.78$\pm$0.67)e-12 & (4.11$\pm$0.73)e-12 & 5.10$\pm$1.36 \\
\hline
\multicolumn{6}{l}{\sgc}\\
\ch & 159   & 1413.750              & 1.08 &  0.21$\pm$0.27 & 8.5 (fixed) & (1.39$\pm$0.18)e-4 & 33.0/21 & (1.80$\pm$0.15)e-13 & (2.26$\pm$0.19)e-13 & 3.11$\pm$0.40 \\
\ch & 1230  & 1413.944              & 1.08 &  0.00$\pm$0.15 & 8.5 (fixed) & (9.06$\pm$0.95)e-5 & 17.3/14 & (1.21$\pm$0.14)e-13 & (1.63$\pm$0.19)e-13 & 2.23$\pm$0.42 \\
\ch & 160   & 1416.131              & 1.08 &  0.00$\pm$0.24 & 8.5 (fixed) & (1.28$\pm$0.18)e-4 & 8.1/10  & (1.70$\pm$0.30)e-13 & (2.30$\pm$0.41)e-13 & 2.23$\pm$0.58 \\
\ch & 158   & 1416.260              & 1.08 &  0.00$\pm$0.14 & 8.5 (fixed) & (8.88$\pm$1.15)e-5 & 3.7/6   & (1.18$\pm$0.20)e-13 & (1.60$\pm$0.27)e-13 & 2.24$\pm$0.61 \\
\ch & 161   & 1416.390              & 1.08 &  0.00$\pm$0.18 & 8.5 (fixed) & (1.59$\pm$0.19)e-4 & 11.1/12 & (2.12$\pm$0.25)e-13 & (2.87$\pm$0.34)e-13 & 2.23$\pm$0.39 \\
\ch & 1442  & 1498.248              & 1.08 &  0.37$\pm$0.36 & 8.5 (fixed) & (1.29$\pm$0.21)e-4 & 7.7/8   & (1.66$\pm$0.23)e-13 & (2.01$\pm$0.28)e-13 & 3.78$\pm$0.84 \\
\ch & 1443  & 1498.373              & 1.08 &  0.00$\pm$0.24 & 8.5 (fixed) & (1.47$\pm$0.21)e-4 & 15.5/11 & (1.98$\pm$0.30)e-13 & (2.68$\pm$0.41)e-13 & 2.25$\pm$0.50 \\
\ch & 1433  & 1498.532              & 1.08 &  0.00$\pm$0.11 & 8.5 (fixed) & (1.75$\pm$0.13)e-5 & 34.7/25 & (2.35$\pm$0.20)e-13 & (3.18$\pm$0.27)e-13 & 2.25$\pm$0.31 \\
\ch & 1434  & 1498.695              & 1.08 &  0.00$\pm$0.06 & 8.5 (fixed) & (1.97$\pm$0.19)e-4 & 12.1/18 & (2.66$\pm$0.33)e-13 & (3.59$\pm$0.45)e-13 & 2.26$\pm$0.50 \\
\xmm & 0122700101/0060 & 1642.242   & 1.08 &  0.10$\pm$0.06 & 8.5 (fixed) & (4.07$\pm$0.13)e-4 &142.8/177& (5.41$\pm$0.15)e-13 & (7.01$\pm$0.19)e-13 & 2.66$\pm$0.12 \\
\xmm & 0122700201/0061 & 1644.226   & 1.08 &  0.06$\pm$0.06 & 8.5 (fixed) & (4.81$\pm$0.18)e-4 &121.4/144& (6.43$\pm$0.21)e-13 & (8.44$\pm$0.28)e-13 & 2.51$\pm$0.13 \\
\xmm & 0122700301/0062 & 1646.228   & 1.08 &  0.00$\pm$0.06 & 8.5 (fixed) & (3.59$\pm$0.14)e-4 & 86.1/93 & (4.83$\pm$0.22)e-13 & (6.52$\pm$0.30)e-13 & 2.24$\pm$0.17 \\
\xmm & 0122700401/0064 & 1650.221   & 1.08 &  0.00$\pm$0.14 & 8.5 (fixed) & (2.51$\pm$0.22)e-4 & 22.6/22 & (3.38$\pm$0.32)e-13 & (4.56$\pm$0.43)e-13 & 2.26$\pm$0.36 \\
\xmm & 0122700501/0065 & 1652.211   & 1.08 &  0.18$\pm$0.16 & 8.5 (fixed) & (4.90$\pm$0.32)e-4 & 60.8/57 & (6.44$\pm$0.35)e-13 & (8.13$\pm$0.44)e-13 & 3.00$\pm$0.27 \\
\ch & 1717  & 1687.947              & 1.08 &  0.39$\pm$0.18 & 8.5 (fixed) & (3.10$\pm$0.30)e-4 & 9.1/17  & (3.97$\pm$0.34)e-13 & (4.79$\pm$0.41)e-13 & 3.87$\pm$0.51 \\
\ch & 1718  & 1688.045              & 1.08 &  0.22$\pm$0.20 & 8.5 (fixed) & (2.90$\pm$0.32)e-4 & 20.1/20 & (3.79$\pm$0.38)e-13 & (4.73$\pm$0.47)e-13 & 3.16$\pm$0.45 \\
\ch & 1719  & 1688.151              & 1.08 &  0.77$\pm$0.46 & 8.5 (fixed) & (2.98$\pm$0.45)e-4 & 9.9/11  & (3.65$\pm$0.46)e-13 & (4.22$\pm$0.53)e-13 & 5.49$\pm$1.27 \\
\ch & 1720  & 1688.256              & 1.08 &  0.42$\pm$0.27 & 8.5 (fixed) & (3.13$\pm$0.37)e-4 & 4.1/14  & (3.99$\pm$0.38)e-13 & (4.79$\pm$0.46)e-13 & 4.03$\pm$0.65 \\
\ch & 1721  & 1688.355              & 1.08 &  0.14$\pm$0.23 & 8.5 (fixed) & (2.97$\pm$0.34)e-4 & 12.1/15 & (3.92$\pm$0.40)e-13 & (5.01$\pm$0.51)e-13 & 2.82$\pm$0.47 \\
\ch & 1722  & 1688.454              & 1.08 &  0.00$\pm$0.15 & 8.5 (fixed) & (2.41$\pm$0.24)e-4 & 14.8/14 & (3.25$\pm$0.40)e-13 & (4.39$\pm$0.54)e-13 & 2.24$\pm$0.42 \\
\ch & 1723  & 1688.552              & 1.08 &  0.36$\pm$0.31 & 8.5 (fixed) & (2.02$\pm$0.28)e-4 & 15.8/11 & (2.59$\pm$0.33)e-13 & (3.15$\pm$0.40)e-13 & 3.75$\pm$0.72 \\
\ch & 1724  & 1688.651              & 1.08 &  0.00$\pm$0.16 & 8.5 (fixed) & (2.65$\pm$0.26)e-4 & 6.0/14  & (3.59$\pm$0.38)e-13 & (4.84$\pm$0.51)e-13 & 2.25$\pm$0.43 \\
\ch & 1725  & 1688.750              & 1.08 &  0.19$\pm$0.20 & 8.5 (fixed) & (3.59$\pm$0.36)e-4 & 19.7/18 & (4.71$\pm$0.45)e-13 & (5.94$\pm$0.57)e-13 & 3.04$\pm$0.40 \\
\ch & 1726  & 1688.848              & 1.08 &  0.39$\pm$0.22 & 8.5 (fixed) & (3.09$\pm$0.33)e-4 & 18.2/16 & (3.95$\pm$0.39)e-13 & (4.77$\pm$0.47)e-13 & 3.92$\pm$0.64 \\
\ch & 1727  & 1688.954              & 1.08 &  0.26$\pm$0.23 & 8.5 (fixed) & (3.05$\pm$0.34)e-5 & 17.4/17 & (3.96$\pm$0.37)e-13 & (4.90$\pm$0.46)e-13 & 3.34$\pm$0.51 \\
\ch & 1728  & 1689.059              & 1.08 &  0.25$\pm$0.22 & 8.5 (fixed) & (2.85$\pm$0.31)e-4 & 9.2/15  & (3.71$\pm$0.36)e-13 & (4.61$\pm$0.45)e-13 & 3.27$\pm$0.50 \\
\ch & 1729  & 1689.164              & 1.08 &  0.00$\pm$0.10 & 8.5 (fixed) & (2.38$\pm$0.21)e-4 & 16.0/14 & (3.21$\pm$0.35)e-13 & (4.33$\pm$0.47)e-13 & 2.26$\pm$0.43 \\
\ch & 1769  & 1730.610              & 1.08 &  0.49$\pm$0.21 & 8.5 (fixed) & (3.24$\pm$0.33)e-4 & 13.7/19 & (4.10$\pm$0.33)e-13 & (4.87$\pm$0.39)e-13 & 4.29$\pm$0.59 \\
\ch & 1770  & 1730.708              & 1.08 &  0.00$\pm$0.06 & 8.5 (fixed) & (2.76$\pm$0.20)e-4 & 12.9/19 & (3.72$\pm$0.38)e-13 & (5.02$\pm$0.51)e-13 & 2.26$\pm$0.34 \\
\ch & 1771  & 1730.801              & 1.08 &  0.07$\pm$0.23 & 8.5 (fixed) & (2.28$\pm$0.29)e-4 & 10.6/15 & (3.04$\pm$0.30)e-13 & (3.99$\pm$0.39)e-13 & 2.53$\pm$0.41 \\
\ch & 1780  & 1730.895              & 1.08 &  0.20$\pm$0.22 & 8.5 (fixed) & (2.28$\pm$0.29)e-4 & 6.7/11  & (3.00$\pm$0.33)e-13 & (3.77$\pm$0.41)e-13 & 3.08$\pm$0.56 \\
\ch & 1781  & 1730.988              & 1.08 &  0.00$\pm$0.19 & 8.5 (fixed) & (1.81$\pm$0.22)e-4 & 10.7/10 & (2.44$\pm$0.38)e-13 & (3.29$\pm$0.51)e-13 & 2.26$\pm$0.51 \\
\ch & 1782  & 1731.081              & 1.08 &  0.09$\pm$0.33 & 8.5 (fixed) & (1.68$\pm$0.28)e-4 & 5.9/9   & (2.23$\pm$0.36)e-13 & (2.90$\pm$0.47)e-13 & 2.64$\pm$0.56 \\
\ch & 1772  & 1731.182              & 1.08 &  0.22$\pm$0.32 & 8.5 (fixed) & (1.85$\pm$0.29)e-4 & 4.5/8   & (2.42$\pm$0.35)e-13 & (3.02$\pm$0.44)e-13 & 3.19$\pm$0.66 \\
\ch & 1773  & 1731.282              & 1.08 &  0.00$\pm$0.20 & 8.5 (fixed) & (1.78$\pm$0.21)e-4 & 11.2/9  & (2.40$\pm$0.35)e-13 & (3.23$\pm$0.47)e-13 & 2.25$\pm$0.57 \\
\ch & 1774  & 1731.375              & 1.08 &  0.00$\pm$0.17 & 8.5 (fixed) & (2.02$\pm$0.21)e-4 & 10.7/10 & (2.72$\pm$0.35)e-13 & (3.68$\pm$0.47)e-13 & 2.26$\pm$0.47 \\
\ch & 1775  & 1731.468              & 1.08 &  0.00$\pm$0.08 & 8.5 (fixed) & (1.97$\pm$0.19)e-4 & 12.4/10 & (2.65$\pm$0.30)e-13 & (3.58$\pm$0.41)e-13 & 2.25$\pm$0.42 \\
\ch & 1776  & 1731.562              & 1.08 &  0.00$\pm$0.11 & 8.5 (fixed) & (2.08$\pm$0.20)e-4 & 15.3/11 & (2.80$\pm$0.31)e-13 & (3.77$\pm$0.42)e-13 & 2.25$\pm$0.36 \\
\ch & 1777  & 1731.655              & 1.08 &  0.00$\pm$0.00 & 8.5 (fixed) & (1.74$\pm$0.18)e-4 & 8.8/10  & (2.34$\pm$0.24)e-13 & (3.16$\pm$0.32)e-13 & 2.25$\pm$0.34 \\
\ch & 1778  & 1731.749              & 1.08 &  0.00$\pm$0.13 & 8.5 (fixed) & (1.96$\pm$0.19)e-4 & 6.0/10  & (2.64$\pm$0.30)e-13 & (3.56$\pm$0.40)e-13 & 2.26$\pm$0.39 \\
\ch & 1779  & 1731.845              & 1.08 &  0.00$\pm$0.33 & 8.5 (fixed) & (1.15$\pm$0.20)e-4 & 22.1/7  & (1.55$\pm$0.27)e-13 & (2.09$\pm$0.36)e-13 & 2.25$\pm$0.61 \\
\ch & 1838  & 1789.607              & 1.08 &  0.46$\pm$0.60 & 8.5 (fixed) & (1.18$\pm$0.27)e-4 & 11.6/7  & (1.50$\pm$0.22)e-13 & (1.79$\pm$0.26)e-13 & 4.22$\pm$0.91 \\
\ch & 1839  & 1789.713              & 1.08 &  0.00$\pm$0.18 & 8.5 (fixed) & (7.74$\pm$1.05)e-5 & 5.3/5   & (1.04$\pm$0.19)e-13 & (1.41$\pm$0.26)e-13 & 2.25$\pm$0.68 \\
\ch & 1840  & 1789.811              & 1.08 &  0.12$\pm$0.42 & 8.5 (fixed) & (7.19$\pm$1.62)e-5 & 2.3/5   & (0.95$\pm$0.12)e-13 & (1.23$\pm$0.16)e-13 & 2.75$\pm$0.86 \\
\ch & 1554  & 2112.341              & 1.08 &  0.00$\pm$0.12 & 8.5 (fixed) & (2.74$\pm$0.22)e-4 & 17.9/24 & (3.69$\pm$0.34)e-13 & (4.98$\pm$0.46)e-13 & 2.25$\pm$0.31 \\
\ch & 2872  & 2531.484              & 1.08 &  0.69$\pm$0.31 & 8.5 (fixed) & (2.22$\pm$0.27)e-4 & 5.7/12  & (2.74$\pm$0.29)e-13 & (3.19$\pm$0.34)e-13 & 5.18$\pm$0.91 \\
\ch & 2873  & 2531.621              & 1.08 &  0.00$\pm$0.11 & 8.5 (fixed) & (1.62$\pm$0.14)e-4 & 9.1/15  & (2.18$\pm$0.23)e-13 & (2.95$\pm$0.31)e-13 & 2.25$\pm$0.39 \\
\ch & 4353  & 2775.347              & 1.08 &  0.00$\pm$0.14 & 8.5 (fixed) & (1.34$\pm$0.15)e-4 & 7.1/13  & (1.82$\pm$0.27)e-13 & (2.45$\pm$0.36)e-13 & 2.25$\pm$0.44 \\
\ch & 3692  & 2776.065              & 1.08 &  0.00$\pm$0.18 & 8.5 (fixed) & (1.02$\pm$0.13)e-4 & 7.3/7   & (1.38$\pm$0.24)e-13 & (1.86$\pm$0.32)e-13 & 2.26$\pm$0.53 \\
\ch & 3699  & 2952.947              & 1.08 &  0.20$\pm$0.41 & 8.5 (fixed) & (1.11$\pm$0.19)e-4 & 5.2/7   & (1.46$\pm$0.18)e-13 & (1.83$\pm$0.23)e-13 & 3.06$\pm$0.69 \\
\ch & 3700  & 2953.078              & 1.08 &  0.02$\pm$0.35 & 8.5 (fixed) & (7.95$\pm$1.51)e-5 & 6.1/6   & (1.07$\pm$0.19)e-13 & (1.43$\pm$0.25)e-13 & 2.34$\pm$0.64 \\
\ch & 5166  & 3079.497              & 1.08 &  0.00$\pm$0.00 & 8.5 (fixed) & (7.75$\pm$0.98)e-4 & 25.1/8  & (1.04$\pm$0.14)e-13 & (1.41$\pm$0.19)e-13 & 2.25$\pm$0.39 \\
\ch & 5165  & 3091.031              & 1.08 &  0.00$\pm$0.00 & 8.5 (fixed) & (1.13$\pm$0.14)e-4 & 9.9/8   & (1.53$\pm$0.21)e-13 & (2.06$\pm$0.28)e-13 & 2.26$\pm$0.40 \\
\ch & 5159  & 3306.131              & 1.08 &  0.00$\pm$0.11 & 8.5 (fixed) & (8.00$\pm$0.94)e-5 & 11.4/7  & (1.08$\pm$0.17)e-13 & (1.46$\pm$0.23)e-13 & 2.26$\pm$0.56 \\
\ch & 5158  & 3427.693              & 1.08 &  0.00$\pm$0.16 & 8.5 (fixed) & (6.91$\pm$1.11)e-5 & 3.2/3   & (0.93$\pm$0.22)e-13 & (1.26$\pm$0.29)e-13 & 2.25$\pm$0.80 \\
\ch & 6070  & 3427.819              & 1.08 &  0.94$\pm$0.76 & 8.5 (fixed) & (7.34$\pm$1.97)e-5 & 4.0/2   & (0.89$\pm$0.21)e-13 & (1.01$\pm$0.23)e-13 & 6.27$\pm$2.64 \\
\ch & 6071  & 3427.946              & 1.08 &  0.10$\pm$0.41 & 8.5 (fixed) & (6.79$\pm$1.46)e-5 & 2.4/4   & (0.90$\pm$0.18)e-13 & (1.17$\pm$0.23)e-13 & 2.66$\pm$0.81 \\
\ch & 6740  & 3788.122              & 1.08 &  0.41$\pm$0.49 & 8.5 (fixed) & (1.08$\pm$0.21)e-4 & 0.7/4   & (1.38$\pm$0.22)e-13 & (1.66$\pm$0.26)e-13 & 3.95$\pm$1.04 \\
\ch & 6741  & 3788.693              & 1.08 &  0.00$\pm$0.16 & 8.5 (fixed) & (1.22$\pm$0.14)e-4 & 9.1/9   & (1.64$\pm$0.23)e-13 & (2.21$\pm$0.31)e-13 & 2.25$\pm$0.49 \\
\ch & 8372  & 4246.075              & 1.08 &  0.25$\pm$0.30 & 8.5 (fixed) & (1.60$\pm$0.22)e-4 & 10.8/13 & (2.08$\pm$0.22)e-13 & (2.59$\pm$0.27)e-13 & 3.29$\pm$0.57 \\
\ch & 8371  & 4248.585              & 1.08 &  0.29$\pm$0.37 & 8.5 (fixed) & (1.74$\pm$0.26)e-4 & 9.4/11  & (2.26$\pm$0.25)e-13 & (2.78$\pm$0.31)e-13 & 3.46$\pm$0.65 \\
\ch & 16421 & 6786.924              & 1.08 &  0.00$\pm$0.26 & 8.5 (fixed) & (8.74$\pm$1.25)e-5 & 2.8/5   & (1.18$\pm$0.22)e-13 & (1.59$\pm$0.30)e-13 & 2.25$\pm$0.75 \\
\xmm & 0804250201/3173 & 7850.674   & 1.08 &  0.44$\pm$0.21 & 8.5 (fixed) & (1.76$\pm$0.15)e-4 & 22.2/30 & (2.24$\pm$0.15)e-13 & (2.68$\pm$0.18)e-13 & 4.10$\pm$0.49 \\
\xmm & 0890200101/3892 & 9283.588   & 1.08 &  0.00$\pm$0.29 & 8.5 (fixed) & (9.06$\pm$1.36)e-5 & 14.6/8  & (1.22$\pm$0.19)e-13 & (1.65$\pm$0.26)e-13 & 2.25$\pm$0.54 \\
\xmm & 0890200301/3892 & 9283.927   & 1.08 &  0.00$\pm$0.47 & 8.5 (fixed) & (6.47$\pm$1.42)e-5 & 8.1/3   & (0.87$\pm$0.18)e-13 & (1.18$\pm$0.25)e-13 & 2.25$\pm$0.78 \\
\ch & 26191 & 9632.455              & 1.08 &  0.00$\pm$0.41 & 8.5 (fixed) & (1.67$\pm$0.24)e-4 & 5.1/7   & (2.26$\pm$0.38)e-13 & (3.06$\pm$0.51)e-13 & 2.26$\pm$0.68 \\
\ch & 26192 & 9635.003              & 1.08 &  0.00$\pm$0.32 & 8.5 (fixed) & (9.98$\pm$1.36)e-5 & 3.9/6   & (1.34$\pm$0.21)e-13 & (1.81$\pm$0.28)e-13 & 2.26$\pm$0.75 \\
\hline
  \end{tabular}
  \end{center}
\end{table*}
\setcounter{table}{4}
 \begin{table*}
  \begin{center}
  \scriptsize
  \caption{Continued.}
  \begin{tabular}{lcccccccccc}
    \hline
Facility & ObsID & mid-HJD & $N_{\rm H}^{ISM}$ & $N_{\rm H}^{add}$ & $kT$ & $norm$ & $\chi^2$(dof) & $F_{tot}^{obs}$ &  $F_{tot}^{unabs}$ & $HR$ \\
& & $-2\,450\,000$ & \multicolumn{2}{c}{($10^{22}$\,cm$^{-2}$)} & (keV) & (cm$^{-5}$) & &  \multicolumn{2}{c}{(erg\,cm$^{-2}$\,s$^{-1}$)} & \\
    \hline
\multicolumn{6}{l}{\cl}\\
\ch & 159   & 1413.750              & 0.84 &  0.57$\pm$0.19 & 14. (fixed) & (3.40$\pm$0.29)e-4 & 38.6/35 & (4.55$\pm$0.33)e-13 & (5.14$\pm$0.37)e-13 & 5.38$\pm$0.69 \\
\ch & 1230  & 1413.944              & 0.84 &  0.20$\pm$0.12 & 14. (fixed) & (4.17$\pm$0.30)e-4 & 34.1/43 & (5.85$\pm$0.28)e-13 & (6.97$\pm$0.33)e-13 & 3.56$\pm$0.33 \\
\ch & 165   & 1414.103              & 0.84 &  0.49$\pm$0.14 & 14. (fixed) & (4.40$\pm$0.32)e-4 & 33.6/37 & (5.94$\pm$0.40)e-13 & (6.76$\pm$0.46)e-13 & 4.98$\pm$0.49 \\
\ch & 160   & 1416.131              & 0.84 &  0.66$\pm$0.18 & 14. (fixed) & (6.14$\pm$0.48)e-4 & 42.4/37 & (8.14$\pm$0.54)e-13 & (9.13$\pm$0.61)e-13 & 5.86$\pm$0.63 \\
\ch & 161   & 1416.390              & 0.84 &  0.59$\pm$0.17 & 14. (fixed) & (5.42$\pm$0.42)e-4 & 45.7/36 & (7.23$\pm$0.49)e-13 & (8.15$\pm$0.55)e-13 & 5.48$\pm$0.62 \\
\ch & 1441  & 1498.120              & 0.84 &  0.38$\pm$0.23 & 14. (fixed) & (3.35$\pm$0.36)e-4 & 25.0/21 & (4.65$\pm$0.42)e-13 & (5.36$\pm$0.48)e-13 & 4.53$\pm$0.61 \\
\ch & 1443  & 1498.373              & 0.84 &  0.80$\pm$0.16 & 14. (fixed) & (5.79$\pm$0.42)e-4 & 31.8/31 & (7.67$\pm$0.50)e-13 & (8.50$\pm$0.55)e-13 & 6.59$\pm$0.91 \\
\ch & 1433  & 1498.532              & 0.84 &  0.24$\pm$0.14 & 14. (fixed) & (3.83$\pm$0.29)e-4 & 42.4/41 & (5.41$\pm$0.38)e-13 & (6.37$\pm$0.45)e-13 & 3.83$\pm$0.37 \\
\ch & 1434  & 1498.695              & 0.84 &  0.29$\pm$0.10 & 14. (fixed) & (3.58$\pm$0.25)e-4 & 40.9/33 & (5.03$\pm$0.35)e-13 & (5.87$\pm$0.41)e-13 & 4.10$\pm$0.39 \\
\xmm & 0122700101/0060 & 1642.242   & 0.84 &  0.28$\pm$0.07 & 14. (fixed) & (5.14$\pm$0.18)e-4 &134.3/158& (7.22$\pm$0.21)e-13 & (8.44$\pm$0.25)e-13 & 4.02$\pm$0.20 \\
\xmm & 0122700401/0064 & 1650.221   & 0.84 &  0.45$\pm$0.06 & 14. (fixed) & (5.80$\pm$0.16)e-4 &209.3/232& (7.98$\pm$0.21)e-13 & (9.12$\pm$0.24)e-13 & 4.85$\pm$0.19 \\
\ch & 1716  & 1687.844              & 0.84 &  0.27$\pm$0.19 & 14. (fixed) & (3.81$\pm$0.38)e-4 & 25.1/22 & (5.36$\pm$0.50)e-13 & (6.28$\pm$0.59)e-13 & 3.98$\pm$0.50 \\
\ch & 1717  & 1687.947              & 0.84 &  0.39$\pm$0.19 & 14. (fixed) & (4.23$\pm$0.40)e-4 & 27.1/23 & (5.86$\pm$0.50)e-13 & (6.74$\pm$0.58)e-13 & 4.57$\pm$0.69 \\
\ch & 1719  & 1688.151              & 0.84 &  0.28$\pm$0.17 & 14. (fixed) & (4.36$\pm$0.39)e-4 & 13.0/23 & (6.13$\pm$0.44)e-13 & (7.17$\pm$0.51)e-13 & 4.01$\pm$0.51 \\
\ch & 1720  & 1688.256              & 0.84 &  0.24$\pm$0.22 & 14. (fixed) & (3.40$\pm$0.37)e-4 & 14.7/18 & (4.80$\pm$0.45)e-13 & (5.66$\pm$0.53)e-13 & 3.84$\pm$0.52 \\
\ch & 1721  & 1688.355              & 0.84 &  0.16$\pm$0.16 & 14. (fixed) & (4.21$\pm$0.39)e-4 & 22.9/23 & (6.01$\pm$0.52)e-13 & (7.20$\pm$0.62)e-13 & 3.44$\pm$0.41 \\
\ch & 1722  & 1688.454              & 0.84 &  0.44$\pm$0.24 & 14. (fixed) & (4.52$\pm$0.47)e-4 & 21.9/22 & (6.23$\pm$0.53)e-13 & (7.13$\pm$0.61)e-13 & 4.80$\pm$0.62 \\
\ch & 1724  & 1688.651              & 0.84 &  0.05$\pm$0.14 & 14. (fixed) & (4.31$\pm$0.36)e-4 & 12.5/26 & (6.25$\pm$0.48)e-13 & (7.74$\pm$0.59)e-13 & 2.89$\pm$0.36 \\
\ch & 1725  & 1688.750              & 0.84 &  0.60$\pm$0.23 & 14. (fixed) & (5.05$\pm$0.48)e-4 & 25.5/25 & (6.84$\pm$0.50)e-13 & (7.69$\pm$0.56)e-13 & 5.63$\pm$0.70 \\
\ch & 1726  & 1688.848              & 0.84 &  0.28$\pm$0.19 & 14. (fixed) & (4.20$\pm$0.40)e-4 & 6.0/22  & (5.90$\pm$0.48)e-13 & (6.91$\pm$0.56)e-13 & 4.01$\pm$0.54 \\
\ch & 1770  & 1730.708              & 0.84 &  0.32$\pm$0.22 & 14. (fixed) & (2.81$\pm$0.33)e-4 & 14.8/14 & (3.93$\pm$0.43)e-13 & (4.57$\pm$0.50)e-13 & 4.21$\pm$0.72 \\
\ch & 1771  & 1730.801              & 0.84 &  0.26$\pm$0.25 & 14. (fixed) & (2.51$\pm$0.33)e-4 & 10.1/12 & (3.54$\pm$0.48)e-13 & (4.15$\pm$0.56)e-13 & 3.92$\pm$0.68 \\
\ch & 1780  & 1730.895              & 0.84 &  0.17$\pm$0.25 & 14. (fixed) & (2.77$\pm$0.36)e-4 & 8.9/13  & (3.95$\pm$0.45)e-13 & (4.73$\pm$0.54)e-13 & 3.45$\pm$0.59 \\
\ch & 1781  & 1730.988              & 0.84 &  0.53$\pm$0.30 & 14. (fixed) & (2.66$\pm$0.35)e-4 & 13.1/12 & (3.62$\pm$0.44)e-13 & (4.10$\pm$0.50)e-13 & 5.26$\pm$0.93 \\
\ch & 1782  & 1731.081              & 0.84 &  0.19$\pm$0.22 & 14. (fixed) & (2.81$\pm$0.34)e-4 & 6.5/13  & (4.00$\pm$0.43)e-13 & (4.76$\pm$0.51)e-13 & 3.58$\pm$0.57 \\
\ch & 1779  & 1731.845              & 0.84 &  0.00$\pm$0.19 & 14. (fixed) & (1.96$\pm$0.25)e-4 & 9.7/9   & (2.86$\pm$0.48)e-13 & (3.60$\pm$0.60)e-13 & 2.67$\pm$0.62 \\
\xmm & 0122701001/0244 & 2009.325   & 0.84 &  1.17$\pm$0.49 & 14. (fixed) & (5.59$\pm$0.91)e-4 & 4.5/7   & (7.16$\pm$1.00)e-13 & (7.80$\pm$1.09)e-13 & 8.57$\pm$2.28 \\
\ch & 2873  & 2531.621              & 0.84 &  0.28$\pm$0.22 & 14. (fixed) & (3.50$\pm$0.38)e-4 & 22.6/19 & (4.92$\pm$0.51)e-13 & (5.75$\pm$0.60)e-13 & 4.04$\pm$0.64 \\
\ch & 4353  & 2775.347              & 0.84 &  0.71$\pm$0.20 & 14. (fixed) & (6.74$\pm$0.55)e-4 & 39.5/35 & (9.02$\pm$0.61)e-13 & (10.1$\pm$0.68)e-13 & 6.13$\pm$0.70 \\
\ch & 4354  & 2780.199              & 0.84 &  0.47$\pm$0.24 & 14. (fixed) & (4.17$\pm$0.43)e-4 & 22.7/25 & (5.72$\pm$0.48)e-13 & (6.52$\pm$0.55)e-13 & 4.93$\pm$0.65 \\
\ch & 3700  & 2953.078              & 0.84 &  0.12$\pm$0.20 & 14. (fixed) & (4.48$\pm$0.47)e-4 & 23.4/22 & (6.42$\pm$0.63)e-13 & (7.77$\pm$0.76)e-13 & 3.26$\pm$0.43 \\
\ch & 5166  & 3079.497              & 0.84 &  0.05$\pm$0.17 & 14. (fixed) & (2.85$\pm$0.29)e-4 & 35.0/21 & (4.13$\pm$0.38)e-13 & (5.12$\pm$0.47)e-13 & 2.88$\pm$0.45 \\
\ch & 5165  & 3091.031              & 0.84 &  0.26$\pm$0.28 & 14. (fixed) & (1.83$\pm$0.24)e-4 & 9.4/12  & (2.58$\pm$0.29)e-13 & (3.02$\pm$0.34)e-13 & 3.93$\pm$0.74 \\
\ch & 5159  & 3306.131              & 0.84 &  0.80$\pm$0.38 & 14. (fixed) & (2.65$\pm$0.38)e-4 & 7.3/11  & (3.51$\pm$0.51)e-13 & (3.89$\pm$0.57)e-13 & 6.63$\pm$1.45 \\
\ch & 5158  & 3427.693              & 0.84 &  0.18$\pm$0.22 & 14. (fixed) & (2.32$\pm$0.27)e-4 & 9.5/15  & (3.30$\pm$0.36)e-13 & (3.94$\pm$0.43)e-13 & 3.53$\pm$0.58 \\
\ch & 6070  & 3427.819              & 0.84 &  0.10$\pm$0.26 & 14. (fixed) & (2.43$\pm$0.31)e-4 & 15.7/15 & (3.49$\pm$0.40)e-13 & (4.26$\pm$0.49)e-13 & 3.14$\pm$0.51 \\
\ch & 6071  & 3427.946              & 0.84 &  0.07$\pm$0.21 & 14. (fixed) & (2.26$\pm$0.26)e-4 & 18.5/17 & (3.27$\pm$0.37)e-13 & (4.03$\pm$0.46)e-13 & 2.98$\pm$0.50 \\
\ch & 6740  & 3788.122              & 0.84 &  0.11$\pm$0.17 & 14. (fixed) & (3.66$\pm$0.34)e-4 & 20.7/23 & (5.26$\pm$0.45)e-13 & (6.38$\pm$0.55)e-13 & 3.20$\pm$0.43 \\
\ch & 6741  & 3788.693              & 0.84 &  0.41$\pm$0.19 & 14. (fixed) & (3.53$\pm$0.34)e-4 & 28.0/22 & (4.88$\pm$0.41)e-13 & (5.60$\pm$0.47)e-13 & 4.66$\pm$0.62 \\
\ch & 8371  & 4248.585              & 0.84 &  0.06$\pm$0.21 & 14. (fixed) & (2.02$\pm$0.24)e-4 & 25.5/16 & (2.92$\pm$0.29)e-13 & (3.61$\pm$0.36)e-13 & 2.94$\pm$0.53 \\
\xmm & 0804250201/3173 & 7850.674   & 0.84 &  0.11$\pm$0.17 & 14. (fixed) & (2.49$\pm$0.23)e-4 & 31.9/23 & (3.59$\pm$0.33)e-13 & (4.36$\pm$0.40)e-13 & 3.16$\pm$0.46 \\
\xmm & 0890200101/3892 & 9283.588   & 0.84 &  0.65$\pm$0.22 & 14. (fixed) & (4.12$\pm$0.33)e-4 & 19.7/31 & (5.55$\pm$0.40)e-13 & (6.21$\pm$0.45)e-13 & 5.86$\pm$0.69 \\
\xmm & 0890200301/3892 & 9283.927   & 0.84 &  0.29$\pm$0.25 & 14. (fixed) & (3.01$\pm$0.33)e-4 & 31.9/20 & (4.22$\pm$0.37)e-13 & (4.93$\pm$0.43)e-13 & 4.06$\pm$0.56 \\
\xmm & 0890200401/3892 & 9284.092   & 0.84 &  1.23$\pm$0.50 & 14. (fixed) & (5.92$\pm$0.85)e-4 & 4.7/9   & (7.54$\pm$0.90)e-13 & (8.20$\pm$0.98)e-13 & 8.93$\pm$2.01 \\
\xmm & 0890200601/3892 & 9284.420   & 0.84 &  0.35$\pm$0.17 & 14. (fixed) & (3.31$\pm$0.25)e-4 & 45.9/42 & (4.61$\pm$0.30)e-13 & (5.34$\pm$0.35)e-13 & 4.38$\pm$0.45 \\
\xmm & 0890200901/3892 & 9284.975   & 0.84 &  0.35$\pm$0.21 & 14. (fixed) & (3.35$\pm$0.30)e-4 & 33.1/27 & (4.67$\pm$0.33)e-13 & (5.40$\pm$0.38)e-13 & 4.35$\pm$0.51 \\
\hline
  \end{tabular}
  \end{center}
\vspace*{-0.5cm}  
\tablefoot{For \xmm\ observations, the Revolution number is provided after the ObsID. For the \sw\ data of \vs, the target ID is mentioned above: the full list of ObsIDs are 00043742001, 00043742002, 00043749001, 00043749002, 00043749003, and 00043749004 for the first epoch, 00010533001 and 00010533002 for the second one, and 00089907001 for the last one. Fluxes are provided in the total energy band (0.5--10.0\,keV), without and with correction for the interstellar absorption. Hardness ratios $HR$ are calculated as the ratios between the ISM-corrected fluxes in the hard (2.0--10.0\,keV) and soft (0.5--2.0\,keV) energy bands. }
\end{table*}

\end{document}